\UseRawInputEncoding
\documentclass[reprint, amsmath,amssymb, aps,superscriptaddress,nofootinbib]{revtex4-2}
\usepackage{wasysym}
\usepackage{amsmath}
\usepackage{graphicx}
\usepackage{tikz-feynman,contour}
\tikzfeynmanset{compat=1.1.0}
\tikzfeynmanset{/tikzfeynman/momentum/arrow shorten = 0.3}
\tikzfeynmanset{/tikzfeynman/warn luatex = false}
\usepackage{dcolumn}
\usepackage{bm}
\usepackage{multirow}
\usepackage{placeins}
\usepackage{subfigure}

\begin{document}

\title{Probing emergent QED in quantum spin ice via Raman scattering of phonons:\\ shallow inelastic scattering and pair production}

\author{Arnab Seth}
\affiliation{International Centre for Theoretical Sciences, Tata Institute of Fundamental Research, Bengaluru 560089, India}

\author{Subhro Bhattacharjee}
\affiliation{International Centre for Theoretical Sciences, Tata Institute of Fundamental Research, Bengaluru 560089, India}

\author{Roderich Moessner}
\affiliation{Max-Planck-Institut f\"{u}r Physik komplexer Systeme, N\"{o}thnitzer Strasse 38, 01187 Dresden, Germany}

\date{\today}

\begin{abstract}
{We present a new mechanism for Raman scattering {of phonons}, which is based on the {\it linear} magnetoelastic coupling present in non-Kramers magnetic ions. This provides a direct  coupling of Raman-active phonons to the magnet's quasiparticles. We propose to use this mechanism to probe the emergent  magnetic monopoles, electric charges, and photons of the emergent quantum electrodynamics (eQED)  of the U(1) quantum spin liquid known as quantum spin ice. Detecting this eQED in candidate rare-earth pyrochlore materials, or indeed signatures of topological magnetic phases more generally, is a challenging task. We show that the Raman scattering cross-section {of the phonons} directly yields relevant information, with the broadening of the phonon linewidth, which we compute,  exhibiting a characteristic  frequency dependence reflecting the two-particle density of states of the emergent excitations. Remarkably, we find that the Raman linewidth is sensitive to the details of the symmetry fractionalisation and hence can reveal information about the projective implementation of symmetry in the quantum spin liquid, thereby providing a diagnostic for a $\pi$-flux phase. The Raman scattering of the phonons thus provides a useful experimental tool to probe the fractionalisation in quantum spin liquids that turns out closely to mirror {pair production in quantum electrodynamics and} the deep inelastic scattering of quantum chromodynamics. Indeed, the difference to the latter is conceptual more than technical: the partons (quarks) emerge from the hadrons at high energies due to asymptotic freedom, while those in eQED arise from fractionalisation of the spins at low energies.
}  
\end{abstract}

\maketitle
%%%%%%%%%%%%%%%%%%%%%%%%
\section{\label{sec_1}Introduction}
The long-range entanglement present in quantum spin liquids (QSLs) lead to novel low-energy quasi-particles with fractionalised quantum numbers~\cite{anderson1987resonating,anderson1973resonating,PhysRevLett.86.1881,wen2002quantum,kitaev2006anyons,balents2010spin,wen2017colloquium,lee2008end,broholm2020quantum,Knolle_ARCMP_2019,takagi2019concept}. Experimental signatures of these fractionalised quasi-particles  can provide direct evidence of the underlying entanglement pattern that characterises the {\it quantum order} in the QSLs. However, detecting experimental signatures of such unconventional fractionalised excitations calls for an array of complementary experimental probes to collectively provide information about the QSL.

In this context, probing the spins through their coupling to phonons--via magnetoelastic interactions-- provides useful  spectroscopic insights into the physics of QSLs. An example of this is the  ultrasonic attenuation~\cite{lee2011,shiralieva2021magnetoelastic,PhysRevResearch.2.033180} and anomalies~\cite{rucl32021,PhysRevB.84.041102} of the acoustic phonons in QSLs. {Magnetoelastic interactions are also believed to play an important role in the large thermal Hall response observed in several correlated insulators including the pseudo-gap phase of lightly doped cuprates~\cite{grissonnanche2019giant} and the magnetic-field induced paramagnetic phase of the honeycomb magnet $\alpha$-RuCl$_3$~\cite{kasahara2018unusual,nasu2016fermionic,banerjee2016proximate,banerjee2017neutron,PhysRevLett.121.147201,vinkler2018approximately,kasahara2018majorana,yamashita2020sample,yokoi2021half,czajka2021oscillations}.}

A related probe for the spin physics are optical phonons, via infrared and Raman scattering experiments where  phonon energy and linewidth encode such effects~\cite{pal2020probing, rucl32021, sandilands, LiIrO32016, satoru, natalia2021}. {Notably, such phonon spectroscopy can sensitively detect  magnetic, superconducting, or charge-density wave ordering, as well as couples to the resultant low-energy quasi-particles in these conventional phases~\cite{luthi2007physical,toth2016electromagnon,aynajian2008energy}.} In the simplest QSLs, however, symmetries are not spontaneously~\cite{anderson1973resonating} broken and  the nature of phonon renormalisation, at low temperatures, is governed by the properties of fractionalised excitations of the QSLs which provide  additional scattering channels for the phonons. This is expected, in particular, to lead to an anomalous broadening of the phonon linewidth at low temperatures whose characterisation can then reveal important information regarding the QSL excitations.

The spin-phonon effects are expected to be particularly strong in spin-orbit coupled magnets where the magnetic moment is sensitive to the real space geometry due to an interlocking of spin and real space \cite{bhattacharjee2012spin,witczak2014correlated,NussinovJeroenRMP,HKK_kitaev}. Indeed such spin-phonon coupling has recently been explored both experimentally and theoretically in candidate Kitaev QSLs such as $\alpha$-RuCl$_3$~\cite{rucl32021,metavitsiadis2021optical,nasu2016fermionic}, Cu$_2$IrO$_3$~\cite{pal2020probing}, $\beta$- and $\gamma$-Li$_2$IrO$_3$~\cite{LiIrO32016,PhysRevB.92.094439} etc. In particular, for Cu$_2$IrO$_3$~\cite{pal2020probing},  the anomalous broadening of the phonon peaks and frequency softening at low temperatures is accounted for by the low-energy Majorana fermions that the spin fractionalises into~\cite{kitaev2006anyons}.  

Another equally interesting family of spin-orbit coupled frustrated magnets are obtained in the rare-earth pyrochlores with magnetic moments  sitting on a three-dimensional network of corner sharing tetrahedra, leading to frustrated spin-spin interactions. These so-called spin ice systems~\cite{BramwellGingras2001,castelnovo2012, gingrass2014, rau, bramwell2020,ramirez1999zero,PhysRevLett.79.2554,erfanifam,yb2016,RevModPhys.82.53,PhysRevB.93.064406,PhysRevLett.119.057203,PhysRevX.1.021002,gingrass2014,PhysRevLett.109.097205,PhysRevB.87.184423,kimura,prhfo,tbtioprincep,HoTbDy,ruminy2014,HoTbDy} are primary candidates to realise both classical cooperative paramagnets~\cite{MoessnerChalker98,chalker,PhysRevB.71.014424,henley2010coulomb} as well as QSLs~\cite{hermele,ramandimer2005,PhysRevLett.91.167004,gingrass2014,savary,sungbin2012,PhysRevLett.108.067204,benton,PhysRevLett.100.047208,PhysRevLett.112.167203,PhysRevB.92.144417,chang2012higgs,PhysRevLett.115.077202}. The magnetic moments result from a very intricate interplay of inter-orbital Coulomb repulsion, atomic spin-orbit coupling, and crystal field effects. 

A rather extreme example of interplay between several competing interactions is seen in an interesting subset among the pyrochlores which are the so-called {\it non-Kramers}  spin ice materials such as Pr$_2$Zr$_2$O$_7$~\cite{kimura,satoru,PhysRevB.94.165153}, Pr$_2$Hf$_2$O$_7$~\cite{prhfo}, Tb$_2$Ti$_2$O$_7$~\cite{tbtioprincep,HoTbDy,ruminy2014}, Ho$_2$Ti$_2$O$_7$~\cite{HoTbDy} etc. In these pyrochlore magnets, the low-energy spin-1/2 magnetic moments arise from even-electron wave functions~\cite{PZO_doublet,tbtioprincep,HoTbDy}. The degeneracy of such a non-Kramers doublet is protected by lattice symmetry, the $D_{3d}$ symmetry at the pyrochlore lattice site, instead of the usual time reversal symmetry for Kramers doublets. Therefore under time reversal symmetry, $\mathcal{T}$, the transformation of the low-energy doublets, $s^\alpha$ ($\alpha=x,y,z$), made out of spin-orbit coupled wave functions is given by
\begin{align}
    \mathcal{T}~:~\{s^x,s^y,s^z\}\rightarrow \{s^x,s^y,-s^z\}
    \label{eq_nonkramers}
\end{align}
This is in stark difference from the usual Kramers case as realised in, {\it e.g.}, Dy$_2$Ti$_2$O$_7$ among others, where all the components of the resultant spin-1/2 are odd under time reversal.

{The non-trivial implementation of time reversal symmetry as in Eq. \ref{eq_nonkramers} opens up the possibility of using experimental probes which are complementary to the conventional ones. For example, the transformation in Eq. \ref{eq_nonkramers} immediately suggests that the transverse components $\{s^x, s^y\}$ can linearly couple to the lattice vibrations of the appropriate space-group symmetry (see Eq. \ref{eq_Eg coupling} and \ref{eq_T2g coupling}) such that this linear coupling makes the above materials ideal candidates to explore the spin physics through the spin-phonon coupling in vibrational IR/Raman spectroscopy of the relevant phonons. The issue assumes particular importance in the context of QSLs since the spin-spin interactions in several of these non-Kramers pyrochlores, such as Pr$_2$Zr$_2$O$_7$~\cite{Comsatoru}, can possibly stabilise a $U(1)$ QSL with gapless emergent photons and gapped bosonic electric and magnetic monopoles~\cite{hermele,ramandimer2005,PhysRevLett.91.167004,gingrass2014,savary,sungbin2012,PhysRevLett.108.067204,benton,PhysRevLett.100.047208,PhysRevLett.112.167203,PhysRevB.92.144417,chang2012higgs,PhysRevLett.115.077202}-- the so-called Quantum spin ice.} \footnote{We note that there are two different assignments of gauge charges found in the literature. In the first assignment and the one that we use here, the magnetic monopoles of a QSL are obtained by violations of the ice-rule on a tetrahedron which are obtained by spin-flips. The electric charges, on the other hand, are the point defects of the compact $U(1)$ gauge field~\cite{sondhi,benton}. In the other convention, the violations of the ice-rule give rise to electric charges (often referred to as spinons in the associated literature) while the point defects associated with the gauge field are dubbed as magnetic monopoles~\cite{hermele}.}

In this paper, we show that indeed such a linear coupling can lead to characteristic experimental signatures of the emergent gauge charges and photons in vibrational Raman spectroscopy of a non-Kramers quantum spin ice, such as those proposed for Pr$_2$Zr$_2$O$_7$. We show that such linear couplings give rise to prominent new interaction channels between the phonon and all the three emergent excitations of the U(1) QSL-- the emergent gapped electric and the magnetic charges as well as the gapless photons. These interactions provide new scattering channels for phonons to decay into and lead to an anomalous broadening of the Raman peaks in the low-temperature regime. Remarkably, as we show,  such Raman signatures are sensitive to the non-trivial symmetry implementation on the emergent degrees of freedom-- the details of the projective representation of the symmetry group~\cite{wen2002quantum} under which the low-energy fractionalised excitations of the QSL transform. In particular, in the context of the quantum spin ice, we discuss the two cases of zero and $\pi$-flux. While in the former, the magnetic monopoles do not see any electric flux, in the latter they see an electric $\pi$-flux through every hexagonal plaquette. As a result, the magnetic monopoles in the $\pi$-flux phase transform under the non-trivial magnetic space group, as opposed to the zero-flux phase, with the magnetic monopoles transforming projectively under lattice translation. The resultant effects for both the QSLs are very different from the phonon renormalisation due to anharmonic contributions or magnetic ordering, and hence might present important signatures of the fractionalisation and the emergent gauge field. 

\begin{figure}
\centering
\subfigure[High energy pair production]{
\begin{tikzpicture}
    \begin{feynman}
    \vertex (a);
    \vertex [right=of a] (b);
    \vertex [above right=of b] (c1);
    \vertex [below right=of b] (c2);
    
    \diagram* {
    (a)-- [gluon, edge label'=\(photon\)] (b)-- [fermion, edge label'=\(e^-\)] (c1), (c2) -- [fermion, edge label'=\(e^+\)] (b) 
     };
    \end{feynman}
    \end{tikzpicture}
}
    \subfigure[Deep Inelastic scattering of hadrons]{
    \begin{tikzpicture}
    \begin{feynman}[every blob={/tikz/fill=gray!30,/tikz/inner sep=2pt}]
    \vertex (a);
    \vertex [blob, right= of a] (b) {};
    \vertex [above right=0.35cm of b] (f1);
    \vertex [right=0.35cm of b] (f2);
    \vertex [below right=0.35cm of b] (f3);
    \vertex [above right=of f1] (c1);
    \vertex [right=of f2] (c2) {q};
    \vertex [right=of f3] (c3) {~q};
    \vertex [right=of c1] (d1) {q};
    \vertex [above left=of c1] (e1);
    \vertex [left=of e1] (e2);
    \vertex [above right=of e1] (e3);
    
    \diagram*[scale=1, transform shape] {
    (a) -- [fermion,red,thick, edge label'=\(hadron\)] (b), (f1) -- [fermion,edge] (c1),
    (f2) -- [fermion,edge] (c2), (f3)--[fermion,edge] (c3), (c1)--[fermion, edge] (d1), (e1)--[gluon, edge label'=\(photon\)] (c1), (e2)--[fermion, blue,edge label'=\(lepton\)] (e1)--[fermion,blue,edge] (e3)
    };
    \end{feynman}
    \draw [line width=0.5mm] (b) --  (f2);
    \end{tikzpicture}
    }
    \subfigure[Pair Production in QSL]{
    \includegraphics[scale=0.5]{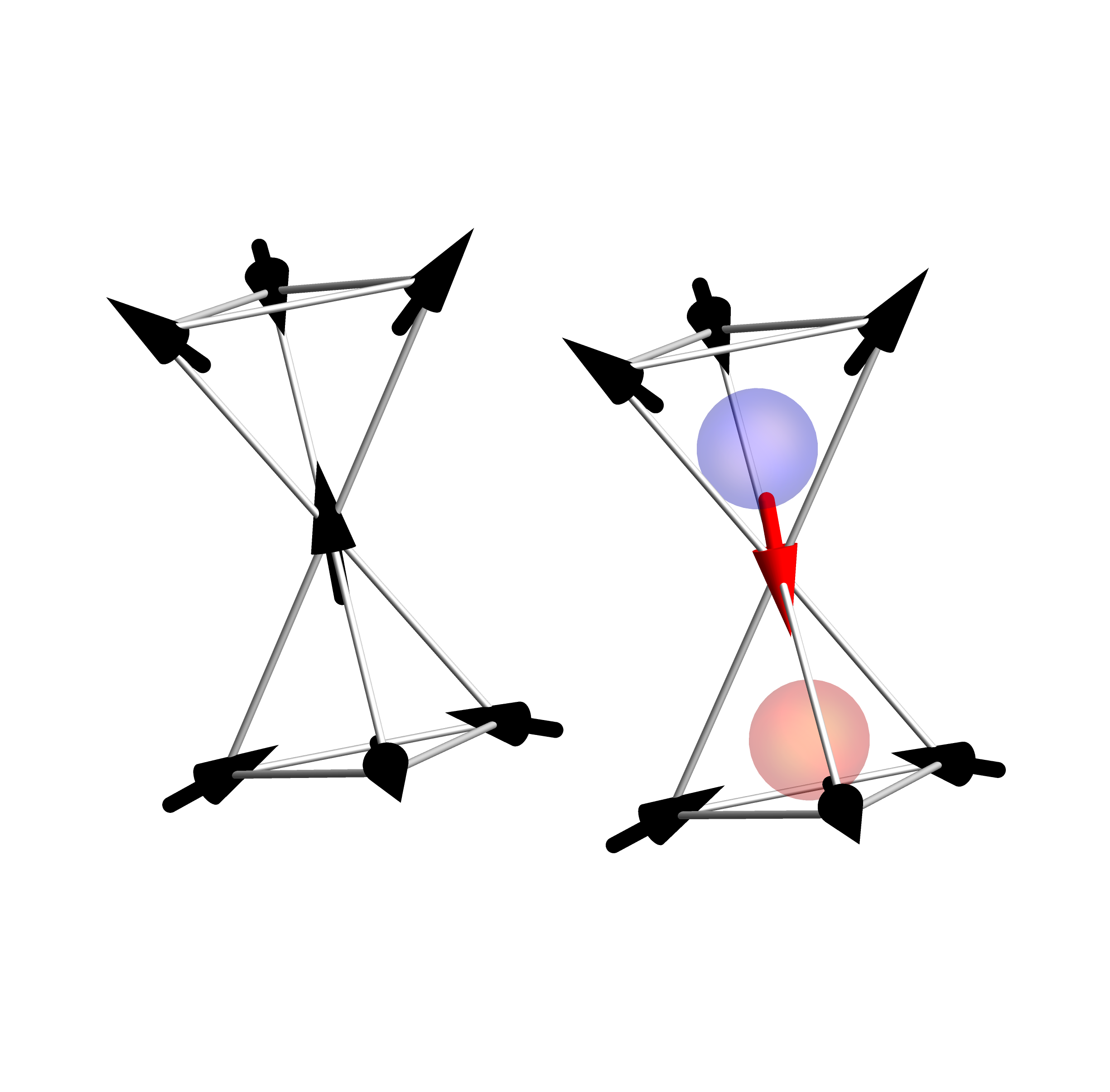}
    }\hspace*{-3.15cm}
    \subfigure{
    \begin{tikzpicture}
    \begin{feynman}
    \vertex (a);
    \vertex [above=1.3cm of a] (a1);
    \vertex [above=2.0cm of a] (a11);
    \vertex [left=1.0cm of a11] (a12);
    \vertex [right=0.5cm of a1] (a2);
    \vertex [above=1.2cm of a1] (a3);
    \vertex [right=0.5cm of a3] (a4);
    \vertex [right= of a] (b);
    \vertex [above=1.3cm of b] (b1);
    \vertex [right=2cm of b1] (x1);
    \vertex [below=of x1] (y1);
    \vertex [above right=0.35cm of b] (f1);
    \vertex [right=0.35cm of b] (f2);
    \vertex [below right=0.35cm of b] (f3);
    \vertex [above right=of a4] (c1);
    \vertex [right=of f2] (c2);
    \vertex [right=of f3] (c3);
    \vertex [right=of c1] (d1);
    \vertex [above=of d1] (d2);
    \vertex [above left=of c1] (e1);
    \vertex [left=of e1] (e2);
    \vertex [above right=of e1] (e3);
    \vertex [left=of a3] (p1);
    \vertex [below=0.5cm of p1] (p2);
    
    \diagram* {
    (a2) -- [dashed,red,thick, edge label=\(monopole\)] (y1), (a4) --[dashed,thick, edge label'=\(monopole\)] (d2), (a12)--[dotted, thick, edge label'=\(phonon\)
    ] (a11)
    };
    \end{feynman}
    \draw [white,line width=0.5mm] (a) --  (a);
    %\draw [line width=0.5mm] (p2) --  (a);
    \end{tikzpicture}
    }
    \addtocounter{subfigure}{-1}
    \subfigure[Shallow inelastic scattering in QSI]{
    \includegraphics[scale=0.5]{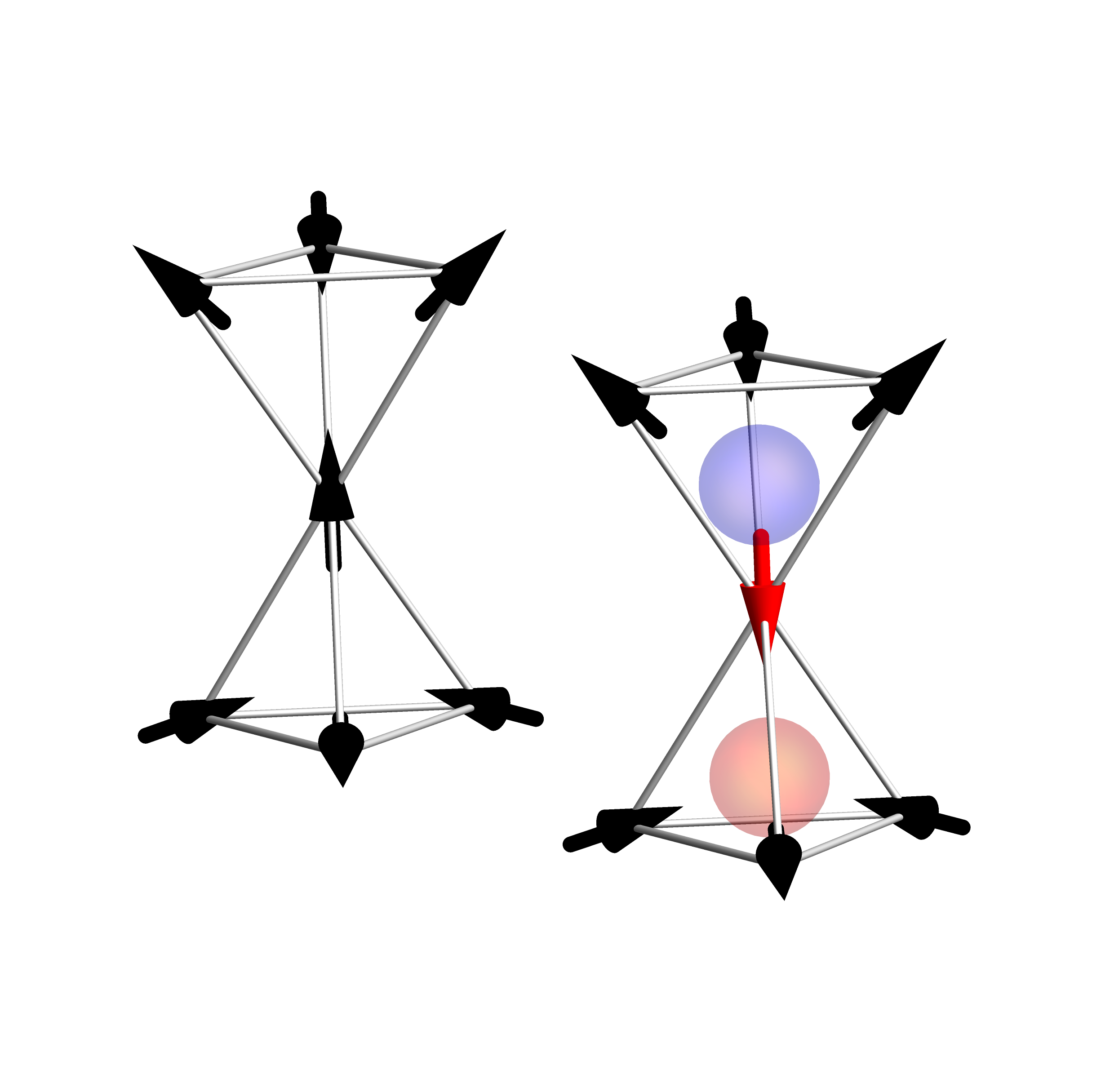}
    }\hspace*{-3.15cm}
    \subfigure{
    \begin{tikzpicture}
    \begin{feynman}
    \vertex (a);
    \vertex [above=1.3cm of a] (a1);
    \vertex [right=0.5cm of a1] (a2);
    \vertex [above=1.2cm of a1] (a3);
    \vertex [right=0.5cm of a3] (a4);
    \vertex [right= of a] (b);
    \vertex [above=1.3cm of b] (b1);
    \vertex [right=2cm of b1] (x1);
    \vertex [above right=0.35cm of b] (f1);
    \vertex [right=0.35cm of b] (f2);
    \vertex [below right=0.35cm of b] (f3);
    \vertex [above right=of a4] (c1);
    \vertex [right=of f2] (c2);
    \vertex [right=of f3] (c3);
    \vertex [right=of c1] (d1);
    \vertex [above left=of c1] (e1);
    \vertex [left=of e1] (e2);
    \vertex [above right=of e1] (e3);
    \vertex [left=of a3] (p1);
    \vertex [below=0.5cm of p1] (p2);
    
    \diagram* {
    (a2) -- [dashed,red,thick, edge label=\(monopole\)] (x1), (a4) -- [dashed,thick,edge] (c1)--[dashed,thick, edge label'=\(monopole\)] (d1), (e1)--[dotted, thick, edge label=\(phonon\)] (c1), (e2)--[gluon, edge label'=\(photon\)] (e1)--[gluon, edge] (e3)
    };
    \end{feynman}
    \draw [white,line width=0.5mm] (a) --  (a);
    %\draw [line width=0.5mm] (p2) --  (a);
    \end{tikzpicture}
    }
    \caption{
    {\bf Correspondence of scattering diagrams}: (a) The high energy photon can lead to creation of a positron and an electron via pair production, (b) In deep inelastic scattering, a photon emitted from a lepton scatters off a parton, a quark $q$, contained in the hadron, a $q\bar{q}$-pion as a free particle at high energies, (c) In non-Kramers quantum spin ice, the phonon flips a spin and creates two magnetic monopoles of opposite charges and, (d) In shallow inelastic scattering, an optical phonon emitted from a photon  scatters off a parton, a magnetic monopole or electric charge, emerging from the spin degrees of freedom from fractionalisation at low energies. 
    }
    \label{fig_shallowvertex}
\end{figure}
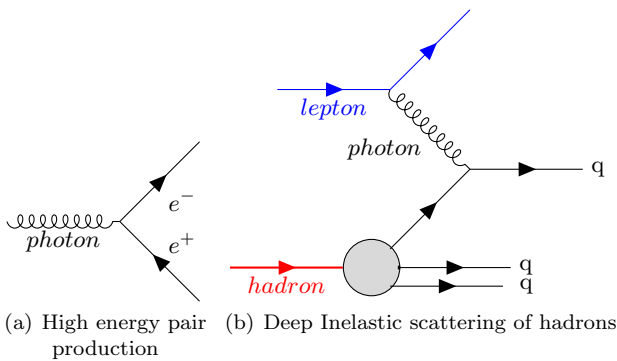
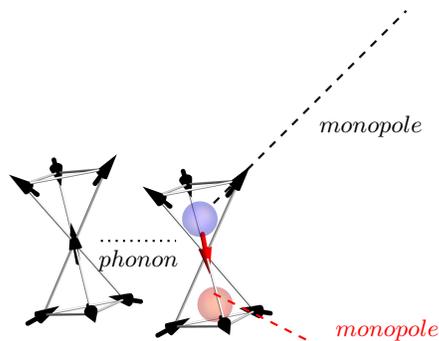
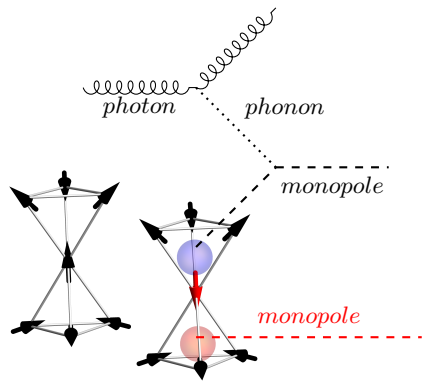

{It turns out that probing the low-energy fractionalised excitations of the QSL via the Raman/infrared scattering of the phonons is  quite similar to-- (a) high energy pair production (Fig. \ref{fig_shallowvertex}(a)), and, (b) the {\it deep inelastic scattering}~\cite{feynman1988behavior,PhysRev.185.1975} of quarks in quantum chromodynamics (QCD) by the leptons as described by the standard model of high-energy particle physics (Fig.~\ref{fig_shallowvertex}(b)). The corresponding two relevant vertices are shown side-by-side in Fig.~\ref{fig_shallowvertex}(c) and (d) respectively. In QCD, the quarks become asymptotically free at high energies and the high energy lepton can then probe them on sub-hadron length-scales~\cite{PhysRevLett.23.930,PhysRevLett.23.935}. In a QSL, however, the non-trivial entanglement leading to fractionalised novel excitations is a low-energy/long-wavelength emergent phenomenon which the phonons can probe via ``shallow" inelastic scattering. In particular, we show below while the first of the two processes dominate for the zero-flux QSL, the latter produces important low signatures of the momentum fractionalisation in the $\pi$-flux case. While our work describes such shallow inelastic scattering of an eQED in the context of the  quantum spin ices, it  readily generalises to other QSLs, and to probe magnetic excitations in quadrupolar systems more broadly.}

We note that the above vibrational Raman signatures of the fractionalisation on the phonons are different from the Loudon-Fleury type of Raman scattering where the external photon scatters directly from the charge fluctuation in the Mott insulating phase~\cite{shastry,PhysRevLett.113.187201,rau2017,PhysRevB.81.024414,PhysRev.166.514}. This kind of coupling has been explored by Cepas et. al. in the context of Kagome spin liquids~\cite{PhysRevB.77.172406}, and more pertinent for us, Fu et al. in the context of generic U(1) quantum spin ice~\cite{rau2017}. These studies already indicate several anomalous peaks in the Raman intensity profile due to the presence of new scattering channels in the QSL phase invariably indicating magnetic monopoles and gauge excitations of the quantum spin ice. However, due to the localized nature of the \textit{4f}-orbitals of Pr$^{3+}$, the scattering via charge fluctuation is significantly suppressed and the Raman probe mediated by the phonons can then provide dominant signatures of the novel excitations of the U(1) QSL phase in non-Kramers quantum spin ice. In fact, as we show here, even if the phonons are not at  resonance with the emergent excitations, the linear magnetoelastic coupling can mediate an effective Loudon-Fleury~\cite{PhysRev.166.514,shastry} type of coupling between emergent QSL excitations and the external Raman photons that form the leading contribution in non-Kramers material realisations of quantum spin ice.

We start with a brief overview of our results before delving into the details.

\subsection{Overview of the results}

The non-Kramers nature of the low-energy doublet in materials such as Pr$_2$Zr$_2$O$_7$ restricts the form of the low-energy spin-spin interactions (Eq. \ref{eq_spinham}) of the non-Kramers doublets as we briefly summarise in Sec. \ref{sec_2}. A further fallout of the unusual implementation of the time reversal symmetry is that the time reversal-even transverse spin components (Eq. \ref{eq_nonkramers}) can couple linearly to the Raman active ${\bf e_g}$ and ${\bf t_{2g}}$ phonons (Eqs. \ref{eq_Eg coupling} and \ref{eq_T2g coupling}) as discussed in Sec. \ref{sec_3}. These linear couplings form the leading order terms that couple the lattice modes with the spins with the latter apparently forming a U(1) QSL state-- the quantum spin ice-- over a sizeable parameter regime. Sec. \ref{sec_4} gives a brief review of this quantum spin ice phase  and their fractionalised low-energy excitations-- the gapped  bosonic electric and magnetic gauge charges and the gapless emergent photons. These excitations are captured via a mean-field description of the parton decomposition of spins leading to a lattice gauge theory. The microscopic couplings of the rare-earth pyrochlore magnets lead to a natural energy-scale separation between the higher energy magnetic sector and lower energy electric sector in quantum spin ice. 

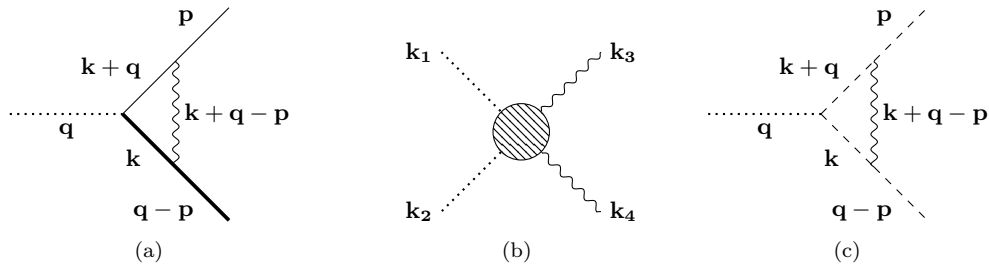
\begin{figure*}
\centering
\subfigure[]{
    {\begin{tikzpicture}
    \begin{feynman}
    \vertex (a);
    \vertex [right= of a] (b);
    \vertex [above right=1cm of b] (f1);
    \vertex [below right=1cm of b] (f2);
    \vertex [above right=1cm of f1] (f3);
    \vertex [below right=1cm of f2] (f4);    
    \diagram* {
    (a) -- [dotted, thick, edge label'=\(\mathbf{q}\)] (b) -- [ edge label=\(\mathbf{k+q}\)] (f1)--[ edge label=\(\mathbf{p}\)](f3),
    (b)-- [ edge label'=\(\mathbf{\textcolor{black}{k}}\)] (f2)--[ edge label'=\(\mathbf{\textcolor{black}{q-p}}\)](f4),
    (f1)--[boson, edge label=\(\mathbf{k+q-p}\)](f2)
    };
    \end{feynman}
    \draw [line width=0.5mm] (b) --  (f2) -- (f4);
    \end{tikzpicture}}
    }\hspace*{1cm}
    \subfigure[]{
 {\begin{tikzpicture}
    \begin{feynman}
    \vertex [blob] (a) {};
    \vertex [above right=of a, label=right:\({\bf k_3}\)] (f1);
    \vertex [below right=of a, label=right:\({\bf k_4}\)] (f2);
    \vertex [above left=of a, label=left:\({\bf k_1}\)] (f3);
    \vertex [below left=of a, label=left:\({\bf k_2}\)] (f4); \diagram* {
    (f3) -- [dotted, thick] (a) -- [dotted, thick] (f4), (f1)--[boson](a)-- [boson] (f2)
        };
    \end{feynman}
    \end{tikzpicture}}
    }\hspace*{0.5cm}
    \subfigure[]{
    {\begin{tikzpicture}
    \begin{feynman}
    \vertex (a);
    \vertex [right=of a] (b);
    \vertex [above right=1cm of b] (f1);
    \vertex [below right=1cm of b] (f2);
    \vertex [above right=1cm of f1] (f3);
    \vertex [below right=1cm of f2] (f4);    
    \diagram* {
    (a) -- [dotted, thick,  edge label'=\(\mathbf{q}\)] (b) -- [dashed, edge label=\(\mathbf{\textcolor{black}{k+q}}\)] (f1)--[dashed, edge label=\(\mathbf{\textcolor{black}{p}}\)](f3),
    (b)-- [dashed, edge label'=\(\mathbf{\textcolor{black}{k}}\)] (f2)--[dashed, edge label'=\(\mathbf{\textcolor{black}{q-p}}\)](f4),
    (f1)--[boson, edge label=\(\mathbf{\textcolor{black}{k+q-p}}\)](f2)
    };
    \end{feynman}
    \end{tikzpicture}}
    }
    \caption{\small {\bf Feynman diagrams for the interactions between the excitations of the quantum spin ice and the Raman active phonons in a non-Kramers system due to the linear spin-phonon coupling (Eqs. \ref{eq_Eg coupling} and \ref{eq_T2g coupling}) :} (a) The vertex corresponds to the phonon-magnetic monopole interaction described by Eq. \ref{eq_eg ph_mon_gauge} and \ref{eq_t2g ph_mon_gauge}. Dotted, solid and curly line denote phonon, monopole and {\it emergent} photon, respectively. {Thin and thick solid lines represent two flavours of monopoles, A and B, respectively. (b) Vertex for the phonon-(emergent) photon interaction described by Eq. \ref{eq_phonon-photon coupling}. The circle represents the dipolar form factors (see Eq. \ref{eq_phonon-photon vertex}) that makes the vertex gauge invariant. (c) {Vertex for the phonon and electric charge interaction.} The dashed line denotes the electric charge.}}
    \label{fig_summary_feynman}
\end{figure*}

The partons naturally allow to re-write the linear  spin-phonon coupling in terms of the coupling of the phonons with the low-energy excitations of the quantum spin ice. The resultant interaction vertices are shown in Fig. \ref{fig_summary_feynman} while the details are discussed in Sec. \ref{sec_5}. In Sec. \ref{sec_6}, the Raman vertex for the phonons is derived. The resulting differential scattering cross-section (Eq. \ref{eq_diffscat}) depends on the phonon Green's function (Eq. \ref{eq_ramanver}) which receives a self-energy contribution due to scattering with the QSL excitations (Fig. \ref{fig_summary_feynman}) via spin-phonon coupling. The extra scattering channels then lead to an anomalous low-temperature broadening of the phonon peaks. The frequency dependence of such phonon linewidth contributions provides information about the QSL excitations, revealing the topologically non-trivial nature of the low-temperature quantum paramagnet.

In Sections \ref{sec_7}, \ref{sec_8}, and \ref{sec_9}, we calculate the resultant self-energy corrections (Figs. \ref{fig_bubble}, \ref{fig_self_photon}, and \ref{fig_self_charge}) within the simplest mean-field approximation for the lattice gauge theory-- the gauge mean-field theory (GMFT)-- where the gauge fluctuations are treated within a weak-coupling perturbation theory with the leading order contributions for the magnetic and electric sectors obtained by neglecting the gauge fluctuations altogether. The resultant frequency dependence for the phonon linewidth is given in Figs. \ref{fig_lw_vs_freq} and \ref{fig_pi} for the magnetic monopoles; Fig. \ref{fig_lw_vs_freq_photon} for emergent photons and Fig. \ref{fig_dos_charge} for the electric charges. The frequency dependence of the phonon linewidth follows the two-particle density of states of the emergent excitations in all three cases and hence provides a direct probe of the different excitations of the QSL. In particular, the energy separation of the electric and the magnetic sectors results in their contributions to the phonon linewidth occurring at separate energies, potentially paving the way for their separate identifications by careful analysis of the frequency and temperature dependence of the spectroscopic data. 

The two-particle density of states are sensitive to the symmetry fractionalisation patterns and in particular the projective symmetry group (PSG) of the QSL. In the case of quantum spin ice, a non-trivial example is the so-called $\pi$-flux state, where each hexagonal closed loop of the pyrochlore threads an electric flux of $\pi$ as opposed to zero in the regular (so-called zero-flux) quantum spin ice phase. The two states can be stabilised for opposite signs of the transverse term in the Hamiltonian in Eq. \ref{eq_spinham}-- $J_\pm>0~(<0)$ leads to the zero ($\pi$-) flux phase. In the $\pi$-flux phase, the momentum is fractionalised due to the larger magnetic unit cell, which is reflected in the two-particle density of states for the monopoles and hence shows up in the Raman linewidth. This can be easily seen by contrasting Figs. \ref{fig_lw_vs_freq} and \ref{fig_pi} for zero and $\pi$-flux, respectively. Therefore, our calculations show that Raman scattering experiments are sensitive to particular aspects of symmetry fractionalisation.

The above application of the mean-field approach to calculate the Raman vertex can be invalidated via strong gauge fluctuations, which couple to the electric charges and the magnetic monopoles. The relevant fine-structure coupling constant for the emergent quantum electrodynamics of quantum spin ice has recently been numerically estimated to be $\lesssim 0.1$~\cite{PhysRevLett.127.117205}. This suggests that the perturbative expansion may provide a leading estimate of the effect of the gauge fluctuations for the magnetic monopoles with their large gap. However, for the lower energy electric charges, the {effect of the coupling to the gauge fluctuations is expected to be even stronger} leading to drastic renormalisation of the two-electric charge density of states. In any case, the perturbative corrections to the phonon self-energy due to the gauge fluctuations are found to be sub-leading at low temperatures as shown in Sec. \ref{sec_7C}.

We also briefly summarise the effect of quadratic spin-phonon coupling terms on the vibrational Raman spectroscopy in Sec. \ref{sec_10}. This will be present both in Kramers and non-Kramers systems. While in non-Kramers systems, they are expected to be sub-leading to the linear coupling discussed above, in the case of Kramers systems, they provide the leading source of magnetoelastic coupling. In Sec. \ref{sec_11}, we calculate the phonon self-energy contribution due to spin-phonon coupling in the high-temperature thermal paramagnet for the spins where the gauge charges are ill-defined. In such a phase, the phonon lifetime is expected to be dominated by anharmonic phonon-phonon interactions, which is qualitatively different from the anomalous low-temperature broadening discussed above.

Finally, we show in Sec. \ref{sec_12} that even in the case of a mismatch of the phonon energy with those of the QSL excitations-- as is likely in some of the present  non-Kramers quantum spin ice candidates~\cite{natalia2021,ruminy2016}-- the above linear coupling contributes (obtained via integrating out the phonon) to the Raman vertex. This leads to a coupling between the {\it external} probe photon with all the excitations of the emergent electrodynamics and provides additional channels for scattering of the phonons that contribute to the Raman linewidth, albeit through the same two-particle density of states.

Finally, the details of various calculations are provided in the appendices.

%%%%%%%%%%%%%%%%%%%%%%%%

\section{\label{sec_2}Magnetism in Non-Kramers rare-earth pyrochlore family}

Several non-Kramers pyrochlore magnets are known in the context of both classical and quantum spin ice physics with substantial spin-lattice effects. The most striking one is possibly Tb$_2$Ti$_2$O$_7$~\cite{tbtioprincep,PhysRevB.78.094418,PhysRevB.64.224416,PhysRevLett.112.017203,PhysRevLett.99.237202,PhysRevB.62.6496,PhysRevLett.82.1012}, where the first crystal field gap is of the order of 10 K and recent neutron scattering experiments suggest that a vibronic bound state arises due to the coupling between acoustic phonon modes and crystal field levels which {is absent} in the paramagnetic phase~\cite{ruminy2014,PhysRevB.92.144417}. However, the exact role of the excited states and the applicability of quantum spin ice physics are currently being debated. Ho$_2$Ti$_2$O$_7$, on the other hand, is a classical spin ice~\cite{HoTbDy}, although it is interesting to note that on integrating out the lattice vibrations, their linear coupling with the transverse spins can induce (presumably very weak) quantum tunneling terms within the classical spin ice. 

The praseodymium pyrochlores, unlike the above extremes, belong to an interesting intermediate regime, where the crystal field gap is reasonably large, but quantum fluctuations are not insignificant~\cite{kimura}. Inelastic neutron scattering by Wen et al. reveals that the existence of quenched structural disorder in Pr$_2$Zr$_2$O$_7$  can act as a transverse field on the non-Kramers Pr$^{3+}$ ion and might lift the degeneracy of the non-Kramers doublet~\cite{wen-disorder}, although X-ray diffraction does not show evidence of any structural distortions. More recently, magnetoelastic  experiments on ultra-pure samples of Pr$_2$Zr$_2$O$_7$ show possibilities of substantial spin-phonon coupling and coupled spin-lattice dynamics~\cite{Comsatoru}. Further, high resolution Raman scattering on the same samples at relatively high temperatures (6 K-100 K) reveals that both the ground state and excited crystal field doublets show a temperature dependent splitting. The  splitting grows more pronounced as temperature is increased and can be accounted for by the dynamical coupling of spins to the phonons~\cite{satoru}. Several other non-Kramers spin-ice candidates such as Pr$_2$Sn$_2$O$_7$~\cite{PhysRevB.88.104421}, Tb$_2$Sn$_2$O$_7$~\cite{PhysRevLett.94.246402} are also known.

\begin{figure}
    \centering
    \includegraphics[width=0.7\linewidth]{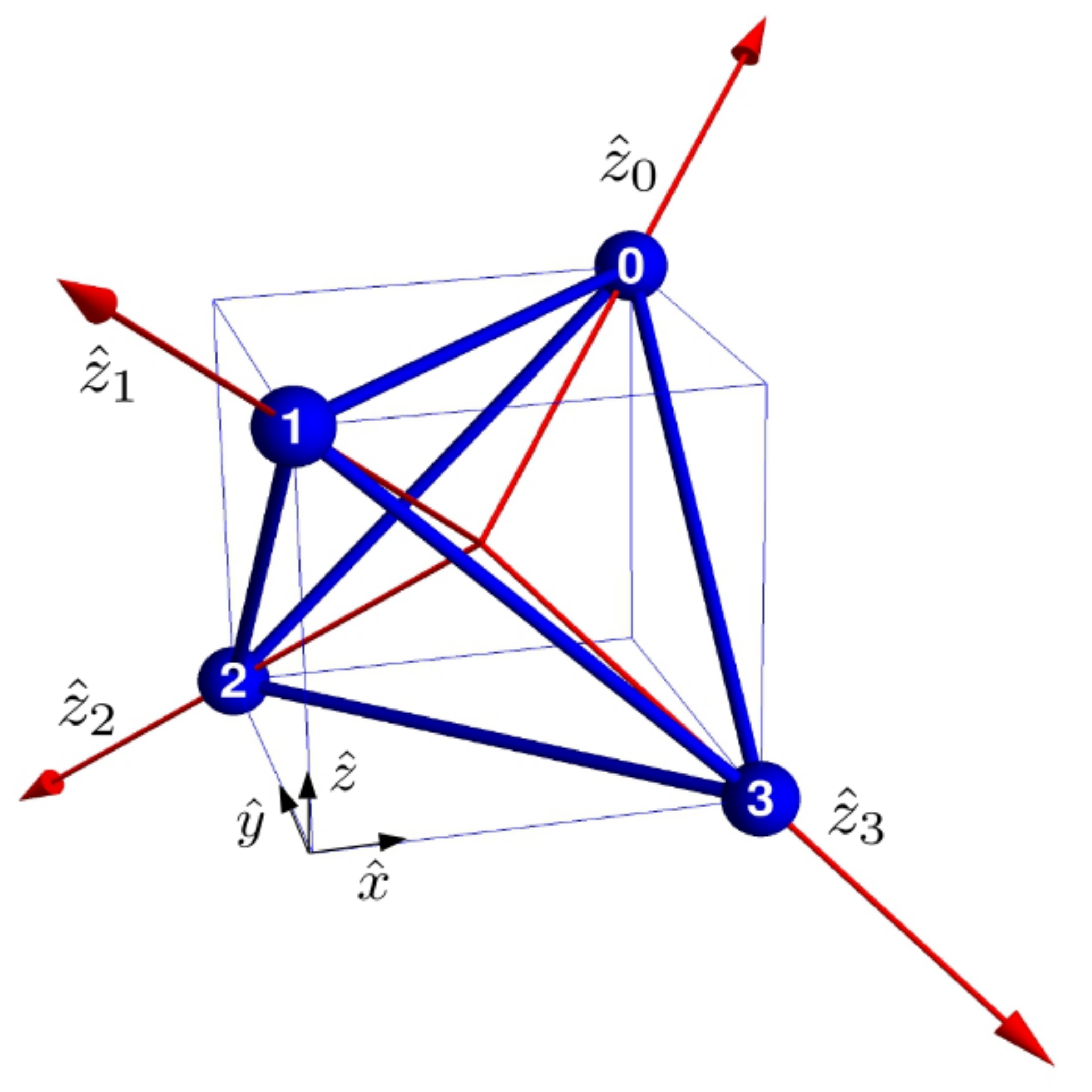}
    \caption{\small {\bf Sublattices of an up tetrahedron:} 0,1,2,3 denote the four sublattices and $\hat{z}_0$, $\hat{z}_1$, $\hat{z}_2$, $\hat{z}_3$ represent the four respective local quantization axes (see Eq. \ref{eq_local basis} in Appendix \ref{appendix_spin axis}).     }
    \label{fig_local basis}
\end{figure}

Therefore, to be concrete, we build our theory using Pr$_2$Zr$_2$O$_7$ as an example, although the results are generically applicable to any non-Kramers quantum spin ice. In Pr$_2$Zr$_2$O$_7$, the magnetic ion is the rare-earth element Pr$^{3+}$, which is in the $4f^2$ electronic configuration.  The ground state manifold is a doublet and given by \cite{PZO_doublet,kimura},
\begin{align}
    \mid\pm\rangle=a\mid\pm4\rangle\mp b\mid\pm1\rangle-c\mid\mp2\rangle
    \label{eq_GS}
\end{align}
where the different states belong to the $J=4$ multiplet with $J^z\mid m\rangle=m\mid m\rangle$.  Notably, characteristic to spin ice, the natural axis of quantization for the spins is along the local $[111]$ axis (see Fig. \ref{fig_local basis} and Appendix \ref{appendix_spin axis}). The ground state doublet is separated from the next crystal field state by almost 10 meV~\cite{kimura}. Due to this large gap, the low-temperature magnetic physics is dominated by the above non-Kramers doublet. The effective low-energy magnetic degrees of freedom are obtained by projecting all the spin operators to the low-energy doublet manifold, and written in terms of the effective pseudo spin-$\frac{1}{2}$ operators as $s^\mu(\equiv\frac{1}{2}\sigma^\mu$)~\cite{PZO_doublet}. 

A central feature of the doublets in Eq. \ref{eq_GS} is that under time reversal ($\mathcal{T}$) they transform as $\mid\pm\rangle\rightarrow\mid\mp\rangle$
such that the pseudo-spins transform as shown in Eq. \ref{eq_nonkramers}.
%%%%%%%%%%%%%%%%%%%%%%%%%%%%%%

\subsection{The spin exchange physics of non-Kramers quantum spin ice}

The pseudo-spins at different sites interact via regular spin exchanges and the minimal symmetry allowed spin Hamiltonian for non-Kramers spin ice is given by~\cite{rau,PhysRevB.83.094411,PhysRevLett.105.047201} 
\begin{align}
    H_0=\sum_{\langle ij\rangle}&\left[J_{zz}s_i^zs_j^z-J_{\pm}(s_i^+s_j^-+s_i^-s_j^+)\right]+\cdots
    %&~~~~~~~~~+\left.J_{\pm\pm}(\gamma_{ij}s_i^+s_j^++\gamma_{ij}^*s_i^-s_j^-)\right]
    \label{eq_spinham}
\end{align}
where $\cdots$ denote other symmetry allowed terms (including further neighbour ones) which do not immediately destabilise the QSL. In fact, their main effect in the QSL phase is to renormalise the dispersion of the excitations of the quantum spin ice~\cite{sungbin2012,PhysRevB.90.214430}. We neglect them here and their effects can be taken into account systematically along the lines discussed in the rest of this work. 

Experiments reveal the exchange coupling to be strongly anisotropic ($J_{zz}\gg J_{\pm}$). Also, $J_{zz}\approx 1.6 K$,~\cite{kimura} which is two orders of magnitude smaller than the single ion crystal field gap. This justifies the use of single ion crystal field states to treat the problem perturbatively.

Interestingly, the transformation of the non-Kramers doublet under $\mathcal{T}$ in Eq. \ref{eq_nonkramers} leads to an unusual Zeeman coupling in such materials. The external magnetic field, being odd under time reversal, can couple linearly only with $s^z$ but not with $s^x$ and $s^y$. The latter, however, can couple to the magnetic field quadratically. The complete onsite Zeeman Hamiltonian can be found in  Ref. \cite{adarsh}.

%%%%%%%%%%%%%%%%%%%%%%%%%%%%%%%

\section{\label{sec_3}Linear Magnetoelastic coupling in non-Kramers systems}

Having discussed the spin physics, we now turn to the linear magnetoelastic coupling in non-Kramers systems. From the point of view of symmetry analysis, the structure of such a linear coupling is quite straightforward. For a single tetrahedron, the linear coupling can be obtained starting with the eight-dimensional vector space spanned by the time reversal even transverse components, $(s_i^x, s_i^y)$, of the spins on four corners of a tetrahedron (Fig. \ref{fig_local basis}). This is then decomposed into the irreducible representations of the tetrahedral group, $T_d$, as
\begin{align}
    \mathbf{e}\oplus \mathbf{t}_1\oplus \mathbf{t}_2
    \label{eq_spin irrep}
\end{align}
where $\mathbf{e}$ denotes the doublet and  $\mathbf{t_1}$, $\mathbf{t_2}$ represent two triplets with different symmetry transformations (see Table \ref{table_spin transformation} in Appendix \ref{appendix_symmetry of spins}). 
Similarly, the (optical) normal vibrational modes of bond distortions of a tetrahedron are decomposed as,

\begin{align}
    \mathbf{a}_1\oplus \mathbf{e}\oplus \mathbf{t}_2 
\end{align}
where $\mathbf{a}_1$ is the singlet. It is evident from the above decomposition that $\mathbf{e}$ and $\mathbf{t}_2$ vibrational (optical) modes of a tetrahedron can linearly couple to the transverse components of the non-Kramers doublet. Since the complete symmetry of the pyrochlore is $T_d\times \mathcal{I}$ (with $\mathcal{I}$ being the inversion), the complete representation is obtained by taking symmetric and antisymmetric combinations of the previous representations to form the 'g' and 'u' modes, which are even and odd under spatial inversion respectively. 

As Raman scattering is insensitive to inversion-odd modes, we only consider the $\mathbf{e_g}$ and $\mathbf{t_{2g}}$ modes. Hence, the symmetry allowed magnetoelastic coupling for the Raman active modes is given by   
\begin{align}
    H^{(e)}_{sp}=\sum_{\mathbf{r},p=1,2}J^{(e)}_{sp}\zeta^{(e)}_{p,g}(\mathbf{r})\left(Q^{(e)}_p(\mathbf{r},A)+Q_p^{(e)}(\mathbf{r},B)\right)
    \label{eq_Eg coupling}
\end{align}
for the $\mathbf{e_g}$ modes and
\begin{align}
    H^{(t_2)}_{sp}=\sum_{\mathbf{r},p=1,2,3}J^{(t_2)}_{sp}\zeta^{(t_2)}_{p,g}(\mathbf{r})\left(Q_{p}^{(t_2)}(\mathbf{r},A)+Q_{p}^{(t_2)}(\mathbf{r},B)\right)
    \label{eq_T2g coupling}
\end{align}
for the $\mathbf{t_{2g}}$ modes. Here $\mathbf{r}$ denotes the centre of an up tetrahedron and $A/B$ denotes the two sublattices of the underlying diamond lattice, dual to the pyrochlore. $Q^{(e)}_{p}(\mathbf{r},A/B)$ and $Q^{(t_2)}_{p}(\mathbf{r},A/B)$ respectively span the $\mathbf{e}$ and $\mathbf{t_2}$ irreducible sector for the spins. For a single up tetrahedron (Fig. \ref{fig_local basis}), they are given by
\begin{align}
    &Q^{(e)}_1=s^x_0+s^x_1+s^x_2+s^x_3\nonumber\\
    &Q^{(e)}_2=s^y_0+s^y_1+s^y_2+s^y_3
    \label{eq_doublet}
\end{align}
and 
\begin{align}
    Q^{(t_2)}_{1}&=\frac{1}{2}(-s^x_0+s^x_1+s^x_2-s^x_3)\nonumber\\
    Q^{(t_2)}_{2}&=\frac{1}{4}(-s^x_0-s^x_1+s^x_2+s^x_3)+\frac{\sqrt{3}}{4}(s^y_0+s^y_1-s^y_2-s^y_3)\nonumber\\
    Q^{(t_2)}_{3}&=\frac{1}{4}(s^x_0-s^x_1+s^x_2-s^x_3)+\frac{\sqrt{3}}{4}(s^y_0-s^y_1+s^y_2-s^y_3)
    \label{eq_triplet}
\end{align}

 Finally, $\zeta^{(e)}_{p,g}(\mathbf{r})$ and $\zeta^{(t_2)}_{p,g}(\mathbf{r})$ are the ${\bf e_g}$ and $\mathbf{t_{2g}}$ normal modes of pyrochlore lattice. These normal modes are given by, $\zeta_{p,g}^{(\rho)}(\mathbf{k})=b^{(\rho)}_{p,\mathbf{k}}+b^{(\rho)\dagger}_{p,\mathbf{-k}}$ where $b^{(\rho)\dagger}_{p,\mathbf{k}}$ is the creation operator of the phonons of the $\rho$ irreducible representation, with the bare phonon Hamiltonian given by
\begin{align}
H_{\zeta}=\sum_{\rho}\sum_{\mathbf{k},p}\omega^{(\rho)}_{\mathbf{k}}\left(b^{(\rho)\dagger}_{p,\mathbf{k}}b^{(\rho)}_{p,\mathbf{k}}+\frac{1}{2}\right).
\label{eq_barephonon}
\end{align} 

An alternate and somewhat more microscopic derivation of the above physics can be obtained by considering the coupling of the doublet wave functions of Eq. \ref{eq_GS} with the phonons, which also gives rise to phonon mediated coupling between different crystal field states. The physics of such couplings will be discussed elsewhere~\cite{j32}. 

The above linear coupling makes the non-Kramers spin ice materials susceptible to {\it spin Jahn-Teller} distortions, where the spin entropy can be quenched by distorting the lattice and thereby splitting the doublet. Indeed, in some samples of Pr$_2$Zr$_2$O$_7$,  signatures of such splitting have been observed \cite{wen-disorder,matsuhira2009spin}, accompanied with random lattice distortions. However, more recent higher quality samples appear devoid of such distortions, suggesting controlled suppression of Jahn-Teller distortions in better quality single crystals~\cite{Comsatoru}.  

In the absence of static deformation of the crystal field environment, the above linear spin-phonon coupling helps to enhance the transverse fluctuations in the spin ice manifold, which could stabilise a U(1) QSL phase via magneto-distortive dynamics~\cite{Comsatoru}. 

To study the effect of linear magnetoelastic coupling (Eq. \ref{eq_Eg coupling} and Eq. \ref{eq_T2g coupling}) via the Raman experiments on quantum spin ice, we need to re-write the above spin-phonon coupling in terms of the coupling of the phonon to the low-energy excitations of the U(1) QSL. To derive this, for completeness we briefly review the well-known mapping between the spins and low-energy gauge theory for quantum spin ice~ \cite{savary,hermele,doron,sungbin2012} next.

%%%%%%%%%%%%%%%%%%%%%%%%%%%%

\section{\label{sec_4}quantum spin ice}

The description of the quantum spin ice is obtained starting with a magnetic monopole charge density operators~\cite{savary,sungbin2012,PhysRevLett.95.217201} $\mathcal{Q}_\mathbf{r}$, defined at the centre of a tetrahedron at $\mathbf{r}$, as
\begin{align}
    \mathcal{Q}_\mathbf{r}=\eta_\mathbf{r}\sum_\mu s^z_{\mathbf{r},\mathbf{r}+\eta_\mathbf{r} \mathbf{e}_{\boldsymbol{\mu}}}
\end{align}
where, $\eta_\mathbf{r}=1~(-1)$ for $\mathbf{r}\in$~up (down) tetrahedra of the pyrochlore lattice and $\mathbf{e}_{\boldsymbol{\mu}}$ is the vector connecting centres of the two nearest neighbour tetrahedra directed from up to down~(see Appendix \ref{appendix_lattice vectors}). We call the positively charged particles monopoles and negatively charged ones antimonopoles. The creation (annihilation) operators for the monopoles are defined as $\phi_\mathbf{r}^\dagger$ ($\phi_\mathbf{r}$) such that it satisfies, $[\mathcal{Q}_\mathbf{r},\phi^\dagger_\mathbf{r'}]= \delta_{\mathbf{r,r'}}\phi^\dagger_\mathbf{r}$.

The relation between the monopole and spin operators is given by
\begin{align}
    s^+_{\mathbf{r},\mathbf{r}+\mathbf{e}_{\boldsymbol{\mu}}}=\frac{1}{2}\phi^\dagger_{\mathbf{r}}~e^{iA_{\mathbf{r},\boldsymbol{\mu}}}\phi_{\mathbf{r}+\mathbf{e}_{\boldsymbol{\mu}}} 
    \label{eq_spin monopole transformation}
\end{align}
where $\mathbf{r}\in$~up tetrahedron and $A_{\mathbf{r},\boldsymbol{\mu}}$ represents the compact $U(1)$ {dual} gauge field on the bond joining $\mathbf{r}$ and $\mathbf{r}+\mathbf{e}_{\boldsymbol{\mu}}$~(In other words, they live on the links of the dual diamond lattice). The spin operators remain invariant under the following $U(1)$ gauge transformation.
\begin{align}
    \phi_\mathbf{r}\rightarrow \phi_\mathbf{r}e^{-i\theta_\mathbf{r}},\indent\indent
    A_{\mathbf{r},\boldsymbol{\mu}}\rightarrow A_{\mathbf{r},\boldsymbol{\mu}} +(\theta_{\mathbf{r}+\mathbf{e}_{\boldsymbol{\mu}}}-\theta_\mathbf{r})
\label{eq_gauge transformation}
\end{align}

The compactness of the gauge field allows for dual electric charge excitations~\cite{hermele} which are gapped in the QSL.   

Using the above mapping the spin Hamiltonian (Eq. \ref{eq_spinham}) can be written in terms of the gauge fields, monopoles, and the charges to obtain the lattice gauge theory description of quantum spin ice. This is given by Eq. \ref{eq_bare monopole hamiltonian} in Appendix \ref{appen_lgt} along with other relevant details.

%%%%%%%%%%%%%%%%

\subsection{\label{sec_4a}The gapless emergent photons}

In the limit $J_{zz}\gg J_\pm$, the magnetic monopoles have a gap of $\mathcal{O}(J_{zz})$ and can be integrated out. The low-energy Hamiltonian is obtained in terms of the fluctuations of the {dual} $U(1)$ gauge field $A_\mathbf{{r},\boldsymbol{\mu}}$. This leads to the well-known ring-exchange Hamiltonian that can be obtained either via degenerate perturbation theory of Eq. \ref{eq_spinham}~\cite{hermele} or equivalently integrating out the magnetic monopoles from Eq. \ref{eq_bare monopole hamiltonian}. This is given by
\begin{align}
    H_{eff}=\frac{U}{2}\sum_{\mathbf{r},\mu}B_{\mathbf{r},\boldsymbol{\mu}}^2-\frac{K}{2}\sum_{\hexagon}\cos\left(\sum_{\mathbf{r},\mu\in\hexagon}A_{\mathbf{r},\boldsymbol{\mu}}\right)
    \label{eq_ring exchange hamiltonian}
\end{align}
where $B_{\mathbf{r},\boldsymbol{\mu}}$($=s^z_{\mathbf{r},\mathbf{r}+\mathbf{e}_{\boldsymbol{\mu}}}$, $\mathbf{r}\in$ up tetrahedron) is the emergent magnetic field that is canonically conjugate to the dual vector potential, {\it i.e.},  $\left[A_{\mathbf{r},\boldsymbol{\mu}},B_{\mathbf{r'},\boldsymbol{\nu}}\right]=i\delta_\mathbf{r,r'}\delta_{\mu,\nu}$, $U$ is a Lagrange multiplier imposing the half-integer constraint on magnetic fields, and $K\sim \frac{J_\pm^3}{J_{zz}^2}$. The emergent electric field is given by ${\bf E}_{\hexagon}=\sum_{\mathbf{r},\mu\in \hexagon}A_{\mathbf{r},\mu}$ where $\sum_{\mathbf{r},\mu\in \hexagon}$ denotes the lattice curl  around the hexagonal loops of the pyrochlore.

The QSL corresponds to the deconfined phase~($\mid K\mid\gg U$) of the above Hamiltonian. {In this limit, the energy for the pure gauge theory can be minimized by setting up zero~($\pi$) electric flux through all the elementary hexagonal plaquettes for $K>0$~($K<0$)~\cite{sungbin2012}. These we shall term as $0$ and $\pi$-flux phases, respectively, since the magnetic monopoles hopping on the diamond lattice (see below) see this electric flux.}

{The low-energy excitations of the gauge theory can then be captured by expanding the \textit{cosine} term up to quadratic order about these static electric flux configurations. This gives rise to a free Maxwell theory with two transverse polarised gapless photon excitations and their dispersion is given by~\cite{rau2017},} 
\begin{align}
    \varepsilon_{\mathbf{k}}=c_e\mid \mathbf{k}\mid
\end{align}
where $c_e=\sqrt{UK}$ is the speed of emergent light.

%%%%%%%%%%%%%%%%%%%%%%%%%

\subsection{\label{sec_4b}The gapped magnetic monopole}

The dynamics of the bare magnetic monopoles, on the other hand can be obtained in a GMFT approximation of Eq. \ref{eq_bare monopole hamiltonian} by freezing the gauge fluctuation~\cite{savary} (see Appendix \ref{appen_lgt} for details). 

{For $K>0$, the ground state of the pure gauge theory is in the zero electric flux sector~(see above) where the gauge mean field ansatz can be chosen as $A_{\mathbf{r},\boldsymbol{\mu}}=0$. The bare band structure for the two flavours ($A$ and $B$) of magnetic monopoles is then given by~\cite{savary}
\begin{align}
    \epsilon^0_{\mathbf{k}}=\sqrt{2J_{zz}\left(\lambda-\frac{J_{\pm}}{2}\sum_{\mu>\nu}\cos\left(\mathbf{k}\cdot(\mathbf{d_\mu-d_\nu})\right)\right)}
    \label{eq_monopole dispersion}
\end{align}
}{where $\lambda$ is a Lagrange multiplier introduced to take into account the unitary constraint of the monopole operators (see Appendix \ref{appen_zero_mon_band}) at the mean-field level.}

{For $K<0$ on the other hand, the monopoles hop in a $\pi$-flux background per hexagonal plaquette. This can be implemented by choosing a suitable gauge~\cite{sungbin2012} (also see Fig. \ref{fig_unit_pi} in Appendix \ref{appen_lgt}) which doubles the size of magnetic unit cell, leading to four flavours of monopoles. The details of their band structure is summarized in Appendix \ref{appendix_pi flux_1}. In contrast to the zero flux phase, two non-degenerate bands  (denoted as $\epsilon^{\pi}_{+}({\bf k})$ and $\epsilon^{\pi}_{-}({\bf k})$) appear due to the presence of non-trivial background flux. It will be shown in Sec. \ref{sec_7B} that this leads to a very different Raman response of these two QSL phases. }

{The bare band structure of monopoles gets further renormalised due to the gauge fluctuations~\cite{PhysRevLett.127.117205}. However, in the following discussion, we will assume the monopole-gauge coupling constant to be small, so that these only lead to a sub-leading corrections of the GMFT results~(see Sec. \ref{sec_7C}). We shall comment on the merits/shortcomings of this approximation in the summary. }

%%%%%%%%%%%%%%%%%%%%%%

\subsection{\label{sec_4c}The gapped electric charge}

The electric charges or the point defects of the gauge field appear due to the $2\pi$ ambiguity of defining the compact vector potential~\cite{hermele,doron}. The fluctuations of the electric field are not small near these excitations and the expansion of the \textit{cosine} term (see Sec. \ref{sec_4a}) is not possible. Unlike the magnetic monopoles and the photons, these excitations are non-local in terms of the underlying spins and their properties are better captured in the dual description~\cite{hermele,doron,motrunich,gang2016} of the emergent gauge theory describing the bosonic electric charges, $\Psi_{\bf \mathsf{r}}$, hopping on the dual diamond lattice, ${\bf \mathsf{r}}$, via~\cite{doron,gang2016}
\begin{align}
    H_{charge}=-\sum_{\langle{\bf \mathsf{r,r'}}\rangle}t~e^{-i2\pi{\bf a_{\mathsf{r,r'}}}}\Psi^\dagger_{\bf \mathsf{r}}\Psi_{\bf \mathsf{r'}}+m\sum_{\mathsf{r}}\Psi^\dagger_{\bf \mathsf{r}}\Psi_{\bf \mathsf{r}}
    \label{eq_electric_hop}
\end{align}
where ${\bf a_{\mathsf{r,r'}}}$ is the vector potential dual to $A_{\mathbf{r},\boldsymbol{\mu}}$;  {$t$ is the effective hopping strength and $m$ is the chemical potential for the electric charges. The vector potential admits only integer values and is defined by, $$\left(\nabla\times{\bf a_{\mathsf{r,r'}}}\right)_{\hexagon^*}=\sum_{{\bf \mathsf{rr'}}\in\hexagon^*}{\bf a_{\mathsf{r,r'}}}= B_{\mathbf{r},\boldsymbol{\mu}}-B^0_{\mathbf{r},\boldsymbol{\mu}}$$
$\hexagon^*$ denotes the dual elementary hexagonal plaquettes and $B^0_{\mathbf{r},\mu}$ is a static divergenceless background field. }Since there is a single spin-$1/2$ on every pyrochlore site, the gauge field has a background $\pi$-flux in every dual hexagonal plaquette~\cite{gang2016} such that in the gauge mean-field limit (where we ignore the fluctuations of ${\bf a_{\mathsf{r,r'}}}$ around the background), the dynamics of electric charges reduces to the problem of bosons hopping on the diamond lattice subject to the background $\pi$-flux in every hexagonal plaquette. This can be solved using a proper gauge choice and gives rise to 12 soft modes~\cite{gang2016,doron}. We denote the soft modes as $\psi_i(i=1,...,12)$. As mentioned earlier (see Sec. \ref{sec_1}), the energy gap of the electric charges, $\Delta_c$, in the QSL phase is $\sim \mid J_{\pm}\mid^3/J^2_{zz}$. Within the hopping model, Eq. \ref{eq_electric_hop}, minimum gap of the electric charges is $\Delta_c=m-2\sqrt{2}t$.

The band structure of the electric charges gets further {renormalised} due to the gauge fluctuations. Compared to magnetic monopoles, they have a much smaller energy gap, and hence on general grounds, their coupling with the emergent photon is expected to be relatively much stronger. However, in order to keep our analysis tractable, we neglect such effects within our GMFT approach and only take them into account perturbatively. 

%%%%%%%%%%%%%%%%

\section{\label{sec_5}Magnetoelastic coupling in non-Kramers quantum spin ice}
The effect of the magnetoelastic coupling in the QSL phase can be analyzed by studying the coupling of the phonon to the emergent excitations using the mapping from spins to gauge charges discussed above. For the linear coupling in Eqs. \ref{eq_Eg coupling} and \ref{eq_T2g coupling}, the resultant interactions are given below. 

\subsection{\label{sec_5A}The magnetic monopole-phonon coupling}
Here we obtain the direct coupling between the phonon and the magnetic monopole. From Eq. \ref{eq_Eg coupling}, we get, for the $\mathbf{e_g}$ phonons :
\begin{align}
    H^{(e)}_{sp}=\frac{J^{(e)}_{sp}}{2}\sum_{\mathbf {r}}\sum_{\mu=0}^3\zeta_{-,g}^{(e)}({\mathbf {r}})\left[\phi_{{\mathbf {r}},A}^{\dagger}e^{iA_{\mathbf{r},\boldsymbol{\mu}}}\phi_{{\bf r+d_{\boldsymbol{\mu}}},B}\right.\nonumber~~~~~~\\
    \left.+\phi_{{\bf r-d_{\boldsymbol{\mu}}},A}^{\dagger}e^{iA_{{\bf r-d_{\boldsymbol{\mu}},\boldsymbol{\mu}}}}\phi_{{\bf r},B}\right]+h.c.
    \label{eq_eg ph_mon_gauge}
\end{align}
where,
\begin{align*}
    \zeta^{(e)}_{\pm,g}({\bf r})=\zeta^{(e)}_{1,g}({\bf r})\pm i\zeta^{(e)}_{2,g}({\bf r})
\end{align*}
are the displacement fields and from Eq. \ref{eq_T2g coupling} for the $\mathbf{t_{2g}}$ phonons,
\begin{align}
    &H^{(t_2)}_{sp}=\nonumber
    \\&\frac{J^{(t_2)}_{sp}}{2}\sum_{\bf r}\sum_{\mu=0}^3\sum_{p=1}^3\left(\zeta^{(t_2)}_{p,g}({\bf r})L^{(t_2)}_{p,x,\mu}-i \zeta^{(t_2)}_{p,g}({\bf r})L^{(t_2)}_{p,y,\mu}\right)\nonumber\\
    &\times\left[\phi^{\dagger}_{{\mathbf{r}},A}e^{iA_{\mathbf {r},\boldsymbol{\mu}}}\phi_{{\bf r+d_\mu},B}+\phi^{\dagger}_{{\mathbf {r}-\mathbf{d}_{\boldsymbol{\mu}}},A}e^{iA_{{\mathbf {r}-\mathbf{d}_{\boldsymbol{\mu}}},{\boldsymbol{\mu}}}}\phi_{{\bf r},B}\right]+h.c.
    \label{eq_t2g ph_mon_gauge}
\end{align}
The form factors $L^{(t_2)}_{p,\alpha,\mu}$ are obtained via the relation $Q^{(t_2)}_{p}=\sum_{\alpha=x,y}~\sum_{\mu=0}^3L^{(t_2)}_{p,\alpha,\mu}s^\alpha_\mu$, their explicit forms can then follow from Eq. \ref{eq_triplet}.

Both these interactions give rise to a Yukawa type coupling between the phonons and the monopole bilinear of the form $\zeta\phi^\dagger e^{iA}\phi$, albeit with different form factors. The corresponding bare vertex is shown in Fig. \ref{fig_summary_feynman}(a). It is clear from the interaction that the above coupling allows for a phonon to decay into a monopole-antimonopole pair: new low-energy scattering channels for the phonons inside the QSL phase open up. Note that while the bare monopole hopping preserves the sub-lattice flavour of the monopole, the above vertex mixes them, keeping only the total monopole number preserved.

\begin{figure}
\centering
    \begin{tikzpicture}
    \begin{feynman}
    \vertex (a);
    \vertex [right= of a] (b);
    \vertex [above right=2cm of b] (f1);
    \vertex [below right=2cm of b] (f2);
    
    \diagram* {
    (a) -- [dotted, thick, edge label'=\(\mathbf{q}\)] (b) -- [edge label=\(\mathbf{k+q}\)] (f1),
    (b) -- [edge label'=\(\mathbf{\textcolor{black}{k}}\)] (f2),
    };
    \end{feynman}
    \draw [line width=0.5mm] (b) --  (f2);
    \end{tikzpicture}
    \caption{\small {GMFT Feynman diagram for phonon and magnetic monopole interaction (described by Eqs. \ref{eq_eg_phonon monopole coupling} and \ref{eq_t2g_phonon monopole coupling}) in the zero flux phase:} ({see Fig. \ref{fig_summary_feynman}(a) for further details}).}
    \label{fig_phonon monopole interaction_1}
\end{figure}
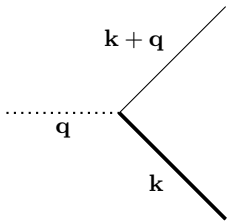

Within GMFT, we assume that the  gauge fluctuations are weak and can be taken into account perturbatively. Thus, within GMFT, the bare vertex for the magnetic monopole-phonon interaction is given by Fig. \ref{fig_phonon monopole interaction_1}, where the  gauge fluctuations have been neglected. Indeed, we shall show that within a perturbative treatment of the gauge field, the {temperature dependence of the corrections due to gauge fluctuations are sub-leading compared to the mean-field results} at low temperatures (see Sec. \ref{sec_7C}). In momentum space, GMFT vertices are given by
\begin{align}
    &H^{(e)}_{sp}=\frac{J^{(e)}_{sp}}{2\sqrt{N}}\sum_{\mathbf{k},\mathbf{k'}}\left[(\alpha^{(e)}_\mathbf{k}+\alpha^{(e)}_{\mathbf{k'}})\zeta^{(e)}_{-,g}(\mathbf{k}-\mathbf{k'})\phi^\dagger_{\mathbf{k},A}\phi_{\mathbf{k'},B}\right.\nonumber\\
    &\hspace{7cm}+h.c.\Big]
    \label{eq_eg_phonon monopole coupling}
\end{align}
\begin{align}
    &H_{sp}^{(t_2)}=\frac{J_{sp}^{(t_2)}}{2\sqrt{N}}\sum_{\mathbf{k},\mathbf{k'}}\sum_{p=1}^3\Big[(\alpha^{(t_2)}_{p,\mathbf{k}}+\alpha^{(t_2)}_{p,\mathbf{k'}})\times\nonumber\\
    &~~~~~~~~~~~~~~~~~~~~~~~~~~\times\zeta^{(t_2)}_{p,g}(\mathbf{k}-\mathbf{k'})\phi_{\mathbf{k},A}^\dagger\phi_{\mathbf{k'},B}+h.c.\Big]
    \label{eq_t2g_phonon monopole coupling}
\end{align}
where $N$ is the total number of unit cells, while $\alpha^{(e)}_{\bf k}$ and $\alpha^{(t_2)}_{p,{\bf k}}$s are vertex functions of the ${\bf e_g}$ and ${\bf t_{2g}}$ coupling, respectively, whose forms are given in Appendix \ref{appendix_zero flux vertex}.

%%%%%%%%%%%%%%%%%%%%%%%%%%%%%%%

\subsection{\label{sec_5B}The (emergent) photon-phonon coupling}

To obtain the coupling between phonon and emergent photon, once again we integrate out the gapped magnetic monopoles (as in Sec. \ref{sec_4a}) in presence of the magnetoelastic coupling described by Eq. \ref{eq_eg ph_mon_gauge} and \ref{eq_t2g ph_mon_gauge}. The leading coupling between phonon and gauge field is obtained in fourth-order of the perturbation theory~\cite{etienne}. For the $\mathbf{e_g}$ phonons, this gives rise to
\begin{align}
    H_{phonon-photon}=-\frac{J_{sp}^{(e)2}J_\pm^2}{2J_{zz}^3}\sum_{\hexagon} \left(\sum_{{\bf r}\in\hexagon}\boldsymbol{\zeta}^{(e)}_{g}({\bf r})\cdot\boldsymbol{\zeta}^{(e)}_g({\bf r})\right)\nonumber\\
    \times\cos\left[{\bf E}_{\hexagon}\right]
    \label{eq_eff_phonon_photon}
\end{align}
In the deconfined QSL phase, the \textit{cosine} term in the Hamiltonian above can be expanded up to quadratic order as $\cos {\bf E}\approx 1-{\bf E}^2/2$. At low energies, the constant term in the expansion leads to a quadratic term in the phonon. This renormalises the frequency of the phonon by a constant shift without affecting its linewidth.

The leading order coupling between the phonon and the emergent photon, in the continuum limit is given by
\begin{align}
    \mathcal{H}_{phonon-photon}=J_{ph-ph}\int d^3{\bf r}~ \boldsymbol{\zeta}^{(e)}_g({\bf r})\cdot\boldsymbol{\zeta}^{(e)}_g({\bf r})~{\bf E}^2({\bf r})
    \label{eq_phonon_photon}
\end{align}
where $J_{ph-ph}\sim \frac{J_{sp}^{(e)2}J_{\pm}^2}{4J_{zz}^2l^3}$ with $l$ being the lattice length-scale. As expected, the phonons cannot simply couple to the dual gauge field since they do not carry the emergent gauge charge. Instead, they couple to the gauge invariant electric field. Further, since the Raman active phonons are even under inversion, they can only couple to the electric field at quadratic order. We note, {in passing}, that the antisymmetric phonon modes ('u' modes) on the other hand are allowed to couple linearly to the emergent electric field. Such interaction effects can be probed using infrared spectroscopy~\cite{etienne}.

In momentum space, Eq. \ref{eq_phonon_photon} takes the form
\begin{align}
    \mathcal{H}_{phonon-photon}=\int &\prod_{i=1}^4~d^3{\bf k}_i~\mathcal{G}^{\alpha\beta}({\bf k_1,k_2,k_3,k_4})\nonumber\\
    &\times\left(\boldsymbol{\zeta}^{(e)}_g({\bf k_1})\cdot\boldsymbol{\zeta}^{(e)}_g({\bf k_2})\right)~A^\alpha_{\bf k_3}A^\beta_{\bf k_4}
    \label{eq_phonon-photon coupling}
\end{align}
where the interaction vertex is given by
\begin{align}
    \mathcal{G}^{\alpha\beta}({\bf k_1,k_2,k_3,k_4})=\frac{J_{ph-ph}}{N}~\left[-{\bf k_3\cdot k_4}\delta_{\alpha\beta}+k_{3}^\beta k_4^\alpha\right]\nonumber\\
    \times\delta({\bf k_1+k_2+k_3+k_4})
    \label{eq_phonon-photon vertex}
\end{align}
The above interaction is shown in Fig. \ref{fig_summary_feynman}(b), {where the circle represents the gauge invariant dipolar vertex function, $\mathcal{G}^{\alpha\beta}$.} Such decay processes for phonons in a QSL phase give rise to an additional contribution to the phonon linewidth similar to that due to the monopoles, albeit at a different energy-scale. 

%%%%%%%%%%%%%%%%%%%%%%

\subsection{\label{sec_5C}The electric charge-phonon coupling}

{Similar to the phonon-magnetic monopole coupling, the phonons also interact with the electric charges via a Yukawa coupling as shown in Fig. \ref{fig_summary_feynman}(c) (again, the electric charge creation/annihilation operators are not gauge invariant and hence cannot couple to the phonons linearly).}

\begin{figure}
\centering
    \begin{tikzpicture}
    \begin{feynman}
    \vertex (a);
    \vertex [right=of a] (b);
    \vertex [above right=2cm of b] (f1);
    \vertex [below right=2cm of b] (f2);
    \diagram* {
    (a) -- [dotted, thick, edge label'=\(\mathbf{q}\)] (b) -- [dashed, edge label=\(\mathbf{\textcolor{black}{k+q}}\)] (f1),
    (b) -- [dashed, edge label'=\(\mathbf{\textcolor{black}{k}}\)] (f2),
    };
    \end{feynman}
    \end{tikzpicture}
    \caption{\small { GMFT Feynman diagram for phonon and electric charge interaction (described by Eq. \ref{eq_phonon_charge coupling}):} ({see Fig. \ref{fig_summary_feynman}(c) for further details}). }
    \label{fig_phonon charge interaction}
\end{figure}
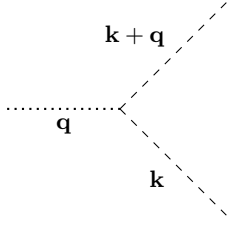

To derive the coupling between the phonons and the electric charges, we construct the bilinears of the soft electric charge modes with appropriate symmetry that can couple to a particular polarisation of the phonon. Here we analyze only the $\mathbf{e_g}$ couplings and the interaction is given by,
\begin{align}
    H_{phonon-charge}=J^{(e)}_{ph-ch}\sum_{\bf r}&\left(\zeta^{(e)}_{1,g}({\bf r})\Theta_1({\bf \mathsf{r}})\right.\nonumber\\
    &~~~\left.+\zeta^{(e)}_{2,g}({\bf r})\Theta_2({\bf \mathsf{r}})\right)
    \label{eq_phonon_charge coupling}
\end{align}
where $(\Theta_1,\Theta_2)$ forms an $\mathbf{e_g}$ doublet and is given by,
\begin{align}
    &\Theta_1=\psi_1^*\psi_1+\psi_{11}^*\psi_{11}-\psi_3^*\psi_3-\psi_9^*\psi_9\nonumber\\
    &\Theta_2=-\psi_1^*\psi_1-\psi_{11}^*\psi_{11}-\psi_3^*\psi_3-\psi_9^*\psi_9+2(\psi_5^*\psi_5+\psi_7^*\psi_7)
\end{align}
where $\psi_i$ ($i=1,2,\cdots 12$) are the soft modes of the electric charges as obtained in Ref. \cite{gang2016} and discussed in the previous section. The above interaction is shown in Fig. \ref{fig_phonon charge interaction}.

Due to the above magnetoelastic coupling, phonons acquire a finite lifetime by scattering with the excitations of the QSL. In the next three sections (Sec. \ref{sec_7}, \ref{sec_8} and \ref{sec_9}), we  compute the lifetime of the phonons and their typical low-temperature behaviour in order to probe the non-Kramers U(1) QSLs.

%%%%%%%%%%%%%%%%%%%%%%%

\section{\label{sec_6}Raman Scattering of the phonons in quantum spin ice phase}

The Raman vertex for the phonon is given by~\cite{porto}
\begin{align}
    H_{Raman}&=\int\mathbf{P}(\mathbf{r})\cdot\mathbf{E}_{ext}(\mathbf{r})~d^3\mathbf{r}
    \label{eq_raman vertex}
\end{align}
where ${\bf P}({\bf r})$ is the electric dipole moment and ${\bf E}_{ext}({\bf r})$ is the external electric field (to be distinguished from the emergent electromagnetism). Relegating details to Appendix \ref{appen_phononraman}, we find the Raman scattering cross-section~\cite{devereaux},
\begin{align}
  \frac{d^2\sigma(\mathbf{q},\omega)}{d\Omega d\omega_s}\propto R(\mathbf{q},\omega), 
  \label{eq_diffscat}
\end{align}
where for a thermal system, by Fermi's Golden rule,
\begin{align}
    R(\mathbf{q},\omega)=\sum_{i,f}\frac{e^{-\beta E_i}}{Z}\mid\langle f\mid H_{Raman}\mid i\rangle\mid^2\delta(E_f-E_i-\omega)
\end{align}

Here $\mathbf{q}=\mathbf{q_{in}-q_{out}}$ is the net momentum transferred to the system by the Raman process and $\omega$ is the difference between the frequency of incident and scattered photons. As the speed of light is very large compared to that of the phonons, only the $\mathbf{q}\rightarrow 0$ regime of the Brillouin zone can be probed by Raman scattering. Further, $\mid i\rangle$, $\mid f\rangle$ denote the initial and final state of the phonons respectively, with energies $E_i$ and $E_f$. Finally, $Z$ is the partition function for a Gibbs distribution at temperature, $T=1/\beta$. 

At low temperatures, the initial state can be approximated by the ground state. Also, we can see from the Raman vertex~(Eqs. \ref{eq_raman vertex} and \ref{eq_ramver_appen}) that the scattering matrix element is non-zero only when $\mid i\rangle$ and $\mid f\rangle$ differ by a single phonon, as higher phonon processes are suppressed at low temperatures. So at low temperatures, $\mid f\rangle$ should be chosen from the single phonon sector leading to
\begin{align}
    R(\mathbf{q},\omega)\propto-\frac{\pi n(\omega)e^{\beta\omega}}{Z}e^{-\beta E_0}\lim_{\delta\rightarrow 0^+}\mathcal{I} m[G_\zeta(\mathbf{q},\omega+i\delta)]
    \label{eq_ramanver}
\end{align}
where, $n(\omega)=\frac{1}{e^{\beta\omega}-1}$ is the Bose-factor and $G_\zeta(\mathbf{q},\omega+i\delta)$ is the retarded Green's function of the phonon. This can be calculated from the analytic continuation of the Matsubara Green's function, $G_{\zeta}({\bf q},i\omega)$, given by 
\begin{align}
    G_\zeta(\mathbf{q},i\omega)&=-\int_0^\beta d\tau\langle\hat{\mathcal{T}}\left(\zeta({\bf q},\tau)\zeta(-{\bf q},0)\right)\rangle e^{i\omega\tau}\nonumber\\
    &=-\frac{2\omega_\mathbf{q}}{\omega^2+\omega_\mathbf{q}^2+2\omega_\mathbf{q}\Sigma_\zeta({\bf q},i\omega)}
\end{align}
where $\omega_{\bf q}$ is the bare dispersion of the phonon (obtained from Eq.~\ref{eq_barephonon}) and $\Sigma_\zeta({\bf q},i\omega)$ is its self-energy arising from the interaction with the QSL excitations. Here, for simplicity of the expression, we have suppressed the superscript denoting the irrep of the phonon. Eq. \ref{eq_ramanver} results in a Lorentzian lineshape. The position of the peak of this curve is shifted from the non-interacting one by 
\begin{align}
    &|\Delta_{Raman}|=\mathcal{R}e\left[\Sigma_\zeta({\bf q},\omega+i\delta)\right]
\end{align}
and the full-width at half maximum of the Lorentzian is given by
\begin{align}
    &\Gamma=2\mid \mathcal{I}m\left[\Sigma_\zeta({\bf q},\omega+i\delta)\right]\mid
\end{align}
which can then be directly compared with experiment.

We now focus on understanding the frequency and temperature dependence of the linewidth, $\Gamma$, in detail, in order to extract the information it contains about the QSL excitations via the linear magnetoelastic coupling. The real part can be computed from the imaginary part using the standard Kramers-Kronig theorem~\cite{mahan}.  Since the three QSL excitations are separated in energy scales, we expect that they dominate the linewidth in different frequency windows. Therefore, we particularly focus on the frequency dependence of the linewidth.

%%%%%%%%%%%%%%%%%%%%%%%

\section{\label{sec_7}Self-energy of the phonon due to phonon-magnetic monopole coupling}
We now calculate the self-energy of the phonons and hence the broadening of the phonon peaks due to the phonon-monopole interaction. We calculate the effect of the coupling in a perturbative approach both in the zero and $\pi$-flux phases. 

\subsection{\label{sec_7A}Zero flux phase}

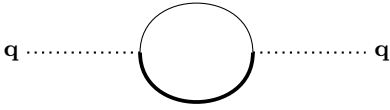
\begin{figure}
    \centering
    {\begin{tikzpicture}
    \begin{feynman}
    \vertex  (a);
    \vertex [left=of a] (b);
    \vertex [right=of a, label=right:\({\bf q}\)] (f1);
    \vertex [left=of b, label=left:\({\bf q}\)] (f3); \diagram* {
    (f3) -- [dotted, thick] (b) -- [half left] (a) -- [dotted, thick] (f1), (a)--[line width=0.5mm, half left](b)
        };
    \end{feynman}
    \end{tikzpicture}}
    \caption{\small Self-energy of the phonon due to the phonon-magnetic monopole interaction (see Fig. \ref{fig_phonon monopole interaction_1})}
    \label{fig_bubble}
\end{figure}

The first non-zero contribution to the self-energy comes at second order, $\mathcal{O}(J_{sp}^{(\rho) 2})$, by computing the bubble diagram of Fig. \ref{fig_bubble}. Within GMFT, for the zero flux phase, the self-energy is given by, 
\begin{align}
    \Sigma^0_{\zeta^{(\rho)}}(\mathbf{q},i\Omega)=
    &-\frac{J^{(\rho)2}_{sp}}{4N\beta}\sum_{\mathbf{k},\omega}\mid\alpha^{(\rho)}_\mathbf{k}+\alpha^{(\rho)}_\mathbf{k+q}\mid^2\nonumber\\
    &\times G^0_\phi(\mathbf{k},A,i\omega)G^0_\phi(\mathbf{k+q},B,i(\omega+\Omega))
    \label{eq_selfenergy_phi}
\end{align}
where the time ordered Green's function ($G_\phi^0$) for monopoles in the zero flux phase is defined as (see Eq. \ref{eq_mon_action_zero} in Appendix \ref{appen_zero_mon_band}), 
\begin{align}
    G^0_\phi(\mathbf{k},A/B,i\omega)&=\int_0^\beta d\tau\langle \hat{\mathcal{T}}\left(\phi_{\mathbf{k},A/B}(\tau)\phi^{\dagger}_{\mathbf{k},A/B}(0)\right) \rangle e^{i\omega\tau}\nonumber\\
    &=\frac{2J_{zz}}{\omega^2+\left(\epsilon^0_\mathbf{k}\right)^2}
    \label{eq_monopole green fn}
\end{align}

To obtain the broadening of the Raman peaks, we calculate the imaginary part of $\Sigma^0_{\zeta^{(\rho)}}(\mathbf{q},i\Omega\rightarrow\Omega+i\delta)$. Performing the frequency summation using standard Matsubara summation techniques~\cite{mahan}, we get
\begin{widetext}
\begin{align}
    \lim_{\delta\rightarrow0}~\mathcal{I}m[\Sigma^0_{\zeta^{(\rho)}}(\mathbf{q},\Omega+i\delta)]=\frac{\pi J_{sp}^{(\rho)2}J_{zz}^2}{4N}\sum_{\mathbf{k}}\mid\alpha^{(\rho)}_\mathbf{k}+\alpha^{(\rho)}_\mathbf{k+q}\mid^2
    \left[\frac{n(\epsilon^0_\mathbf{k})-n(\epsilon^0_\mathbf{k+q})}{\epsilon^0_\mathbf{k}\epsilon^0_\mathbf{k+q}}\Big(\delta(\Omega+\epsilon^0_\mathbf{{k+q}}-\epsilon^0_{\mathbf{k}})-\delta(\Omega+\epsilon^0_\mathbf{k}-\epsilon^0_\mathbf{k+q})\Big)\right.\nonumber\\
    \left.+\frac{n(\epsilon^0_\mathbf{k})+n(\epsilon^0_\mathbf{{k+q}})+1}{\epsilon^0_\mathbf{k}\epsilon^0_\mathbf{k+q}}\Big(\delta(\Omega+\epsilon^0_\mathbf{k}+\epsilon^0_\mathbf{k+q})-
    \delta(\Omega-\epsilon^0_\mathbf{k}-\epsilon^0_\mathbf{k+q})\Big)\right]
    \label{eq_freq_summed}
\end{align}
\end{widetext}
where, $n(\epsilon^0_\mathbf{k})=\frac{1}{e^{\beta\epsilon^0_\mathbf{k}}-1}$ is the Bose occupation for the magnetic monopole with $\epsilon^0_\mathbf{k}$ being the single-particle dispersion within GMFT as given by Eq. \ref{eq_monopole dispersion}.

{The first two delta functions of Eq. \ref{eq_freq_summed} imply processes where a monopole scatters by absorption of a phonon~({\it absorption} process). The prefactor $(n(\epsilon^0_\mathbf{k})-n(\epsilon^0_\mathbf{k+q}))$ {represents the net probability} of such processes. On the other hand, the last two delta functions in Eq. \ref{eq_freq_summed} arise due to the conversion of a phonon into a monopole-antimonopole pair or vice-versa~({\it pair production} process, Fig. \ref{fig_shallowvertex}(c)). The prefactor $(1+n(\epsilon^0_\mathbf{k})+n(\epsilon^0_{\mathbf{k+q}}))$ represents the net probability of two competing processes- the first(second) is the annihilation (creation) of a phonon followed by creation (annihilation) of the monopole-antimonopole pair.}

Eq. \ref{eq_freq_summed} is one of the central results of this work. It shows {that} the self-energy correction of the phonons arises from its coupling to the magnetic monopoles. We now analyze the self-energy, in particular, its frequency dependence, which can be detected in Raman scattering experiments. For Raman scattering, only the ${\bf q}\approx 0$ regime of the Brillouin zone is accessible. {In this limit, clearly the probability of the absorption process of the phonons vanishes since the difference of the two Bose factors go to zero as ${\bf q}\rightarrow 0$, leading to}
\begin{widetext}
\begin{align}
    \Gamma(\Omega, T)=2 \mid\mathcal{I}m[\Sigma^0_{\zeta^{(\rho)}}(\mathbf{q}=0,\Omega)]\mid=\frac{2\pi J_{sp}^{(\rho)2}J_{zz}^2}{N}\sum_{\mathbf{k}}\mid\alpha^{(\rho)}_\mathbf{k}\mid^2
    \Big[\frac{2n(\epsilon^0_\mathbf{k})+1}{(\epsilon^0_\mathbf{k})^2}\mid\Big(\delta(\Omega+2\epsilon^0_\mathbf{k})-
    \delta(\Omega-2\epsilon^0_\mathbf{k})\Big)\mid\Big]
    \label{eq_selfenergy q=0}
\end{align}
\end{widetext}

\begin{figure*}
    \centering
    \subfigure[]{
    \includegraphics[width=0.475\linewidth]{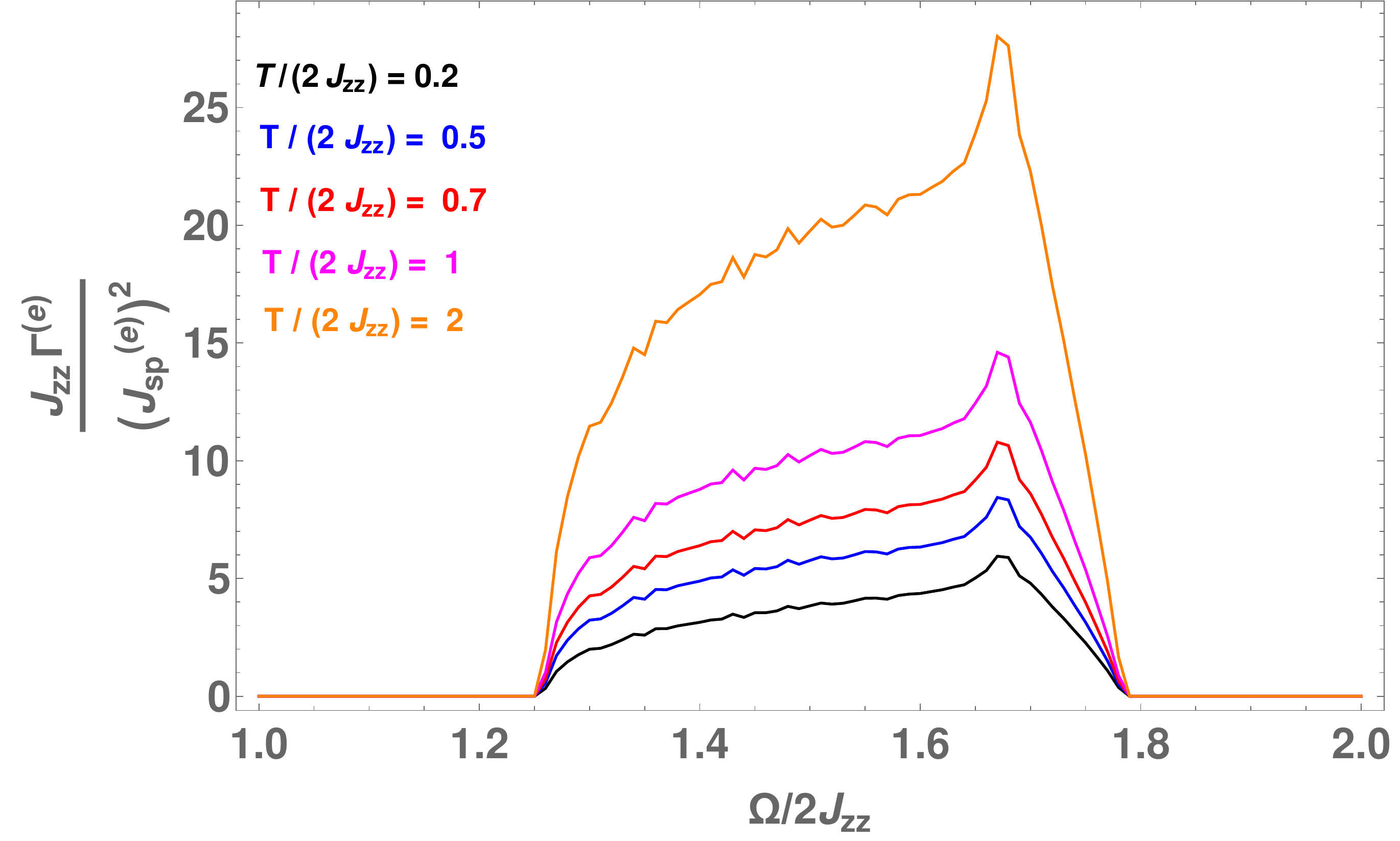}
    }
    \subfigure[]{
    \includegraphics[width=0.475\linewidth]{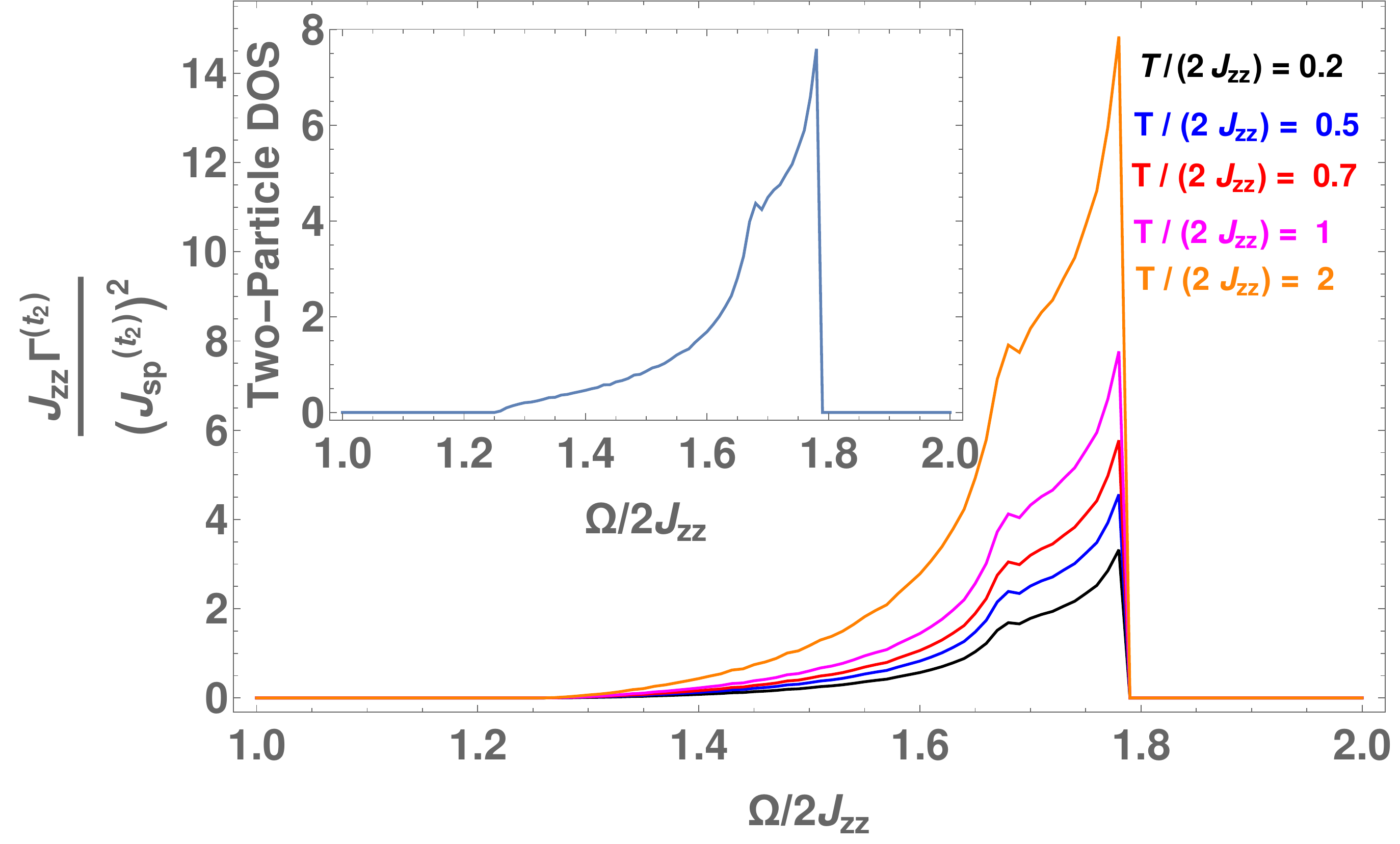}
    }
    \caption{\small \textbf{Frequency dependence of linewidth of (a) $\mathbf{e_g}$ and (b) $\mathbf{t_{2g}}$ phonons in the zero flux phase due to the phonon-magnetic monopole coupling:} For both plots, we have chosen $\frac{\lambda}{2J_{zz}}=0.7$ and $\frac{J_{\pm}}{2J_{zz}}=0.1$ for illustrative purpose. The inset of (b) shows the two-particle density of states (DOS) of magnetic monopoles for the same values of $\lambda/2J_{zz}$ and $J_\pm/2J_{zz}$.  }
    \label{fig_lw_vs_freq}
\end{figure*}

From Eq. \ref{eq_monopole dispersion}, we see that the bare monopole band structure is gapped with its minima at $\mathbf{k}=0$ and the energy gap, $\Delta_0=\sqrt{2J_{zz}\left(\lambda-3J_{\pm}\right)}$. It is evident from Eq. \ref{eq_selfenergy q=0} that {the splitting of phonons into a monopole-antimonopole pair}  occurs only if the phonon frequency is larger than the pair creation energy ($2\Delta_0$) such that $\Gamma\sim \Theta\left(\mid\Omega\mid-2\Delta_0\right)$. This is visible in Fig. \ref{fig_lw_vs_freq}, where we plot the linewidth, $\Gamma(\Omega,T)$ versus the frequency, $\Omega$, for various temperatures, $T$, for both the $\mathbf{e_g}$ and the $\mathbf{t_{2g}}$ modes, {for $\Delta_0=1.26J_{zz}$ as an illustrative example for plotting. The profile of the curve remains qualitatively same as long as the constraint $\lambda>3J_{\pm}$ is satisfied, which defines the extent of the QSL. } Apart from the dependence on the form factors, $\alpha_{\bf k}^{(\rho)}$, and the Bose factor, both these curves reflects the two-particle  density of states profile of monopoles, shown in the inset of Fig. \ref{fig_lw_vs_freq}(b). The effect of the form factors can be noted from the qualitative difference of the two plots. Since $\alpha_{\mathbf{k}}^{(t_2)}\rightarrow 0$ as $\mathbf{k}\rightarrow0$ (from Eq. \ref{eq_vertex functions}), the linewidth for $\mathbf{t_{2g}}$ smoothly vanishes for $\Omega\rightarrow2\Delta_0$. By contrast, for $\mathbf{e_g}$, the vertex function ($\alpha_{\mathbf{k}}^{(e)}$) tends to a nonzero constant as $\mathbf{k}\rightarrow0$ (from Eq. \ref{eq_vertex functions}) and the linewidth shows a sharp behaviour even at zero momentum. 

%%%%%%%%%%%%%%%%%%%

%\subsection{\label{sec_7B} \texorpdfstring{$\pi-$}{}flux phase}

\subsection{\label{sec_7B} {$\pi-$} flux phase}

{The phonon-magnetic monopole coupling in the $\pi-$flux phase is obtained from the linear spin-phonon coupling of Eq. \ref{eq_Eg coupling} and Eq. \ref{eq_T2g coupling} via parton decomposition of the spins and freezing the gauge fluctuations to a suitable GMFT ansatz as described in Sec. \ref{sec_4b}. Focusing only on the $\mathbf{e_g}$ phonons, the phonon-monopole coupling is given by, }

\begin{align}
    H^{(e)}_{sp}=\frac{J^{(e)}_{sp}}{2\sqrt{N}}\sum_{\mathbf{k}}\sum_{\mu,\nu=1,2}\left(M^{\mu\nu}_{\mathbf{k}}\zeta^{(e)}_{-,g}(\mathbf{q}=0)\phi^\dagger_{\mathbf{k},A\mu}\phi_{\mathbf{k},B\nu}\right.\nonumber\\
    +h.c.\Big)
    \label{eq_phonon monopole_pi}
\end{align}
The details of the vertex functions $M^{\mu\nu}_{\bf k}$ are given in Appendix \ref{appendix_pi flux_2}. {The Feynman diagram of the above interaction is again represented by the Yukawa vertex which is very similar to Fig. \ref{fig_phonon monopole interaction_1}  except for the fact that \textit{four} distinct diagrams are possible due to the extended sublattice structure.} The phonon self-energy in this phase is given by,

\begin{align}
    \Sigma^{\pi}_{\zeta^{(e)}}(\mathbf{q}=0,&i\Omega)=-\frac{J_{sp}^{(e)2}}{4N\beta}\sum_{\mathbf{k},\omega}\sum_{\mu,\nu,\alpha,\beta}M^{\mu\alpha}_\mathbf{k}M^{\nu\beta}_\mathbf{-k}\nonumber\\
    &\times\left[G^{\pi}_\phi\right]_{\mu\nu}(\mathbf{k},A,i\Omega+i\omega)\left[G^{\pi}_\phi\right]_{\alpha\beta}(\mathbf{k},B,i\omega)
\end{align}
where, $[G^{\pi}_\phi]_{\mu\nu}(\mathbf{k},A/B,i\omega)$ is the Green's function for the $A/B$ monopoles in the $\pi$-flux phase~(see Eq. \ref{eq_pi_green} in Appendix \ref{appendix_pi flux_1} for the detailed expressions). Computing the imaginary part of the above expression, we obtain the linewidth of the phonons in the $\pi-$flux phase. The contribution where the phonon creates into two monopoles, is given by, 
\begin{widetext}
\begin{align}
    \Gamma(\Omega,T)=\frac{\pi J_{sp}^{(e)2}}{2N}\sum_{\mathbf{k}}&\left[\frac{1+2n(\epsilon^\pi_+(\mathbf{k}))}{\epsilon^\pi_+(\mathbf{k})^2}\mathcal{P}_1(\mathbf{k})\delta\left(\Omega-2\epsilon^\pi_+(\mathbf{k})\right)+\frac{1+2n(\epsilon^\pi_-(\mathbf{k}))}{\epsilon^\pi_-(\mathbf{k})^2}\mathcal{P}_2(\mathbf{k})\delta\left(\Omega-2\epsilon^\pi_-(\mathbf{k})\right)+\left(\mathcal{P}_3(\mathbf{k})+\mathcal{P}_4(\mathbf{k})\right)\times\right.\nonumber\\
    &\left.\left(\frac{1+n(\epsilon^\pi_+(\mathbf{k}))+n(\epsilon^\pi_-(\mathbf{k}))}{\epsilon^\pi_+(\mathbf{k})\epsilon^\pi_-(\mathbf{k})}\delta\left(\Omega-\epsilon^\pi_+(\mathbf{k})-\epsilon^\pi_-(\mathbf{k})\right)+\frac{n(\epsilon^\pi_+(\mathbf{k}))-n(\epsilon^\pi_-(\mathbf{k}))}{\epsilon^\pi_+(\mathbf{k})\epsilon^\pi_-(\mathbf{k})}\delta\left(\Omega+\epsilon^\pi_+(\mathbf{k})-\epsilon^\pi_-(\mathbf{k})\right)\right)\right]
    \label{eq_lw_pi}
\end{align} 
\end{widetext}
{where, $\mathcal{P}_1({\bf k})$, $\mathcal{P}_2({\bf k})$, $\mathcal{P}_3({\bf k})$, $\mathcal{P}_4({\bf k})$ are real functions of momentum whose detailed forms are given by  Eq. \ref{eq_self energy_pi} in Appendix \ref{appendix_pi flux_3} and  $\epsilon^\pi_{\pm}({\bf k})$ are the bare monopole dispersions in the $\pi$-flux phase as discussed above. The detailed forms are given by Eq. \ref{eq_band_pi_1} and \ref{eq_band_pi_2} in Appendix \ref{appendix_pi flux_1}.}

The above expression should be contrasted with that for zero flux (Eq. \ref{eq_selfenergy q=0}). There are four distinct delta functions appearing in the expressions. The first two terms are closely related to the two-particle density of states for the $\epsilon^\pi_+({\bf k})$ and $\epsilon^\pi_-({\bf k})$ bands, implying the decay of a phonon into monopole-antimonopole pair with respective energy in the two bands, $\pm$. On the other hand, the last two entries represent the processes where a phonon scatters into monopole-antimonopole pair of different energy bands. Consequently, unlike the zero flux case, both the pair production and absorption processes show non-zero amplitude even at $\mathbf{q}=0$. 

{As an aside, we briefly comment on the connection of this `shallow inelastic scattering' referred to in Fig. \ref{fig_shallowvertex}(d) to the deep inelastic scattering familiar from QCD. In the latter, a photon scatters off a quark which, when it is highly relativistic, is possible with only a minor momentum contribution from other quarks. By contrast, with fractionalisation being a low-energy phenomenon, the kinematics works out differently despite the topological correspondence between the two diagrams. It is the capacity of the scattering between the two bands for the $\pi$-flux phase  to absorb energy and momentum which provides the non-vanishing cross-section even at low $\mathbf{q}$ for the shallow scattering.}

\begin{figure}
    \centering
    \includegraphics[width=0.9\linewidth]{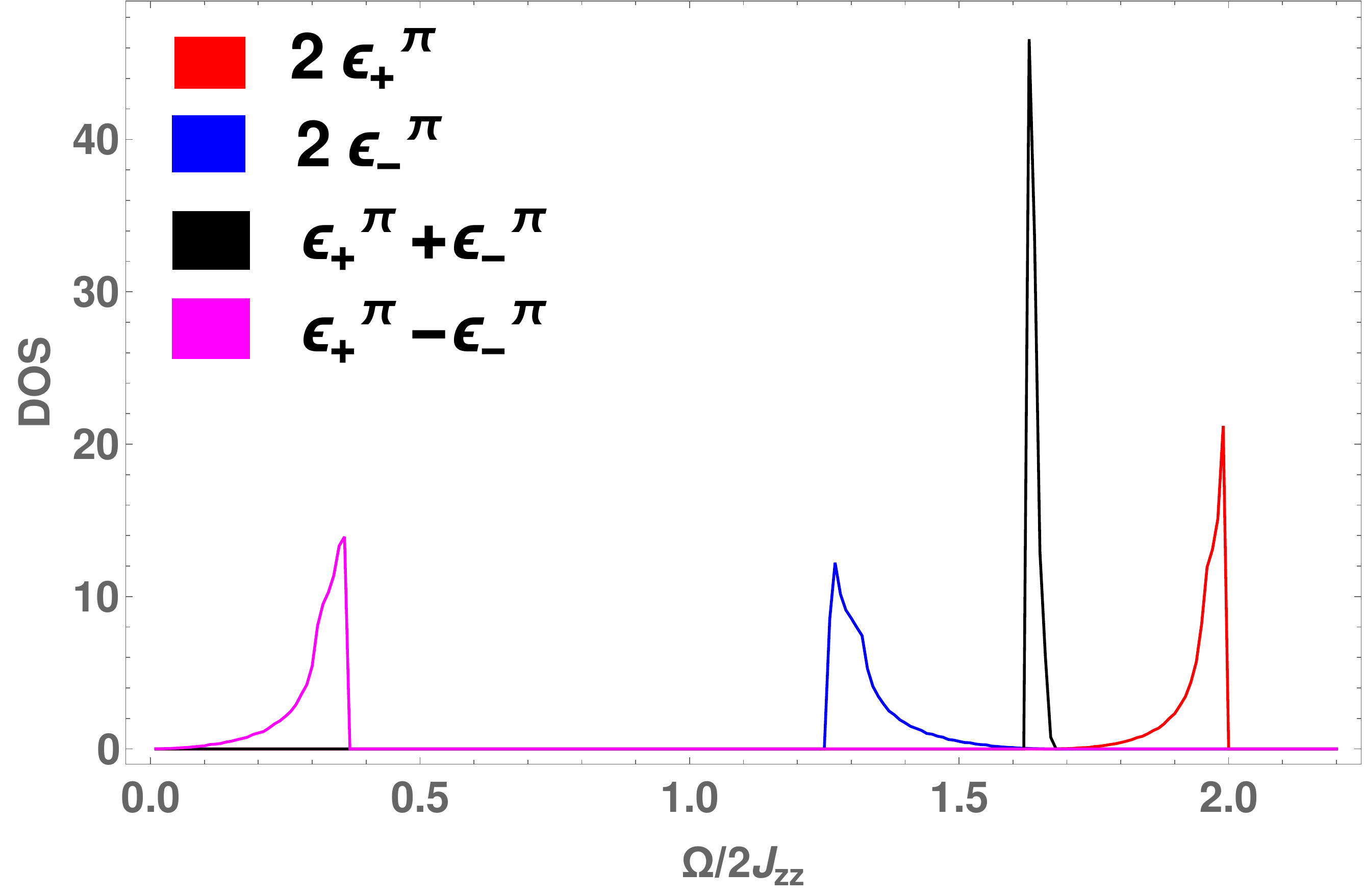}
    \caption{\small \textbf{Density of states of different bands contributing to the phonon linewidth in the $\pi$-flux phase:} We have chosen $\lambda/2J_{zz}=0.7,~ J_{\pm}/2J_{zz}=0.3$ for illustrative purpose. Red and blue curves denote the two-particle density of states for upper($\epsilon^\pi_+$) and lower($\epsilon^\pi_-$) bands, respectively. Black and magenta curves denote density of states of $\epsilon^\pi_++\epsilon^\pi_-$ and $\epsilon^\pi_+-\epsilon^\pi_-$, respectively. }
    \label{fig_pi}
\end{figure}

{In Fig. \ref{fig_pi}, we plot various contributions to the two-particle density of states of magnetic monopoles in the $\pi-$flux phase, which represent the four distinct delta functions of Eq. \ref{eq_lw_pi}. The phonon linewidth is obtained from the sum of these delta functions weighted by appropriate momentum dependent form factors ($\mathcal{P}_1({\bf k}),\mathcal{P}_2({\bf k}),\mathcal{P}_3({\bf k}),\mathcal{P}_4({\bf k})$) and the Bosonic distribution functions at finite temperature. It is evident from the figure that, unlike the zero flux case, {the Raman linewidth shows a non-zero signal even at very low energy compared to the monopole gap. {Availability of the two non-degenerate bands allow a non-zero probability of the process where a monopole (say with energy $\epsilon^{\pi}_{-}({\bf k})$) absorbs the phonon and converts into another monopole of different band structure ($\epsilon^{\pi}_+({\bf k})$) even at $\mathbf{q}=0$.} }  Also, the enlargement of the magnetic unit cell compared to that of the zero flux case-- leading to the {momentum} fractionalisation-- is very well captured in such a Raman response profile, which is a signature of the non-trivial projective implementation of symmetry. Hence, the phonon linewidth measurements via Raman experiments can be an extremely useful tool to identify the non-trivial projective symmetry group of a QSL phase.}

%%%%%%%%%%%%%%%%%%%

\subsection{\label{sec_7C}Beyond GMFT : Gauge fluctuations}

The above Raman cross-section was obtained within GMFT neglecting the gauge fluctuations. We now consider the effect of long-wavelength gauge fluctuations within a weak-coupling approach for the emergent electrodynamics. At present, it is not clear that such a weak-coupling approach is valid for treating the gauge fluctuations. {In fact the coupling parameter-- the fine structure constant-- for the emergent electrodynamics is generically expected to be sizeable}. However, recent numerical calculations~\cite{PhysRevLett.127.117205} on quantum spin ice (via Eq. \ref{eq_ring exchange hamiltonian}) suggest that the emergent fine-structure constant is $\lesssim 0.1$ which may suggest that the perturbative expansion could still provide an estimate of the effect of gauge fluctuations.

\begin{figure}
    \centering
    {\begin{tikzpicture}
    \begin{feynman}
    \vertex  (a);
    \vertex [left=of a] (b);
    \vertex [right=of a, label=right:\({\bf q}\)] (f1);
    \vertex [left=of b, label=left:\({\bf q}\)] (f3); \diagram* {
    (f3) --  (b) -- (a) -- (f1), (a)--[ boson, half right](b)
        };
    \end{feynman}
    \end{tikzpicture}}
    \caption{\small Self-energy of the magnetic monopoles due to the gauge fluctuation.}
    \label{fig_rainbow}
\end{figure}
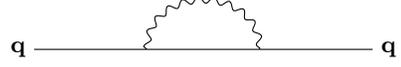

For the zero flux case, this is captured by the expansion, $e^{\pm iA_{\mathbf{r},{\boldsymbol{\mu}}}}\approx \left(1\pm iA_{\mathbf{r},{\boldsymbol{\mu}}}\right)$. Hence, (from Eq. \ref{eq_bare monopole hamiltonian}) the interaction between monopole and gauge field is given by,  
\begin{align}
    H_{GF}=\frac{iJ_{\pm}}{4\sqrt{N}}\sum_{\mathbf{k,k'},\mu\neq\nu}\Big{[}&\gamma_B^{\mu\nu}(\mathbf{k,k'})A_{\mathbf{k-k'},\mu}\phi_{\mathbf{k},B}^\dagger\phi_{\mathbf{k'},B}\nonumber\\
    &+\gamma_A^{\mu\nu}(\mathbf{k,k'})A_{\mathbf{k-k'},\mu}\phi^\dagger_{\mathbf{k},A}\phi_{\mathbf{k'},A}\Big{]}
\label{eq_gauge fluctuation hamiltonian}
\end{align}
where $A_{\mathbf{k},\mu}=\frac{1}{\sqrt{N}}\sum_{\mathbf{r}\in I} A_{\mathbf{r},{\boldsymbol{\mu}}}e^{i\mathbf{k}\cdot \mathbf{r}}$. The details of the vertex functions, $\gamma^{\mu\nu}_{A/B}({\bf k,k'})$, are given in Appendix \ref{appendix_beyond gmft} for the zero flux phase. The $\pi$-flux phase can be treated in a similar way. There are two (related by Ward identities) effects of the gauge fluctuations-- renormalisation of the vertex (Fig. \ref{fig_summary_feynman}(a)) and renormalisation of the monopole propagator (Fig. \ref{fig_rainbow})-- which we discuss in turn.

In presence of such gauge fluctuations, the vertex functions for the bare phonon-monopole interactions get dressed via the virtual photon exchange processes as described by Fig. \ref{fig_summary_feynman}(a). This effect can be taken into account by calculating the modified vertices, $\alpha_{\bf k}^{(\rho)}+\delta\alpha_{\bf k}^{(\rho)}$. We compute the leading order corrections by expanding the bare monopole energy about the band minima at ${\bf k}=0$ (Eq. \ref{eq_monopole_quadratic}). Similarly, all the bare vertex functions ($\alpha^{(e)}_{\bf k}, \alpha^{(t_2)}_{\bf k}, \gamma^{\mu\nu}_{A/B}({\bf k,k'})$) are also Taylor expanded in polynomials of momentum and only the leading terms are considered. We note that the terms with higher powers of momentum contribute to more {sub-leading (in temperature) corrections to the mean-field vertices at low temperatures.} With the above approximations, the leading frequency independent corrections to the ${\bf e_g}$ and ${\bf t_{2g}}$ vertices are obtained as (see Appendix \ref{appendix_beyond gmft} for further details),

{
\begin{align}
    &\delta\alpha_{\bf k}^{(e)}\approx a_0+\frac{a_1}{\beta^4}+\frac{a_2e^{-\beta\Delta_0}}{\beta^{\frac{3}{2}}}+k^2\left(\frac{a_3}{\beta^2}+\frac{a_4e^{-\beta\Delta_0}}{\beta^{\frac{1}{2}}}\right)\nonumber\\
    &\delta\alpha_{\bf k}^{(t_2)}\approx \Tilde{a}_0+\frac{\Tilde{a}_1}{\beta^5}+\frac{\Tilde{a}_2e^{-\beta\Delta_0}}{\beta^2}+k^2\left(\frac{\Tilde{a_3}}{\beta^3}+\frac{\Tilde{a_4}e^{-\beta\Delta_0}}{\beta}\right)
    \label{eq_vertex correction_final}
\end{align}
where $a_i$ and $\tilde a_i$ are temperature independent constants. The correction to the linewidth can now be obtained by incorporating these vertex corrections to Eq. \ref{eq_selfenergy q=0}. We note that such contributions do not change the dependence of the Raman response on the two-particle density of states of the monopoles. Instead, they modify the temperature dependence and overall profile of the linewidth vs frequency plots (see Fig. \ref{fig_lw_vs_freq}) obtained from the GMFT ansatz by renormalisation of the form factors. However, since the QSL phase is stabilised only at low temperatures, the temperature dependent vertex corrections merely give rise to a sub-leading correction to Eq. \ref{eq_selfenergy q=0} as $T\rightarrow 0$. }

{Apart from the vertex corrections, the virtual photon exchange due to the gauge fluctuations also renormalises the monopole self-energy, via processes shown in Fig. \ref{fig_rainbow}. Such contributions renormalise {the bare monopole linewidth as well as its band structure. The broadening of the linewidth is sub-leading in the low-temperature regime. On the other hand, the renormalisation of the band structure modifies the two-particle density of states of monopoles by an amount proportional to the speed of emergent light ($c_e$)}. As a result, the Raman linewidth gets renormalised compared to the GMFT results described in Fig. \ref{fig_lw_vs_freq} via the dressed two-monopole density of states. However, since the large anisotropy of the exchange coupling ($J_{zz}\gg J_{\pm}$) ensures $\Delta_0\gg c_e$~\cite{PhysRevLett.127.117205,rau2017,benton}, such effects are small. The large gap of the magnetic monopoles in QSL phase preserves the essential features of the Raman response obtained in the GMFT ansatz.}

%%%%%%%%%%%%%%%%%%%%%%%%%%

\section{\label{sec_8}Self-energy of the phonon due to phonon-photon coupling}
\begin{figure}
    \centering
    {\begin{tikzpicture}
    \begin{feynman}
    \vertex  (a);
    \vertex [left=of a] (b);
    \vertex [right=of a, label=right:\({\bf q}\)] (f1);
    \vertex [left=of b, label=left:\({\bf q}\)] (f3); \diagram* {
    (f3) -- [dotted, thick] (b) -- [dotted, thick] (a) -- [dotted, thick] (f1), (a)--[ boson, half left](b)-- [ boson, half left] (a)
        };
    \end{feynman}
    \end{tikzpicture}}
    \caption{\small Self-energy of the phonon due to the phonon-(emergent) photon interaction (see Fig. \ref{fig_summary_feynman}(b)).}
    \label{fig_self_photon}
\end{figure}
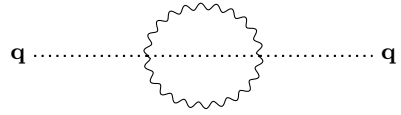

Similar to the Raman response due to the phonon-monopole coupling, the leading contribution to the phonon linewidth due to phonon-photon interaction (see Eq. \ref{eq_phonon-photon coupling}) can be computed from the Feynman diagram shown in Fig. \ref{fig_self_photon} appearing in the second-order perturbation theory. The phonon self-energy is given by,

\begin{widetext}
\begin{align}
    \Sigma_{\zeta^{(e)}}({\bf q},i\Omega)=\frac{1}{\beta^2}\sum_{\bf k_2,k_3,k_4}\sum_{\Omega_2\Omega_3\Omega_4}\mathcal{G}^{\beta\gamma}({\bf q,k_2,k_3,k_4})&\mathcal{G}^{\mu\nu}({\bf q,k_2,k_3,k_4})\delta(\Omega+\Omega_2+\Omega_3+\Omega_4)\nonumber\\
    &\times~G_\zeta({\bf k_2},\Omega_2) D_{\beta\mu}({\bf k_3},\Omega_3)D_{\gamma\nu}({\bf k_4},\Omega_4)\delta({\bf q+k_2+k_3+k_4})
    \label{eq_self_photon}
\end{align}

\end{widetext} 

{Here $D_{\mu\nu}({\bf q}, i\omega)$ denotes the photon propagator which can be calculated from the effective low-energy Hamiltonian of the pure gauge theory given in Eq. \ref{eq_ring exchange hamiltonian}, {\it i.e.},
\begin{align}
    D_{\mu\nu}({\bf q},i\omega)&=-\int_{0}^\beta d\tau\langle\hat{\mathcal{T}}\left(A_{{\bf q},\mu}(\tau)A_{{\bf -q},\nu}(0)\right)\rangle e^{i\omega\tau}\nonumber\\
    &=-\frac{U\delta_{\mu\nu}}{\omega^2+\varepsilon_{\bf q}^2}
\end{align}
Eq. \ref{eq_self_photon} can be further simplified by performing the frequency summation~\cite{mahan}. For the Raman scattering experiments discussed earlier, we consider only the ${\bf q}\rightarrow 0$ limit and focus on the imaginary part. Typically, the dispersion for the optical phonon can be approximated as, $\omega_{\bf q}\rightarrow\omega_0$. Also, the energy scale of the emergent photon is much smaller than the optical phonon excitations of the pyrochlores~\cite{moon,rau2017,natalia2021}. Hence, at the low temperatures of the QSL phase, it is fair to consider $n(\omega_0)\ll n(\varepsilon_{\bf k})$. Setting $n(\omega_0)=0$ in the leading approximation, the contribution to the phonon linewidth is obtained as,}

\begin{widetext}
\begin{align}
    \Gamma(0,E)=\frac{\pi(J_{ph-ph} U)^2}{2N^2}\sum_{\bf k}\left[{\bf k\cdot k}\delta_{\beta\gamma}-k^\beta k^\gamma\right]^2~\frac{1}{4\varepsilon_{\bf k}^2}\bigg[\delta({E+\varepsilon_{\bf k}})~[n(-E)]^2&+\delta({E-\varepsilon_{\bf k}})~[1+n(E)]^2\nonumber\\
    &+\delta(E)~\left\{2[n(\varepsilon_{\bf k})][n(\varepsilon_{\bf k})+1] \right\}\bigg]
    \label{eq_linewidth_phonon_photon}
\end{align}
\end{widetext}
where, $E=(\Omega-\omega_0)/2$. It is clear from the above expression that the Raman response occurs around $\Omega=\omega_0$ due to the gaplessness of the photons, which is different from the frequency window at which the magnetic monopole signatures occur. {For small positive energies $E$}, the above expression is further simplified to, 

\begin{align}
\Gamma(0,E)\propto E^4(1+n(E))^2
\end{align}

\begin{figure}
    \centering
    \includegraphics[width=0.9\linewidth]{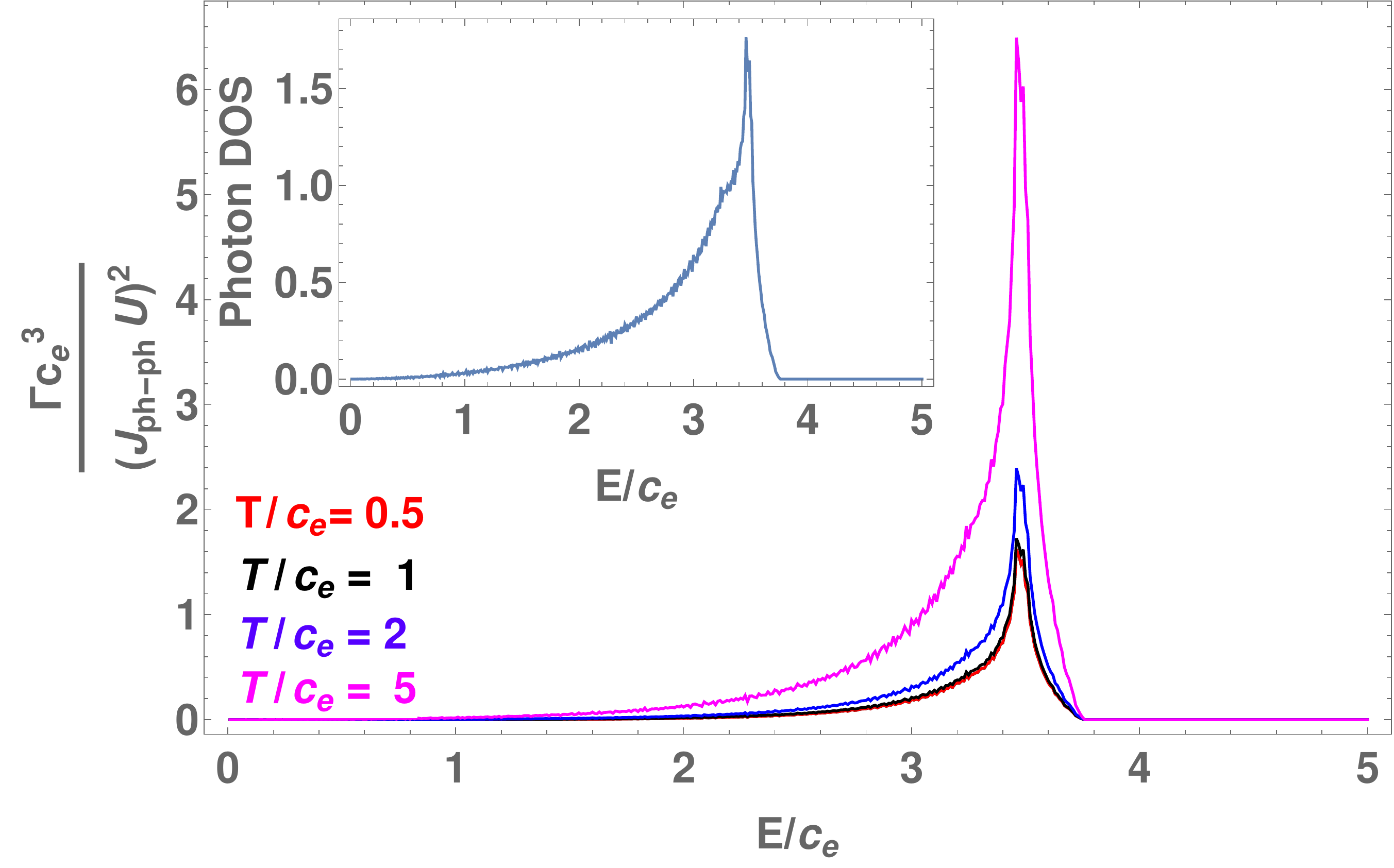}
    \caption{\textbf{Energy dependence of the linewidth of the $\mathbf{e_g}$ phonons due to phonon-(emergent) photon coupling:}  The energy dependence is shown at different temperatures. $c_e=\sqrt{UK}$ is the velocity of the emergent photons and its density of states is plotted in the inset.}
    \label{fig_lw_vs_freq_photon}
\end{figure}

For the higher energy regime, the photon band structure starts deviating from the linear behaviour and the above form is no longer valid. The complete energy dependence of the above contribution to the linewidth is shown in Fig. \ref{fig_lw_vs_freq_photon} for different temperatures, where we have used the lattice regularized dispersion for the emergent photons~\cite{rau2017,benton}. Apart from the usual dipolar form factor, the linewidth profile is mostly sensitive to the photon density of states, which is shown in the inset of Fig. \ref{fig_lw_vs_freq_photon}.

%%%%%%%%%%%%%%%%%%%%%%%%

\section{\label{sec_9}Self-energy of the phonon due to phonon-electric charge coupling} 

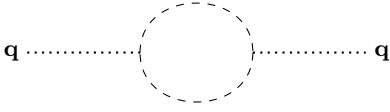
\begin{figure}
    \centering
    {\begin{tikzpicture}
    \begin{feynman}
    \vertex  (a);
    \vertex [left=of a] (b);
    \vertex [right=of a, label=right:\({\bf q}\)] (f1);
    \vertex [left=of b, label=left:\({\bf q}\)] (f3); \diagram* {
    (f3) -- [dotted, thick] (b) -- [dashed, half left] (a) -- [dotted, thick] (f1), (a)--[dashed, half left](b)
        };
    \end{feynman}
    \end{tikzpicture}}
    \caption{\small Self-energy of the phonon due to the phonon-electric charge interaction (see Fig. \ref{fig_phonon charge interaction}) }
    \label{fig_self_charge}
\end{figure}

\begin{figure}
    \centering
    \includegraphics[width=0.9\linewidth]{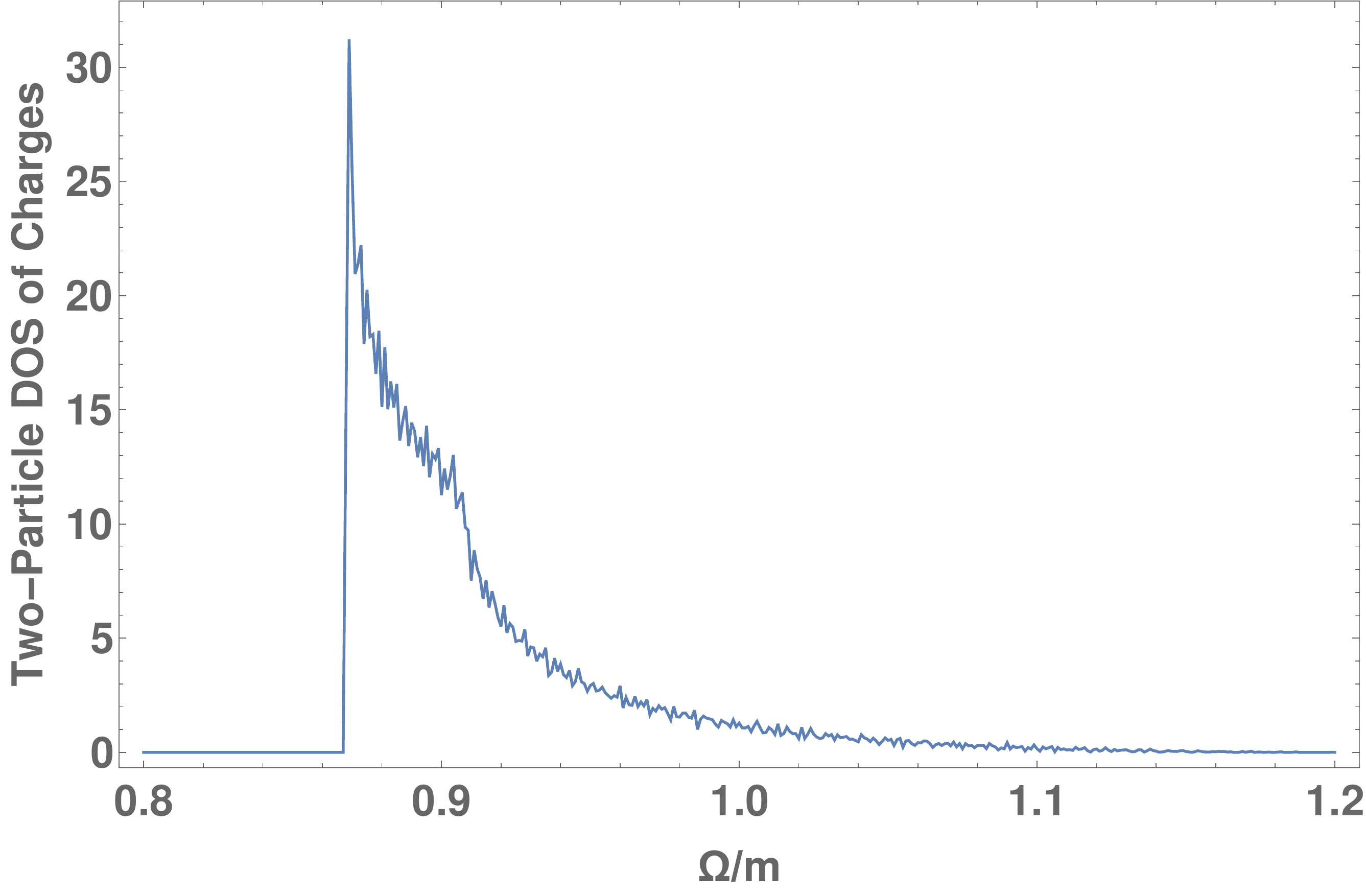}
    \caption{\small \textbf{Two-particle density of states of the charges}: For illustrative purpose, we have chosen $t/m=0.2$ where $\Delta_c/m=0.43$.}
    \label{fig_dos_charge}
\end{figure}

The final contribution to the phonon self-energy in the QSL phase arises from scattering of the phonons off the electric charges. Again, assuming weak coupling between the charges and the gauge field, we compute the phonon linewidth due to Eq. \ref{eq_phonon_charge coupling} using GMFT. As we have already seen, this interaction is very similar to that between phonons and monopoles. Hence, the contribution to the phonon self-energy also comes from similar Feynman diagrams as shown in Fig. \ref{fig_self_charge}. There are two possible scattering channels for electric charge-phonon interactions-- absorption of a phonon by a charge, or, {annihilation of a phonon followed by pair production of charges (with charge $\pm 1$)}. Similar to monopoles, only the second process is relevant here. Therefore, $\Gamma\sim \Theta(\mid\Omega\mid-2\Delta_c)$, and the linewidth vs frequency profile closely follows the two-particle density of states of the electric charges. This is shown in Fig. \ref{fig_dos_charge} for $t/m=0.2$ as an illustrative example ({However, it can be chosen from any value that satisfies, $m>2\sqrt{2}t$, defining the validity of the QSL description, and the profile remains qualitatively unchanged}). Clearly, the Raman response due to the phonon-charge coupling has a threshold energy scale of $\sim2\Delta_c$ which is a different energy scale compared to the response due to magnetic monopoles and photons.

%%%%%%%%%%%%%%%%%%%%%%%%%%%%%%%%%%%%%%%

\section{\label{sec_10}Bilinear coupling}

Having discussed the effects of the linear magnetoelastic coupling in the QSL phase, we now briefly discuss the more familiar contribution to the Raman response of the phonons arising due to magnetoelastic interaction. This is present both in Kramers and non-Kramers systems, as it arises due to modulation of the spin-exchange interactions via the phonons and  can be obtained from the bare spin Hamiltonian of Eq. \ref{eq_spinham} by Taylor expanding the exchange coupling constants in powers of lattice displacements ($\boldsymbol{\delta_{\mu\nu}}$) from the ionic equilibrium position ($\mathbf{\bar{R}}_{\boldsymbol{\mu\nu}}$)~\cite{cdcro} as
\begin{align}
    J^{\mu\nu}_{\alpha}({\bf r})=J_{\alpha}&+\frac{\partial J^{\mu\nu}_{\alpha}({\bf r})}{\partial R^a_{\mu\nu}}\delta^a_{\mu\nu}({\bf r})\nonumber\\
    &+\frac{1}{2}\frac{\partial^2J_{\alpha}^{\mu\nu}({\bf r})}{\partial R^a_{\mu\nu}\partial R^b_{\mu\nu}}\delta^a_{\mu\nu}({\bf r})\delta^b_{\mu\nu}({\bf r})
    \label{eq_exchange coupling constant}
\end{align}
Here, $J^{\mu\nu}_{\alpha}({\bf r})$ denotes the generic bond dependent exchange coupling constant on the bond of the pyrochlore connecting the sites $({\bf r},\boldsymbol{\mu})$ and $({\bf r},\boldsymbol{\nu})$. Here, $(\mathbf{r},{\boldsymbol{\mu}})$ denotes the position vector of the four spins sitting on the corners of the tetrahedron with its centre at $\mathbf{r}$ for $\mu=0,1,2,3$, with $\alpha$ representing $zz$ or $\pm$ interactions, and, 
\begin{align}
    &R^a_{\mu\nu}=(\mathbf{r},{\boldsymbol{\mu}})^a-(\mathbf{r},{\boldsymbol{\nu}})^a, \indent (a=x,y,z) \nonumber\\ 
    &\delta^a_{\mu\nu}=R^a_{\mu\nu}-\bar{R}^a_{\mu\nu}\nonumber
\end{align}

\begin{figure}
\centering
\subfigure[]{
    {\begin{tikzpicture}
    \begin{feynman}
    \vertex (a);
    \vertex [right=of a] (b);
    \vertex [above right=2cm of b] (f1);
    \vertex [below right=2cm of b] (f2);
    \vertex [above right=2cm of f1] (f3);
    \vertex [below right=1cm of f2] (f4); \diagram* {
    (a) -- [dotted, thick, edge label'=\(\mathbf{q}\)] (b) -- [edge label=\(\mathbf{k+q}\)] (f1),
    (b) -- [ edge label'=\(\mathbf{\textcolor{black}{k}}\)] (f2),
    };
    \end{feynman}
    \end{tikzpicture}}
    }
    \subfigure[]{
    \begin{tikzpicture}
    \begin{feynman}
    \vertex (a);
    \vertex [right=of a] (b);
    \vertex [above right=2cm of b] (f1);
    \vertex [below right=2cm of b] (f2);
    
    \diagram* {
    (a) -- [dotted, thick, edge label'=\(\mathbf{q}\)] (b) -- [line width=0.5mm, edge label=\(\textcolor{black}{\mathbf{k+q}}\)] (f1),
    (b) -- [line width=0.5mm, edge label'=\(\mathbf{\textcolor{black}{k}}\)] (f2),
    };
    \end{feynman}
    \end{tikzpicture}
    }
    \subfigure[]{
    \begin{tikzpicture}
    \begin{feynman}
    \vertex (a);
    \vertex [above left=2cm of a] (f1);
    \vertex [below left=2cm of a] (f2);
    \vertex [above right=2cm of a] (f3);
    \vertex [below right=2cm of a] (f4);
    
    \diagram* {
    (f1) -- [dotted, thick, edge label'=\(\mathbf{q}\)] (a) -- [ edge label'=\(\textcolor{black}{\mathbf{q+q'+k}}\)] (f3),
    (f2) -- [dotted, thick, edge label=\(\mathbf{q'}\)] (a)-- [ edge label=\(\mathbf{\textcolor{black}{k}}\)] (f4),
    };
    \end{feynman}
    \end{tikzpicture}
    }
    \subfigure[]{
    \begin{tikzpicture}
    \begin{feynman}
    \vertex (a);
    \vertex [above left=2cm of a] (f1);
    \vertex [below left=2cm of a] (f2);
    \vertex [above right=2cm of a] (f3);
    \vertex [below right=2cm of a] (f4);
    
    \diagram* {
    (f1) -- [dotted, thick, edge label'=\(\mathbf{q}\)] (a) -- [line width=0.5mm, edge label'=\(\textcolor{black}{\mathbf{q+q'+k}}\)] (f3),
    (f2) -- [dotted, thick, edge label=\(\mathbf{q'}\)] (a)-- [line width=0.5mm, edge label=\(\mathbf{\textcolor{black}{k}}\)] (f4),
    };
    \end{feynman}
    \end{tikzpicture}
    }
    \caption{\small {\bf Feynman diagram for phonon and magnetic monopole interaction due to the spin-phonon coupling quadratic in spin operators:} (a) and (b) are the contributions from $H_1$, and, (c) and (d) are the contributions from $H_2$ (see Eq. \ref{eq_quadratic_qsl_1} and \ref{eq_quadratic_qsl_2}). }
    \label{fig_phonon monopole interaction_quad}
\end{figure}
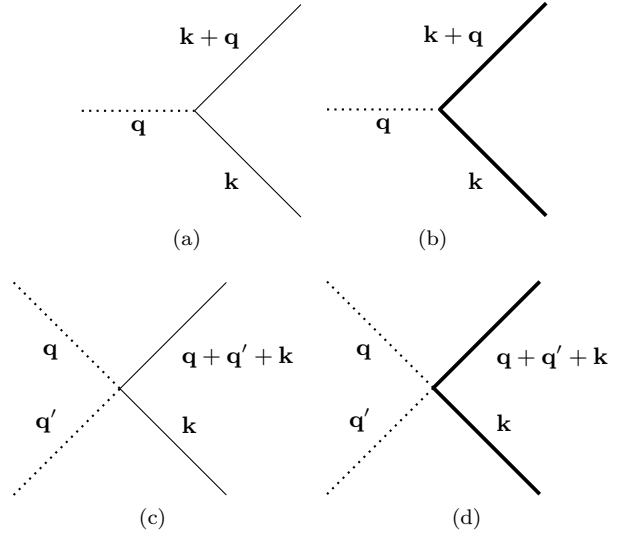

Substituting Eq. \ref{eq_exchange coupling constant} in the spin-Hamiltonian of Eq. \ref{eq_spinham}, we get the coupling between the phonons and spin-bilinears, 
\begin{align}
    H^{quad}_{sp}=H_1+H_2
    \label{eq_quadratic spin phonon coupling_0}
\end{align}
where $H_1$ and $H_2$ represent the interaction vertices linear and quadratic in phonons, respectively. Their detailed forms are given by Eq. \ref{eq_quadratic spin phonon coupling_1} and \ref{eq_quadratic spin phonon coupling_2} in Appendix \ref{appendix_quadratic}. A unitary transformation can be performed on the displacement operators, $\boldsymbol{\delta}_{\mu\nu}(\mathbf{r})$, to re-write it in the normal mode coordinates, $\boldsymbol{\zeta}^{(\rho)}(\mathbf{r})$, described in Sec. \ref{sec_3}. The above interaction is re-written in terms of the fractionalised degrees of freedom in a QSL phase using the parton decomposition of spins as described in Sec. \ref{sec_4}. Within GMFT, the quadratic magnetoelastic coupling between the phonons and emergent excitations of the QSL is described by Fig. \ref{fig_phonon monopole interaction_quad} and \ref{fig_phonon photon interaction_quad} (also see Eq. \ref{eq_quadratic_qsl_1} and \ref{eq_quadratic_qsl_2} in Appendix \ref{appendix_quadratic}).

{The phonon-magnetic monopole vertex arising from the quadratic coupling is shown in Fig \ref{fig_phonon monopole interaction_quad} where (a) and (b) panels show the contribution from $H_1$ , and, (c) and (d) panels show the contribution from $H_2$. It is clear from these diagrams that such magnetoelastic coupling gives rise to the hopping of the monopoles which preserves the monopole flavour, i.e., monopoles on A and B sublattices do not mix under this dynamics. This feature can be contrasted with the monopole dynamics due to the linear magnetoelastic coupling described earlier in Eq. \ref{eq_eg_phonon monopole coupling} and \ref{eq_t2g_phonon monopole coupling}.} 

{The quadratic coupling also generates a coupling between phonons and emergent photons which is shown in Fig. \ref{fig_phonon photon interaction_quad} with (a) and (b) panels depicting contributions from $H_1$ and $H_2$, respectively. In contrast to the linear coupling case, the phonons now couple to the gauge invariant magnetic field. As expected from time reversal invariance of the phonons, the magnetic field appears only at quadratic order in such couplings.} We also note that the process shown in Fig. \ref{fig_phonon photon interaction_quad}(a) is in the single phonon scattering channel, which was not present in the previous case. 
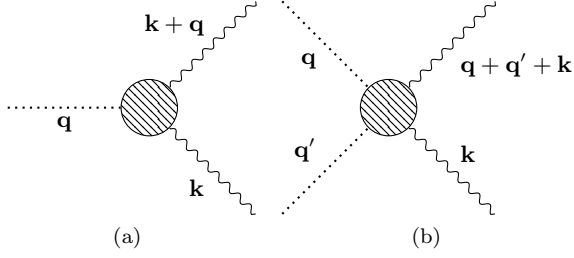
\begin{figure}
\centering
\subfigure[]{
    {\begin{tikzpicture}
    \begin{feynman}
    \vertex (a);
    \vertex [blob, right=of a] (b) {};
    \vertex [above right=2cm of b] (f1);
    \vertex [below right=2cm of b] (f2);
    \vertex [above right=2cm of f1] (f3);
    \vertex [below right=1cm of f2] (f4); \diagram* {
    (a) -- [dotted, thick, edge label'=\(\mathbf{q}\)] (b) -- [boson, edge label=\(\mathbf{k+q}\)] (f1),
    (b) -- [boson, edge label'=\(\mathbf{\textcolor{black}{k}}\)] (f2),
    };
    \end{feynman}
    \end{tikzpicture}}
    }
    \subfigure[]{
    \begin{tikzpicture}
    \begin{feynman}
    \vertex [blob] (a) {};
    \vertex [above left=2cm of a] (f1);
    \vertex [below left=2cm of a] (f2);
    \vertex [above right=2cm of a] (f3);
    \vertex [below right=2cm of a] (f4);
    
    \diagram* {
    (f1) -- [dotted, thick, edge label'=\(\mathbf{q}\)] (a) -- [boson, edge label'=\(\textcolor{black}{\mathbf{q+q'+k}}\)] (f3),
    (f2) -- [dotted, thick, edge label=\(\mathbf{q'}\)] (a)-- [boson, edge label=\(\mathbf{\textcolor{black}{k}}\)] (f4),
    };
    \end{feynman}
    \end{tikzpicture}
    }
    \caption{\small {\bf Feynman diagram for phonon and photon interaction due to the spin-phonon coupling quadratic in spin operators:} The circles represent the form factor that makes the vertex gauge invariant. (a) and (b) are contributions due to $H_1$ and $H_2$, respectively (see Eq. \ref{eq_quadratic_qsl_1} and \ref{eq_quadratic_qsl_2}).}
    \label{fig_phonon photon interaction_quad}
\end{figure}

{Similar to the linear magnetoelastic interaction, the quadratic coupling also renormalises the phonon frequency and linewidth by opening up the decay channels for phonons depicted in Fig. \ref{fig_phonon monopole interaction_quad} and \ref{fig_phonon photon interaction_quad}. However, these new scattering channels do not change the essential features of the Raman linewidth, and its frequency dependence on the density of states of emergent excitations remains unchanged. In the non-Kramers materials, this contribution is expected to be sub-dominant compared to the phonon renormalisation due to the linear spin-phonon coupling. However, we note that the quadratic coupling is the only component of the magnetoelastic coupling present in the Kramers materials.}

%%%%%%%%%%%%%%%%%%%%%%%%%%%%%%%%%%%%%%

\section{\label{sec_11}Self-energy calculation in (thermal) paramagnetic regime}

Finally, to contrast the case of the QSL to an ordinary paramagnet, we compute the self-energy of phonon in the high-temperature paramagnetic regime, where $T\gg J_{zz}$ such that the thermal fluctuations predominate.  In this thermal paramagnet, {due to the presence of abundant thermally excited both electric charges and magnetic monopoles, they cease to be well-defined (sparse) quasiparticles, instead presenting randomly fluctuating background fields. In such a case, individual monopoles or charges cannot propagate coherently, and, the deconfined $U(1)$ gauge theory no longer is a valid description of the system.} Instead of using the emergent excitations, the dressed self-energy due to the spin-phonon interaction (as described in Eq. \ref{eq_Eg coupling} and \ref{eq_T2g coupling}) is now computed in terms of the original short-range correlated spin degrees of freedom. The phonon self-energy due to the linear spin-phonon coupling is given, {\it e.g.} for ${\bf e_g}$ modes, by
\begin{widetext}
\begin{align}
    &\Sigma_{\zeta^{(e)}}(\mathbf{q},i\Omega)=-J^{(e)2}_{sp}\sum_{\alpha=x,y}\sum_{\mu,\nu=0}^3\left[\eta^{(e)}_{\mu\nu}(\mathbf{q})\kappa^{\alpha\alpha}_{\mu\nu}(\mathbf{q},i\Omega)+\eta^{(e)}_{\mu\nu}(-\mathbf{q})\kappa^{\alpha\alpha}_{\mu\nu}(-\mathbf{q},-i\Omega)\right]
    \label{eq_self energy in paramagnet}
\end{align}
\noindent where
\end{widetext}
\begin{align}
&\kappa_{\mu \nu}^{\alpha\beta}(\mathbf{q},i\omega)=\int_0^\beta~d\tau\langle \hat{\mathcal{T}}\left(s_\mu^{\alpha}(\mathbf{q},\tau)s_{\nu}^{\beta}(-\mathbf{q},0)\right)\rangle_0 e^{i\omega \tau}
\end{align}
is the time ordered spin correlation function and $\eta^{(e)}_{\mu\nu}(\mathbf{q})$ 
is the form factor for the $\mathbf{e_g}$ mode. Similar expressions hold for $\mathbf{t_{2g}}$ modes as discussed in Appendix \ref{appendix_paramagnetic}.

From the bare spin Hamiltonian, we expect the spin-correlations to be diagonal {in the spin indices (defined using the local quantisation axes given by Eqs. \ref{eq_spin quantisation} and \ref{eq_local basis})}, {\it i.e.},  $\kappa^{\alpha\beta}_{\mu\nu}(\mathbf{q},i\omega)=\delta_{\alpha\beta}\kappa^{\alpha\beta}_{\mu\nu}(\mathbf{q},i\omega)$. Further, in this thermal  paramagnetic phase, the spins are incoherent and, therefore, the spin correlations are dominated by the short time values which we replace by the equal time correlators, which in turn can be computed from the high-temperature expansion using the bare spin-exchange Hamiltonian. The leading contribution is given by,
\begin{align}
    \langle s^x_\mathbf{r}s^x_{\mathbf{r'}}\rangle_0=\langle s^y_\mathbf{r}s^y_{\mathbf{r'}}\rangle_0\approx e^{-\frac{\mid \mathbf{r-r'}\mid}{\xi}}
\end{align}

\noindent where $\xi\sim 1/\ln{\left(\frac{T}{J_{\pm}}\right)}$ is the finite correlation length in the paramagnetic phase. Taking the Fourier transform and substituting it in in Eq. \ref{eq_self energy in paramagnet}, we obtain,

\begin{align}
    &\Sigma_{\zeta^{(\rho)}}(\mathbf{q},i\Omega)\propto-\frac{J^{(\rho)2}_{sp}\xi^3\beta}{\sqrt{N}(1+q^2\xi^2)^2}
    \label{eq_temp dependence at paramegnet}
\end{align}
The above expression is purely real, and hence contributes only to a Raman frequency shift that decays inversely with temperature. 

Therefore the leading effect of the spin-phonon coupling is to renormalise the phonon energy while its lifetime receives sub-leading contributions. Therefore, the Raman linewidth for the phonons acquires an anomalous broadening while going from the high-temperature paramagnetic phase to the low-temperature QSL.  This leads to the question what happens to the linewidth at the thermal confinement-deconfinement phase transition between the low-temperature quantum and high-temperature thermal paramagnets? This is an interesting and experimentally relevant question which will be very useful to understand in the future. 

%%%%%%%%%%%%%%%%%%%%

\section{\label{sec_12}Phonon mediated Loudon-Fleury vertex}

In addition to the renormalisation of optical phonons, the magnetoelastic coupling can further mediate interaction between the external Raman photons and the magnetic degrees of freedom. Such interactions are of particular interest in those materials where the phonon has a very different energy scale compared to the QSL excitations~\cite{natalia2021,ruminy2016}. In such a scenario, the renormalised Raman vertex is obtained by integrating out the phonons leading to a \textit{phonon mediated Loudon-Fleury} vertex.

As explained in the earlier sections, the external photons of the Raman experiment probe the phonons of the system  (via Eq. \ref{eq_raman vertex}), which further couple to the fractionalised excitations via the magnetoelastic coupling (see Eq. \ref{eq_eg_phonon monopole coupling}, \ref{eq_t2g_phonon monopole coupling}, \ref{eq_phonon-photon coupling}, \ref{eq_phonon_charge coupling}). Therefore, integrating out the phonons leads to an interaction between the external photons and the emergent electrodynamics (see Eqs. \ref{eq_LF1} and \ref{eq_LF2} in Appendix \ref{appendix_loudon-fleury} for further details)  as shown schematically in Fig. \ref{fig_LF vertex} for the leading interaction between external photons and magnetic monopoles. The vertices for the other emergent excitations are detailed in Appendix \ref{appendix_loudon-fleury}.

\begin{figure}
\centering
    {\begin{tikzpicture}
    \begin{feynman}
    \vertex  (a);
    \vertex [left=of a] (b);
    \vertex [above right=1.2cm of a] (f1);
    \vertex [below right=1.2cm of a] (f2);
    \vertex [above left=1.2cm of b] (f3);
    \vertex [below left=1.2cm of b] (f4);
    \diagram* {
    (f3) -- [gluon] (b) -- [ half left] (a) -- [gluon] (f1), 
    (f4) -- [gluon] (b) -- [line width=0.5mm, half right] (a) -- [gluon] (f2)     };
    \end{feynman}
    \end{tikzpicture}}
    \caption{\small{ Feynman diagram contributing to the Raman intensity due to the {\it phonon mediated Loudon-Fleury} vertex for magnetic monopoles.}}
    \label{fig_LF}
\end{figure}
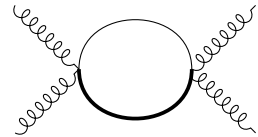

 Typically, all the phonon mediated vertices are suppressed by the energy scale of an optical phonon and would lead to corrections to the usual Loudon-Fleury vertices described in Ref. \cite{rau2017}. Raman intensity due to such processes is obtained by calculating the imaginary part of the bubble diagram shown in Fig. \ref{fig_LF}. It is clear from the diagram that the resulting monopole bubble is exactly same as the one obtained (see Fig. \ref{fig_bubble}) earlier. Therefore, the Raman intensity due to the \textit{phonon mediated Loudon-Fleury} processes are sensitive to the two-monopole density of states in the QSL phase and it can in principle also characterise the physics of spin fractionalisation even if the phonon is off-resonant to the quasiparticles of the QSL phase.

\begin{figure}
\centering
\begin{align*}
    \vcenter{\hbox{%
    \begin{tikzpicture}
    \begin{feynman}
    \vertex (a) at (0,0);
    \vertex [right=1.2cm of a] (i1);
    \vertex [below left=1.5cm of a] (f1);
    \vertex [above left=1.5cm of a] (f2);
    \diagram* {
    (i1) -- [dotted, thick] (a) -- [gluon] (f1),
    (a) -- [gluon] (f2),};
    \end{feynman}
    \end{tikzpicture}
    }}
    &~+~
    \vcenter{\hbox{%
    \begin{tikzpicture}
    \begin{feynman}
    \vertex (a) at (0,0);
    \vertex [left=1.2cm of a] (i1);
    \vertex [below right=1.5cm of a] (f1);
    \vertex [above right=1.5cm of a] (f2);
    \diagram* {
    (i1) -- [dotted, thick] (a) -- [line width=0.5mm] (f1),
    (a) -- [] (f2),
    };
    \end{feynman}
    \end{tikzpicture}
    }}
    &\longrightarrow 
    &\vcenter{\hbox{%
    \begin{tikzpicture}
    \begin{feynman}
    \vertex (a) at (0,0);
    \vertex [above left=1.5cm of a] (i1);
    \vertex [below left=1.5cm of a] (i2);
    \vertex [above right=1.5cm of a] (f1);
    \vertex [below right=1.5cm of a] (f2);
    \diagram* {
    (i1) -- [gluon] (a) -- [] (f1),
    (i2) -- [gluon] (a) -- [line width=0.5mm] (f2),
    };
    \end{feynman}
    \end{tikzpicture}}}
\end{align*}
    \caption{\small {\bf Feynman diagram for the {phonon mediated Loudon-Fleury} vertex for magnetic monopoles:} The curly lines denote the external Raman photons. The first and second figure show the coupling of phonon to external photon (see Eq. \ref{eq_raman}) and magnetic monopole, respectively. Integrating out the phonons, the external photon-monopole (see Eq. \ref{eq_LF1}) vertex is obtained which is shown in the rightmost panel.}
    \label{fig_LF vertex}
\end{figure}
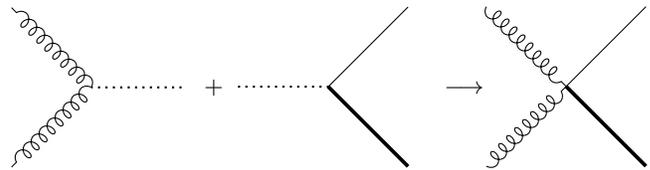

\section{Summary and Outlook}

To summarise, our results emphasise that magnetoelastic coupling can provide a very useful tool to probe the novel low-energy excitations of a  QSL via spectroscopic methods such as Raman spectroscopy. We provide an explicit example of such a case in the context of non-Kramers candidate quantum spin ices, such as the recently studied material Pr$_2$Zr$_2$O$_7$. In such systems, the spin-phonon interaction is enhanced due to the presence of {\it linear} spin-phonon coupling, which is an essential consequence of the non-Kramers nature of the low-energy magnetic degrees of freedom. We show that in the U(1) QSL phase of the quantum spin ice, all the emergent excitations--the emergent gapped magnetic and electric charges as well as gapless emergent photons--interact with the phonons leading to new scattering channels for the latter and resulting in an anomalous renormalisation of its frequency and lifetime. Such renormalisations are very different from those arising due to the anharmonic effects or spin-wave excitations of a magnetically ordered state. 

We characterise all the three types of excitations by studying the frequency dependence of the Raman linewidth of the relevant phonon modes, which in turn depend on the two-particle density of states of the respective excitations. Therefore, they carry characteristic signatures of the fractionalisation. Since in the quantum spin ice phase, the magnetic sectors and the electric sectors are naturally separated in energy, the Raman response also appears in different energy windows for these degrees of freedom alongside the gapless emergent photon to which both the charges couple. 

Moreover, it is further shown that such probes can also distinguish between zero flux and $\pi-$flux phases of a QSL, and hence the PSG implementation realised in the QSL. The results remain valid even if the phonon frequencies are much larger than those of QSL excitations via renormalisation of the Raman vertex for the spins. Such {\it phonon mediated Loudon-Fleury} contributions are expected to be the leading contributor to the Raman response in non-Kramers quantum spin ice. Given the recent development in synthesising high-quality single crystals of Pr$_2$Zr$_2$O$_7$ and obtaining their Raman response, albeit so far only at high temperatures, we hope that our work will contribute to the uncovering of the experimental signatures of QSLs in the context of the search for fractionalised quantum phases of matter in $d=3$.

The present calculations use a generalised mean field theory applied to lattice gauge theory. While such approaches can generally provide the correct description of the physics of the QSL qualitatively, fluctuations of the emergent U(1) gauge field will affect the quantitative comparison of the present results with experiments. While a perturbative (in the gauge-matter coupling) calculation, presented here, shows that such effects are sub-dominant, the premise of the smallness of the coupling is an assumption of the present work. In fact, there are  indications that the coupling between matter (monopoles) and light (photons) is larger in spin ice than in ordinary QED, as well as tunable from material to material, calling for a research programme addressing the phenomenology of eQED at intermediate to strong-coupling~\cite{PhysRevLett.127.117205}. The deviation of the present results due to strong gauge-matter interactions provides an important and interesting theoretical as well as experimental context to study strong-coupling eQED. In this regard, experimental deviations of the above Raman signatures in candidate materials will provide a concrete motivation to understand concrete experimental consequence of such a strongly-coupled emergent QED. Such a research program would obviously be of interest well beyond the spin ice setting.

%%%%%%%%%%%%%%%%%%
\acknowledgments
We thank S. Nakatsuji, A. Sood, N. Drichko and J. Knolle for discussion and related collaborations. A.S. and S.B. thank S. Pal for various discussions. The authors acknowledge funding from Max Planck Partner group Grant at ICTS. A.S. and S.B. acknowledge the support of SERB-DST (India) for funding through project Grant No. ECR/2017/000504 and the Department of Atomic Energy, Government of India, under Project No. RTI4001. This work was in part supported by the Deutsche Forschungsgemeinschaft under grants SFB 1143 (project-id 247310070) and the cluster of excellence ct.qmat (EXC 2147, project-id 390858490).

%%%%%%%%%%%%%%%%%%%
\appendix

\section{Details of the pyrochlore lattice}

\subsection{\label{appendix_spin axis}Local basis for the spins}

The spins on a tetrahedron are described by,
\begin{align}
\mathbf{s}_i=s^x_i\hat{x}_i+s^y_i\hat{y}_i+s^z_i\hat{z}_i
\label{eq_spin quantisation}
\end{align}
where, $(\hat{x}_i,\hat{y}_i,\hat{z}_i)$ is the set of local basis defined at site $i$ of a tetrahedron(see Fig. \ref{fig_local basis}). In terms of the global coordinates, these local basis vectors for an up tetrahedron are given by,

\begin{align}
    &\hat{z}_0=\frac{(1,1,1)}{\sqrt{3}}~,~
    &\hat{x}_0=\frac{(\bar{2},1,1)}{\sqrt{6}}~,~
    &\hat{y}_0=\frac{(0,\bar{1},1)}{\sqrt{2}} \nonumber\\
    &\hat{z}_1=\frac{(\bar{1},\bar{1},1)}{\sqrt{3}}~,~
    &\hat{x}_1=\frac{(2,\bar{1},1)}{\sqrt{6}}~,~
    &\hat{y}_1=\frac{(0,1,1)}{\sqrt{2}} \nonumber\\
    &\hat{z}_2=\frac{(\bar{1},1,\bar{1})}{\sqrt{3}}~,~
    &\hat{x}_2=\frac{(2,1,\bar{1})}{\sqrt{6}}~,~
    &\hat{y}_2=\frac{(0,\bar{1},\bar{1})}{\sqrt{2}} \nonumber\\
    &\hat{z}_3=\frac{(1,\bar{1},\bar{1})}{\sqrt{3}}~,~
    &\hat{x}_3=\frac{(\bar{2},\bar{1},\bar{1})}{\sqrt{6}}~,~
    &\hat{y}_3=\frac{(0,1,\bar{1})}{\sqrt{2}} 
    \label{eq_local basis}
\end{align}

%%%%%%%%%%%%%%%%%%%%

\subsection{\label{appendix_lattice vectors}Lattice vectors}

The four nearest neighbour vectors, which connect the centre of an up tetrahedron to that of its adjacent down tetrahedra, are given by,

\begin{align}
    &\bf{e_0}=\frac{(1,1,1)}{\sqrt{3}}~~&~~\bf{e_1}=\frac{(\bar{1},\bar{1},1)}{\sqrt{3}}\nonumber\\
    &\bf{e_2}=\frac{(\bar{1},1,\bar{1})}{\sqrt{3}}~~&~~\bf{e_3}=\frac{(1,\bar{1},\bar{1})}{\sqrt{3}}
\end{align}

The FCC lattice vectors are given by,
\begin{align}
    {\bf d_\mu}={\bf e_0-e_\mu}~,~~~~~~ for ~~~\mu=1,2,3
\end{align}

%%%%%%%%%%%%%%%%%%%%%%%%%%%%%%%%%

%%%%%%%%%%%%%%%%%%%%%%%%%%%%%%%%%%%%%%%%%
\subsection{\label{appendix_symmetry of spins}Symmetry table for spins}

The tetrahedral group, $T_d$, is made out of 24 symmetry elements which can further be classified into 5 conjugacy classes. To decompose the vector space of $(s^x_i,s^y_i)$ operators into the irreducible representations (see Sec. \ref{sec_3}), we compute their transformations under one representative symmetry transformation from each non-trivial class : $C_3[111]$ (three-fold rotation about the global $(1,1,1)$ axis), $C_2[\hat{z}]$ (two-fold rotation about the global $\hat{z}$ axis), $\sigma_d[x=y]$ (reflection about the $x=y$ plane) and $S_4[\hat{z}]$ (reflection about the $z=0$ plane followed by four-fold rotation about the global $\hat{z}$ axis). This is given in Table \ref{table_spin transformation} for the transverse spin components, $s^x_i$ and $s^y_i$. Here we do not consider time reversal odd $s^z_i$ operators, since these are not relevant to the linear magnetoelastic coupling.

\begin{table}[h!]
\begin{center}
\begin{tabular}{ | c | c| }
\hline
Symmetry &  Transformation of spin operators \\ 
\hline
$C_3[111]$  
& ~$s_0^x\rightarrow\frac{1}{2}s_0^x+\frac{\sqrt{3}}{2}s_0^y ~~,~~ s_0^y\rightarrow-\frac{\sqrt{3}}{2}s_0^x-\frac{1}{2}s_0^y$~~\\
& ~$s_1^x\rightarrow -\frac{1}{2}s_2^x+\frac{\sqrt{3}}{2}s_2^y~~,~~ s_1^y\rightarrow-\frac{\sqrt{3}}{2}s_2^x-\frac{1}{2}s_2^y$~~\\
& ~$s_2^x\rightarrow-\frac{1}{2}s_3^x+\frac{\sqrt{3}}{2}s_3^y ~~,~~ s_2^y\rightarrow-\frac{\sqrt{3}}{2}s_3^x-\frac{1}{2}s_3^y$~~\\
& ~$s_3^x\rightarrow -\frac{1}{2}s_1^x+\frac{\sqrt{3}}{2}s_1^y ~~,~~ s_3^y\rightarrow-\frac{\sqrt{3}}{2}s_1^x-\frac{1}{2}s_1^y$~~\\

\hline

$C_2[\hat{z}]$ 
& $s_0^x\rightarrow s_1^x ~~;~~ s_0^y\rightarrow s_1^y$\\
& $s_1^x\rightarrow s_0^x ~~;~~ s_1^y\rightarrow s_0^y$\\
& $s_2^x\rightarrow s_3^x~~;~~ s_2^y\rightarrow s_3^y$\\
& $s_3^x\rightarrow s_2^x~~;~~ s_3^y\rightarrow s_2^y$\\ 
\hline
$\sigma_d[x=y]$ 
& ~$s_0^x\rightarrow -\frac{1}{2}s_0^x-\frac{\sqrt{3}}{2}s_0^y ~~,~~ s_0^y\rightarrow -\frac{\sqrt{3}}{2}s_0^x+\frac{1}{2}s_0^y$~~\\
& ~$s_1^x\rightarrow -\frac{1}{2}s_1^x-\frac{\sqrt{3}}{2}s_1^y~~,~~ s_1^y\rightarrow -\frac{\sqrt{3}}{2}s_1^x+\frac{1}{2}s_1^y$~~\\
& ~$s_2^x\rightarrow -\frac{1}{2}s_3^x-\frac{\sqrt{3}}{2}s_3^y~~,~~ s_2^y\rightarrow -\frac{\sqrt{3}}{2}s_3^x+\frac{1}{2}s_3^y$~~\\
& ~$s_3^x\rightarrow -\frac{1}{2}s_2^x-\frac{\sqrt{3}}{2}s_2^y~~,~~ s_3^y\rightarrow -\frac{\sqrt{3}}{2}s_2^x+\frac{1}{2}s_2^y$~~\\ 
\hline
$S_4[\hat{z}]$ 
& ~$s_0^x\rightarrow -\frac{1}{2}s_2^x-\frac{\sqrt{3}}{2}s_2^y~~,~~ s_0^y\rightarrow -\frac{\sqrt{3}}{2}s_2^x+\frac{1}{2}s_2^y$~~\\
& ~$s_1^x\rightarrow -\frac{1}{2}s_3^x-\frac{\sqrt{3}}{2}s_3^y~~,~~ s_1^y\rightarrow -\frac{\sqrt{3}}{2}s_3^x+\frac{1}{2}s_3^y$~~\\
& ~$s_2^x\rightarrow -\frac{1}{2}s_1^x-\frac{\sqrt{3}}{2}s_1^y~~,~~ s_2^y\rightarrow -\frac{\sqrt{3}}{2}s_1^x+\frac{1}{2}s_1^y$~~\\
& ~~$s_3^x\rightarrow -\frac{1}{2}s_0^x-\frac{\sqrt{3}}{2}s_0^y ~~,~~ s_3^y\rightarrow -\frac{\sqrt{3}}{2}s_0^x+\frac{1}{2}s_0^y$~~\\ 
\hline
$\mathcal{T}$
& $s_0^x\rightarrow s_0^x~~;~~ s_0^y\rightarrow s_0^y$\\
& $s_1^x\rightarrow s_1^x~~;~~ s_1^y\rightarrow s_1^y$\\
& $s_2^x\rightarrow s_2^x~~;~~ s_2^y\rightarrow s_2^y$\\
& $s_3^x\rightarrow s_3^x~~;~~ s_3^y\rightarrow s_3^y$\\ 
\hline
\end{tabular}
\caption{Transformation of the transverse components of spins under lattice symmetries and time reversal}
\label{table_spin transformation}
\end{center}
\end{table}

%%%%%%%%%%%%%%%%%%%%%%%%%%

%%%%%%%%%%%%%%%%%%%%%%%%%%%%%%%%
\section{\label{appen_lgt}Details of the GMFT of quantum spin ice}

The lattice gauge theory description of the spin Hamiltonian in Eq. \ref{eq_spinham} is given by  
\begin{align}
    H_0=&\sum_\mathbf{r}\frac{J_{zz}}{2}(\mathcal{Q}_{\mathbf{r},A}^2+\mathcal{Q}_{\mathbf{r},B}^2)\nonumber\\
    &-\frac{J_{\pm}}{4}\sum_{\mathbf{r}, \mu\neq\nu}\phi^\dagger_{\mathbf{r}+\mathbf{d}_{\boldsymbol{\mu}},B}~e^{i(A_{\mathbf{r},\boldsymbol{\nu}}-A_{\mathbf{r},\boldsymbol{\mu}})}~\phi_{\mathbf{r}+\mathbf{d}_{\boldsymbol{\nu}},B}\nonumber\\
    &-\frac{J_{\pm}}{4}\sum_{\mathbf{r}, \mu\neq\nu}\phi^\dagger_{\mathbf{r}-\mathbf{d}_{\boldsymbol{\mu}},A}~e^{i(A_{\mathbf{r}-\mathbf{d}_{\boldsymbol{\mu}},\boldsymbol{\mu}}-A_{\mathbf{r}-\mathbf{d}_{\mathbf{\nu}},\boldsymbol{\nu}})}\phi_{\mathbf{r}- \mathbf{d}_{\boldsymbol{\nu}},A}~
    \label{eq_bare monopole hamiltonian}
\end{align}
subject to the hard-core constraint 
\begin{align}
    \phi^\dagger_\mathbf{r}\phi_\mathbf{r}=1
    \label{eq_lagrange}
\end{align}
arising from the spin-1/2 Hilbert space dimension of the non-Kramers doublet. In Eq. \ref{eq_bare monopole hamiltonian}, $A, B$ denote two sublattices of the diamond lattice. $\mathbf{d_\mu}$s ($\mu=1,2,3$) are the lattice vectors and $\mathbf{d_0}=0$ (see Appendix \ref{appendix_lattice vectors}).

%%%%%%%%%%%%%%%%%

\subsection{{\label{appen_zero_mon_band}}GMFT of monopole dynamics in the zero-flux phase}

To implement the unitary constraint of the monopole operators described by Eq. \ref{eq_lagrange}, we introduce a new term to the Hamiltonian with a global Lagrange multiplier, $\lambda$.
\begin{align}
    \lambda\sum_{\bf r}\left(\phi^\dagger_{\bf r}\phi_{\bf r}-1\right)
\end{align}
The constraint is imposed softly if we consider $\lambda$ to be a large number (more precisely, it needs to be the largest energy scale of the problem). With this term, the constraint can be relaxed by rewriting monopole operators as $\phi^\dagger_{\mathbf{r}}=e^{i\chi_{\mathbf{r}}}$~(where $\chi_{\mathbf{r}}$ takes real eigenvalues from $(0,2\pi]$ and satisfies $\left[\chi_{\mathbf{r}},\mathcal{Q}_{\mathbf{r'}}\right]=i\delta_{\mathbf{r,r'}}$) and expanding it up to linear order of $\chi_{\bf r}$.
\begin{align}
    \phi^\dagger_{\bf r}\approx 1+i\chi_{\bf r}
\end{align}
Substituting the above expansion in the bare monopole Hamiltonian (obtained by freezing the dual gauge fluctuations in Eq. \ref{eq_bare monopole hamiltonian}) along with the Lagrange multiplier term, we obtain,
\begin{align}
    H_{0}&\approx \frac{J_{zz}}{2}\sum_{\bf r}\mathcal{Q}_{{\bf r},B}^2-\frac{J_{\pm}}{4}\sum_{{\bf r},\mu\neq\nu}\chi_{{\bf r+d_\mu},B}\chi_{{\bf r+d_\nu},B}\nonumber\\
    &\hspace{3cm}+\lambda\sum_{\bf r}\chi_{\bf r,B}\chi_{{\bf r},B}+~~B\rightarrow A\nonumber\\
    &=\sum_{\bf k}\left[\frac{J_{zz}}{2}\mid \mathcal{Q}_{{\bf k},B}\mid^2+\frac{\left(\epsilon^0_{\bf k}\right)^2}{2J_{zz}}\mid\chi_{{\bf k},B}\mid^2\right]~~+~~B\rightarrow A
\end{align}
where $\epsilon^0_{\bf k}$ is the bare monopole dispersion in the zero flux sector and given by Eq. \ref{eq_monopole dispersion}. In the above equation, we ignore the unimportant additive constant. We note that the above Hamiltonian describes a bunch of decoupled Harmonic oscillators which can be easily diagonlised using the standard ladder operator formalism.

\begin{align*}
   &\mathcal{Q}_{{\bf k},A/B}=-i \sqrt{\frac{\epsilon^0_{\bf k}}{2J_{zz}}}(a_{{\bf k},A/B}-a_{-{\bf k},A/B}^\dagger)\nonumber\\
   &\chi_{{\bf k},A/B}= \sqrt{\frac{J_{zz}}{2\epsilon^0_{\bf k}}}(a_{{\bf k},A/B}+a_{-{\bf k},A/B}^\dagger)\indent
\end{align*}

The monopole Hamiltonian is simplified to,
\begin{align}
    H_0=\sum_{\bf k}\epsilon^0_{\bf k}\left(a_{{\bf k},A}^\dagger a_{{\bf k},A}+a_{{\bf k},B}^\dagger a_{{\bf k},B}+1\right)   
\end{align}
%%%%%%%%%%%%%%%

\subsubsection{\label{appendix_action of monopoles}Action for the magnetic monopoles}

We can obtain the action for the monopoles corresponding to the mean field Hamiltonian $H_0$ using standard Trotter decomposition technique and implementing the unitary constraint of $\phi_{\bf r}$ field via the Lagrange multiplier term in the path integral formulation. We note that since A and B monopoles are decoupled in the mean field Hamiltonian, their action is additive.
\begin{align}
    &S_0=S_0^A+S_0^B\nonumber\\
    &S_0^A=\sum_{{\bf k},\omega}\phi^*_{\bf{k}\omega,A}\left[\frac{\omega^2}{2J_{zz}}+\lambda-\frac{J_{\pm}}{2}\sum_{\mu>\nu}\cos\left(\bf{k\cdot(d_\mu-d_\nu)}\right)\right]\nonumber\\
    &\hspace{7.1cm}\times\phi_{\bf{k}\omega,A}\nonumber\\
    &S_0^B=\sum_{{\bf k},\omega}\phi^*_{\bf{k}\omega,B}\left[\frac{\omega^2}{2J_{zz}}+\lambda-\frac{J_{\pm}}{2}\sum_{\mu>\nu}\cos\left(\bf{k\cdot(d_\mu-d_\nu)}\right)\right]\nonumber\\
    &\hspace{7.1cm}\times\phi_{\bf{k}\omega,B}
    \label{eq_mon_action_zero}
\end{align}
We can further compute the Green's function for monopoles from the above action which is given by Eq. \ref{eq_monopole green fn} of the main text.
%%%%%%%%%%%%%%%%%

%\subsection{\label{appendix_pi flux_1} GMFT of monopole  dynamics in the \texorpdfstring{$\pi-$}{} flux phase}

\subsection{\label{appendix_pi flux_1} GMFT of monopole  dynamics in the {$\pi-$} flux phase}

\begin{figure}
    \centering
    \includegraphics[width=0.7\linewidth]{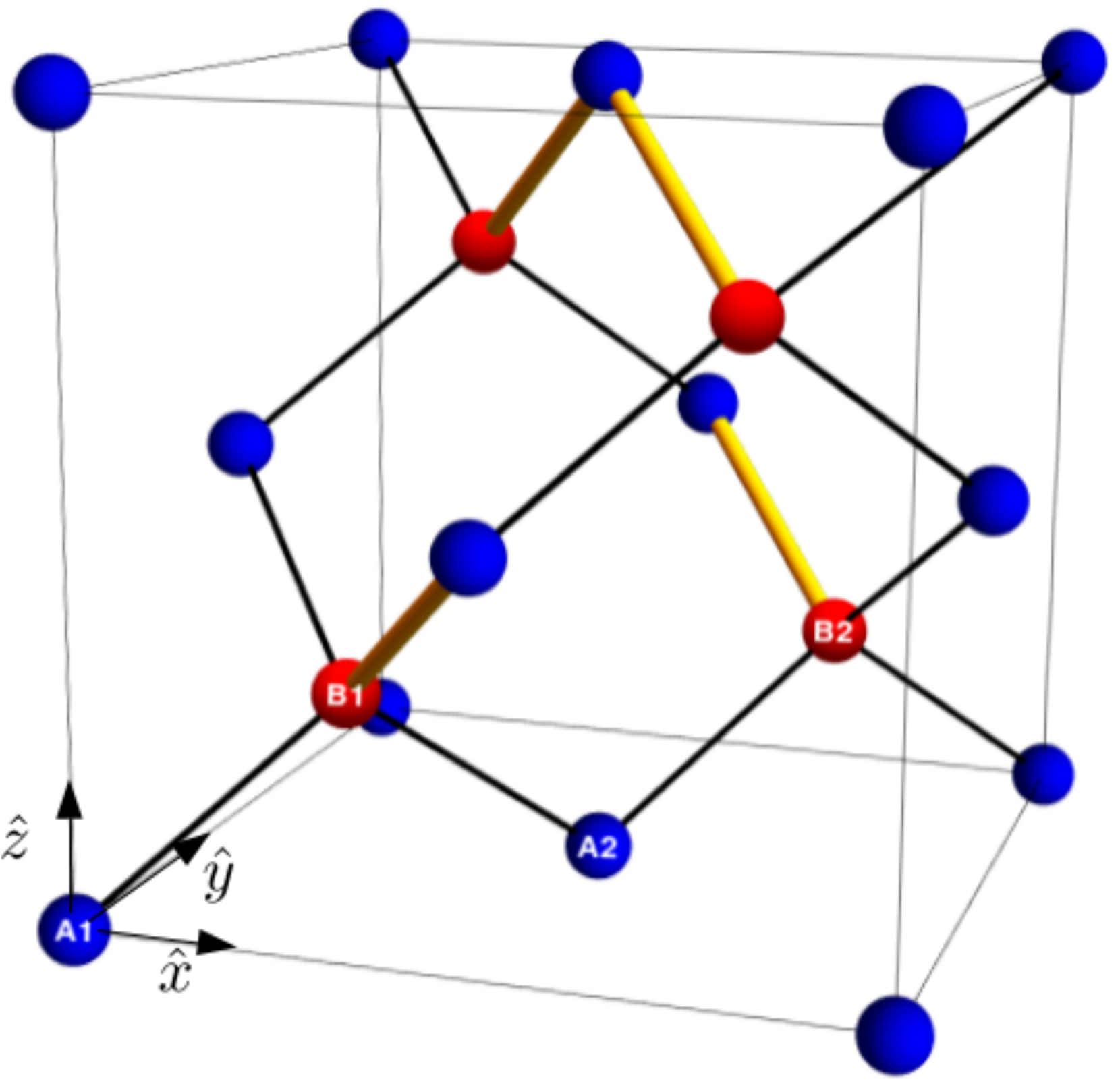}
    \caption{\small {\bf Unit cell of the diamond lattice with $\pi$-flux:} The gauge mean field is chosen such that $A_\mathbf{r,r+e_\mu}=\pi$ on the yellow bonds and $A_\mathbf{r,r+e_\mu}=0$ on the black bonds. $A1, A2, B1, B2$ denote the four sublattices in the enlarged unit cell of the $\pi-$flux phase.}
    \label{fig_unit_pi}
\end{figure}

The bare dynamics of the magnetic monopoles in the $\pi$-flux phase can be obtained by choosing a suitable gauge fixing condition shown in Fig. \ref{fig_unit_pi}. It can further be written as $A_{\mathbf{r},\boldsymbol{\mu}}=\epsilon_\mu\mathbf{Q}\cdot\mathbf{r}$ with $\epsilon_\mu=\{0,1,1,0\}$ and $\mathbf{Q=\frac{\sqrt{3}\pi}{2}}(1,0,0)$. Similar to the zero flux case, the monopoles can hop only inside the $A$ or $B$ sublattice. Hence, the monopole dynamics can be expressed in terms of the following action.

\begin{widetext}
\begin{align}
    &S_{\pi}=S^A_{\pi}+S^B_{\pi}\nonumber \\
    &S^A_{\pi}=\sum_{\mathbf{k},\omega}(\phi^*_{\mathbf{k}\omega,A1}~~\phi^*_{\mathbf{k}\omega,A2})\begin{pmatrix} 
\frac{\omega^2}{2J_{zz}}+\lambda+\frac{J_{\pm}}{4}d_A(\mathbf{k}) & \frac{J_{\pm}}{4}f_A(\mathbf{k}) \\
\frac{J_{\pm}}{4}f^*_A(\mathbf{k}) &  \frac{\omega^2}{2J_{zz}}+\lambda-\frac{J_{\pm}}{4}d_A(\mathbf{k})
\end{pmatrix}
\begin{pmatrix} 
\phi_{\mathbf{k}\omega,A1} \\
\phi_{\mathbf{k}\omega,A2}
\end{pmatrix}\\
    &S^B_{\pi}=\sum_{\mathbf{k},\omega}(\phi^*_{\mathbf{k}\omega,B1}~~\phi^*_{\mathbf{k}\omega,B2})\begin{pmatrix} 
\frac{\omega^2}{2J_{zz}}+\lambda+\frac{J_{\pm}}{4}d_B(\mathbf{k}) & \frac{J_{\pm}}{4}f_B(\mathbf{k}) \\
\frac{J_{\pm}}{4}f^*_B(\mathbf{k}) &  \frac{\omega^2}{2J_{zz}}+\lambda-\frac{J_{\pm}}{4}d_B(\mathbf{k})
\end{pmatrix}
\begin{pmatrix} 
\phi_{\mathbf{k}\omega,B1} \\
\phi_{\mathbf{k}\omega,B2}
\end{pmatrix}
\end{align}

\noindent where,
\begin{align}
    &d_A(\mathbf{k})=2 \left(\cos\left(\mathbf{k}\cdot \mathbf{d_1}\right)+\cos\left(\mathbf{k}\cdot(\mathbf{d_1-d_3})\right)\right)\nonumber\\
    &f_A(\mathbf{k})=1+e^{-i\mathbf{k\cdot d_1}}+e^{-i\mathbf{k\cdot d_2}}-e^{-i\mathbf{k\cdot d_3}}+e^{-i\mathbf{k\cdot(d_1+d_2)}}-e^{-i\mathbf{k\cdot(d_3-d_1)}}+e^{-i\mathbf{k\cdot(d_2+d_3)}}+e^{-i\mathbf{k\cdot(d_2+d_3-d_1)}}  \nonumber\\
    &d_B(\mathbf{k})=2 \left(\cos\left(\mathbf{k}\cdot \mathbf{d_1}\right)-\cos\left(\mathbf{k}\cdot(\mathbf{d_1-d_3})\right)\right)\nonumber\\
    &f_B(\mathbf{k})=1-e^{-i\mathbf{k\cdot d_1}}+e^{-i\mathbf{k\cdot d_2}}+e^{-i\mathbf{k\cdot d_3}}+e^{-i\mathbf{k\cdot(d_1+d_2)}}-e^{-i\mathbf{k\cdot(d_3-d_1)}}+e^{-i\mathbf{k\cdot(d_2+d_3)}}+e^{-i\mathbf{k\cdot(d_2+d_3-d_1)}}  \nonumber
\end{align}
\end{widetext}
\noindent $\lambda$ is the global Lagrange multiplier introduced to take into account the constraint $\phi^\dagger_{\bf r}\phi_{\bf r}=1$. $(A1,A2,B1,B2)$ denotes four sublattices of the enlarged unit cell. The above action can further be diagonalised to obtain the dispersion for four monopole bands. 
\begin{align}
    \epsilon^{\pi}_{A\pm}(\mathbf{k})=\sqrt{2J_{zz}\left(\lambda\pm\frac{J_{\pm}}{4}\sqrt{\mid d_A(\mathbf{k})\mid^2+\mid f_A(\mathbf{k})\mid^2}\right)}\nonumber\\
    \epsilon^{\pi}_{B\pm}(\mathbf{k})=\sqrt{2J_{zz}\left(\lambda\pm\frac{J_{\pm}}{4}\sqrt{\mid d_B(\mathbf{k})\mid^2+\mid f_B(\mathbf{k})\mid^2}\right)}
    \label{eq_band_pi_1}
\end{align}
where $A\pm$($B\pm$) denote two bands made out of linear combination of $A1$ and $A2$ ($B1$ and $B2$) to diagonalise the $S^A_\pi$ ($S^B_\pi$). Since A and B monopoles do not mix under the above dynamics, their bands are degenerate.
\begin{align}
    \epsilon^\pi_{A\pm}({\bf k})=\epsilon^\pi_{B\pm}({\bf k})\equiv\epsilon^\pi_{\pm}({\bf k})
    \label{eq_band_pi_2}
\end{align}

\begin{widetext}

We can compute different Green's function for monopole from the above action of the monopoles. The Green's function is defined as,
\begin{align}
    \left[G^{\pi}_{\phi}\right]_{\mu\nu}(\mathbf{k},A/B,i\omega)=\int_0^\beta d\tau \left\langle \hat{\mathcal{T}}\left(\phi_{\mathbf{k},A/B,\mu}(\tau)\phi^\dagger_{\mathbf{k},A/B,\nu}(0)\right)\right\rangle e^{i\omega\tau}
\end{align}
where, $\mu,\nu=1,2$. The different Green's function are given by,
\begin{align}
    &\left[G^{\pi}_{\phi}\right]_{11}(\mathbf{k},A/B,i\omega)=\frac{J_{zz}}{\sqrt{\mid d_{A/B}(\mathbf{k})\mid^2+\mid f_{A/B}(\mathbf{k})\mid^2}}\left[\frac{\sqrt{\mid d_{A/B}(\mathbf{k})\mid^2+\mid f_{A/B}(\mathbf{k})\mid^2}+d_{A/B}(\mathbf{k})}{\omega^2+\left(\epsilon^\pi_{+}(\mathbf{k})\right)^2}\right.\nonumber\\
    &\hspace{10cm}\left.+\frac{\sqrt{\mid d_{A/B}(\mathbf{k})\mid^2+\mid f_{A/B}(\mathbf{k})\mid^2}-d_{A/B}(\mathbf{k})}{\omega^2+\left(\epsilon^\pi_{-}(\mathbf{k})\right)^2}\right]  \nonumber\\
    &\left[G^{\pi}_{\phi}\right]_{22}(\mathbf{k},A/B,i\omega)=\frac{J_{zz}}{\sqrt{\mid d_{A/B}(\mathbf{k})\mid^2+\mid f_{A/B}(\mathbf{k})\mid^2}}\left[\frac{\sqrt{\mid d_{A/B}(\mathbf{k})\mid^2+\mid f_{A/B}(\mathbf{k})\mid^2}-d_{A/B}(\mathbf{k})}{\omega^2+\left(\epsilon^\pi_{+}(\mathbf{k})\right)^2}\right.\nonumber\\
    &\hspace{10cm}\left.+\frac{\sqrt{\mid d_{A/B}(\mathbf{k})\mid^2+\mid f_{A/B}(\mathbf{k})\mid^2}+d_{A/B}(\mathbf{k})}{\omega^2+\left(\epsilon^\pi_{-}(\mathbf{k})\right)^2}\right]  \nonumber\\
    &\left[G^{\pi}_{\phi}\right]_{12}(\mathbf{k},A/B,i\omega)=\frac{J_{zz}f_{A/B}(\mathbf{k})}{\sqrt{\mid d_{A/B}(\mathbf{k})\mid^2+\mid f_{A/B}(\mathbf{k})\mid^2}}\left[\frac{1}{\omega^2+\left(\epsilon^\pi_{+}(\mathbf{k})\right)^2}-\frac{1}{\omega^2+\left(\epsilon^\pi_{-}(\mathbf{k})\right)^2}\right] 
    \label{eq_pi_green}
\end{align}

\end{widetext}

%%%%%%%%%%%%%%%%%%%%%%%%%%%%%

\subsection{\label{appendix_zero flux vertex}The GMFT vertex functions for magnetoelastic coupling in zero-flux case}

The vertex functions for the magnetic monopole-phonon interaction vertices of Eq. \ref{eq_eg_phonon monopole coupling} and \ref{eq_t2g_phonon monopole coupling} are given by,

\begin{align}
    &\alpha^{(e)}_\mathbf{k}=\frac{1}{2}\sum_{\mu}e^{i\mathbf{k}\cdot\mathbf{d_\mu}}\nonumber\\
    &\alpha^{(t_2)}_{1,\mathbf{k}}=\frac{1}{4}(-e^{i\mathbf{k}\cdot\mathbf{d_0}}+e^{i\mathbf{k}\cdot\mathbf{d_1}}+e^{i\mathbf{k}\cdot\mathbf{d_2}}-e^{i\mathbf{k}\cdot\mathbf{d_3}})\indent\nonumber\\
    &\alpha^{(t_2)}_{2,\mathbf{k}}=\frac{e^{i\frac{\pi}{3}}}{4}(-e^{i\mathbf{k}\cdot\mathbf{d_0}}-e^{i\mathbf{k}\cdot\mathbf{d_1}}+e^{i\mathbf{k}\cdot\mathbf{d_2}}+e^{i\mathbf{k}\mathbf{d_3}})\nonumber\\
    &\alpha^{(t_2)}_{3,\mathbf{k}}=\frac{e^{-i\frac{\pi}{3}}}{4}(e^{i\mathbf{k}\cdot\mathbf{d_0}}-e^{i\mathbf{k}\cdot\mathbf{d_1}}+e^{i\mathbf{k}\cdot\mathbf{d_2}}-e^{i\mathbf{k}\cdot\mathbf{d_3}})
    \label{eq_vertex functions}
\end{align}

%%%%%%%%%%%%%%%%%%%%%%%%%%%%%%%%%
\section{Vibrational Raman spectroscopy}
\label{appen_phononraman}

Necessary condition for a phonon mode to be Raman active is that it should have even parity and the phonon contribution to the polarizability tensor ($\Lambda$) should oscillate as a function of time. For small amplitude of vibration, we can expand $\Lambda$ as powers of normal modes~\cite{porto}.
\begin{align}
  \Lambda=\Lambda_0+\boldsymbol{\zeta^{(\rho)}}\cdot\left[\boldsymbol{\nabla_{\zeta^{(\rho)}}}\Lambda\right]_{\boldsymbol{\zeta^{(\rho)}}=0}  
  \label{eq_raman}
\end{align}
where ${\boldsymbol{ \zeta^{(\rho)}}}$ is the phonon modes belonging to $\rho$ irreducible representation of the symmetry group. $\Lambda_0$ is the time independent part, hence do not contribute to the Raman scattering. The Raman coupling in Eq. \ref{eq_raman vertex} is then given by~\cite{porto}

\begin{align}
    H_{Raman}&=\sum_{p}\int d\mathbf{k}~d\mathbf{k'}\left[\nabla_{\zeta^{(\rho)}_p}\Lambda\right]^{ij}_{\zeta^{(\rho)}_p=0}\omega^{in}_{\bf k}\omega^{out}_{-\bf k'}\nonumber\\
    &\hspace{2cm}\times\zeta^{(\rho)}_p({\bf k-k'})\mathcal{A}^{in}_i({\bf k})\mathcal{A}^{out}_j(\mathbf{k'})
\label{eq_ramver_appen}
\end{align}

where $\mathbf{\mathcal{A}(r)}$ is the vector potential corresponding to the external electric field, ${\bf E}_{ext}({\bf r})$,-- again not to be confused with emergent electromagnetism. $\boldsymbol{\nabla_{\zeta^{(\rho)}}}\Lambda$ forms a set of symmetric $3\times3$ matrices which have the same symmetry properties as $\boldsymbol{\zeta^{(\rho)}}$. In other words, this set forms an irreducible representation ($\rho$) of the symmetry group. The detailed structure of the matrices are given below.

\subsection{\label{appendix_raman}Raman Matrices}
Structures of the Raman matrices are obtained by decomposing six dimensional space of second order~\cite{porto} polynomials($x^2,y^2,z^2,xy,yz,zx$) into the irreducible representations of symmetry group $T_d$ and constructing the Hessian matrices for different components. The decomposition is as follows:  
$$\mathbf{a_1}\oplus\mathbf{e}\oplus\mathbf{t_2}$$
where the basis for the irreducible subspaces are
\begin{align}
    &{\bf a_1}~ :~ x^2+y^2+z^2\\
    &{\bf e}~:~ (2z^2-x^2-y^2,x^2-y^2)\\
    &{\bf t_2}~:~ (xy,yz,zx)
\end{align}
Hence, the relevant polarizability matrices of the Raman scattering are given by, 

\begin{align}
{\bf e_g}:~
&[\nabla_{\zeta^{(e)}_{1,g}}\Lambda]_{\zeta^{(e)}_{1,g}=0}\propto
\begin{pmatrix} 
-1 & 0 & 0 \\
0 & -1 & 0\\
0 & 0 & 2
\end{pmatrix},\nonumber\\
&[\nabla_{\zeta^{(e)}_{2,g}}\Lambda]_{\zeta^{(e)}_{2,g}=0}\propto
\begin{pmatrix} 
1 & 0 & 0 \\
0 & -1 & 0\\
0 & 0 & 0
\end{pmatrix}\\
{\bf t_{2g}} ~:~
&[\nabla_{\zeta^{(t_2)}_{1,g}}\Lambda]_{\zeta^{(t_2)}_{1,g}=0}\propto
\begin{pmatrix} 
0 & 1 & 0 \\
1 & 0 & 0\\
0 & 0 & 0
\end{pmatrix},\nonumber\\
&[\nabla_{\zeta^{(t_2)}_{2,g}}\Lambda]_{\zeta^{(t_2)}_{2,g}=0}\propto
\begin{pmatrix} 
0 & 0 & 1 \\
0 & 0 & 0\\
1 & 0 & 0
\end{pmatrix},\nonumber\\
&[\nabla_{\zeta^{(t_2)}_{3,g}}\Lambda]_{\zeta^{(t_2)}_{3,g}=0}\propto
\begin{pmatrix} 
0 & 0 & 0 \\
0 & 0 & 1\\
0 & 1 & 0
\end{pmatrix}
\end{align}
\\

%%%%%%%%%%%%%%%%%%%%%%%%%%%%%%%%%%%%%%%%%%

%\section{\label{appendix_pi flux}Raman response in \texorpdfstring{$\pi-$}{}flux phase}

\section{\label{appendix_pi flux}Raman response in {$\pi-$} flux phase}

\subsection{\label{appendix_pi flux_2}Vertex functions of magnetoelastic coupling}

The vertex functions ($M^{\mu\nu}_{\bf k}$) of the magnetic monopole-phonon coupling in the $\pi$-flux phase (see in Eq. \ref{eq_phonon monopole_pi}) are given by,
\begin{align}
    &M^{11}_\mathbf{k}=2(1+e^{i\mathbf{k\cdot d_1}}) \nonumber\\
    &M^{12}_\mathbf{k}=2(1-e^{i\mathbf{k}\cdot(\mathbf{d_1-d_3})})\nonumber\\
    &M^{21}_\mathbf{k}=2(e^{i\mathbf{k}\cdot(\mathbf{d_1+d_2})}+e^{i\mathbf{k}\cdot(\mathbf{d_2+d_3})})\nonumber\\
    &M^{22}_\mathbf{k}=2(1-e^{i\mathbf{k\cdot d_1}})
\end{align}

%%%%%%%%%%%%%%%%%%%%%%%

\subsection{\label{appendix_pi flux_3}Self-energy of phonons due to magnetic monopoles}
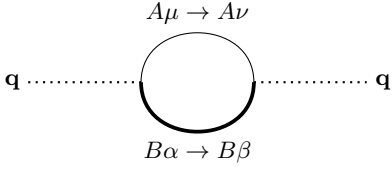
\begin{figure}
\centering
    {\begin{tikzpicture}
    \begin{feynman}
    \vertex  (a);
    \vertex [left=of a] (b);
    \vertex [right=of a, label=right:\({\bf q}\)] (f1);
    \vertex [left=of b, label=left:\({\bf q}\)] (f3); \diagram* {
    (f3) -- [dotted, thick] (b) -- [half left,edge label=\(A\mu\rightarrow A\nu\)] (a) -- [dotted, thick] (f1), (a)--[line width=0.5mm, half left,edge label=\(\textcolor{black}{B\alpha\rightarrow B\beta}\)](b)
        };
    \end{feynman}
    \end{tikzpicture}}
    \caption{\small {\bf Self-energy bubble diagrams for phonon in $\pi$-flux phase:} The label $A\mu\rightarrow A\nu$ ($B\alpha\rightarrow B\beta$ ) implies the $A\mu$ ($B\alpha$) monopole is created in the left vertex and $A\nu$ ($B\beta$) monopole is annihilated at the right vertex. $\mu,\nu,\alpha,\beta=1,2$, hence, there are 16 possible distinct diagrams.} 
    \label{fig_self energy_pi}
\end{figure}

Similar to the zero flux case, the self-energy can be obtained by calculating the bubble diagrams appearing in the second-order perturbation theory. The only difference is that due to the larger unit cell, \textit{sixteen} nonequivalent diagrams (see Fig. \ref{fig_self energy_pi}) need to be taken care of. 

For convenience, we introduce the following convention.

\begin{equation*}
\begin{aligned}[c]
\begin{tikzpicture}
\begin{feynman}
    \vertex  (a);
    \vertex [left=of a] (b);
    \diagram* {
     (b) -- [half left,edge label=\(A\mu\rightarrow A\nu\)] (a), (a)--[line width=0.5mm, half left,edge label=\(\textcolor{black}{B\alpha\rightarrow B\beta}\)](b)
    };
\end{feynman}
\end{tikzpicture}
\end{aligned}
~~~~~~~~~~
\begin{aligned}[c]
&\equiv~\begin{pmatrix} 
A\mu\rightarrow A\nu  \\
B\alpha\rightarrow B\beta 
\end{pmatrix}
\end{aligned}
\end{equation*}

We now compute all the distinct contributions to the phonon self-energy. In the following equations, we group the distinct diagrams along with their Hermitian conjugate.

\begin{widetext}

\begin{align*}
&\begin{pmatrix} 
A1\rightarrow A1  \\
B1\rightarrow B1 
\end{pmatrix}
=\sum_{\mathbf{k}}\frac{M^{11}_{\mathbf{k}}M^{11}_{\mathbf{-k}}J_{zz}^2}{\sqrt{\mid d_A\mid^2+\mid f_A\mid^2}\sqrt{\mid d_B\mid^2+\mid f_B\mid^2}}
\left[\left(\sqrt{\mid d_A\mid^2+\mid f_A\mid^2}+d_A\right)\left(\sqrt{\mid d_B\mid^2+\mid f_B\mid^2}+d_B\right)W_{++}\right. \nonumber\\
&\hspace{8.5cm}\left.+\left(\sqrt{\mid d_A\mid^2+\mid f_A\mid^2}+d_A\right)\left(\sqrt{\mid d_B\mid^2+\mid f_B\mid^2}-d_B\right)W_{+-}\right. \nonumber\\
&\hspace{8.5cm}\left.+\left(\sqrt{\mid d_A\mid^2+\mid f_A\mid^2}-d_A\right)\left(\sqrt{\mid d_B\mid^2+\mid f_B\mid^2}+d_B\right)W_{-+}\right. \nonumber\\
&\hspace{8.5cm}\left.+\left(\sqrt{\mid d_A\mid^2+\mid f_A\mid^2}-d_A\right)\left(\sqrt{\mid d_B\mid^2+\mid f_B\mid^2}-d_B\right)W_{--}\right] \nonumber\\
&\begin{pmatrix} 
A1\rightarrow A1  \\
B1\rightarrow B2
\end{pmatrix}+
\begin{pmatrix} 
A1\rightarrow A1  \\
B2\rightarrow B1
\end{pmatrix}
=\sum_{\mathbf{k}}\frac{\mathcal{R}e\left(M^{11}_{\mathbf{k}}M^{12}_\mathbf{-k}f_B(\mathbf{k})\right)J_{zz}^2}{\sqrt{\mid d_A\mid^2+\mid f_A\mid^2}\sqrt{\mid d_B\mid^2+\mid f_B\mid^2}}\left[\left(\sqrt{\mid d_A\mid^2+\mid f_A\mid^2}+d_A\right)\left(W_{++}-W_{+-}\right)\right.\\
&\hspace{11cm}\left.+\left(\sqrt{\mid d_A\mid^2+\mid f_A\mid^2}-d_A\right)\left(W_{-+}-W_{--}\right)\right]\\
&\begin{pmatrix}
A1\rightarrow A1 \\
B2\rightarrow B2
\end{pmatrix}
=\sum_{\mathbf{k}}\frac{M^{12}_{\mathbf{k}}M^{12}_{\mathbf{-k}}J_{zz}^2}{\sqrt{\mid d_A\mid^2+\mid f_A\mid^2}\sqrt{\mid d_B\mid^2+\mid f_B\mid^2}}\left[\left(\sqrt{\mid d_A\mid^2+\mid f_A\mid^2}+d_A\right)\left(\sqrt{\mid d_B\mid^2+\mid f_B\mid^2}-d_B\right)W_{++}\right. \\
&\hspace{8.5cm}\left.+\left(\sqrt{\mid d_A\mid^2+\mid f_A\mid^2}+d_A\right)\left(\sqrt{\mid d_B\mid^2+\mid f_B\mid^2}+d_B\right)W_{+-}\right. \\
&\hspace{8.5cm}\left.+\left(\sqrt{\mid d_A\mid^2+\mid f_A\mid^2}-d_A\right)\left(\sqrt{\mid d_B\mid^2+\mid f_B\mid^2}-d_B\right)W_{-+}\right. \\
&\hspace{8.5cm}\left.+\left(\sqrt{\mid d_A\mid^2+\mid f_A\mid^2}-d_A\right)\left(\sqrt{\mid d_B\mid^2+\mid f_B\mid^2}+d_B\right)W_{--}\right]\\
&\begin{pmatrix}
A1\rightarrow A2\\
B2\rightarrow B1
\end{pmatrix}+
\begin{pmatrix}
A2\rightarrow A1\\
B1\rightarrow B2
\end{pmatrix}
=\sum_{\mathbf{k}}\frac{J_{zz}^2\left(M^{12}_{\mathbf{k}}M^{21}_{\mathbf{-k}}f_A^*f_B+M^{21}_{\mathbf{k}}M^{12}_{\mathbf{-k}}f_Af_B^*\right)}{\sqrt{\mid d_A\mid^2+\mid f_A\mid^2}\sqrt{\mid d_B\mid^2+\mid f_B\mid^2}}\left[W_{--}-W_{-+}-W_{+-}+W_{++}\right]\\
&\begin{pmatrix}
A2\rightarrow A2\\
B2\rightarrow B2
\end{pmatrix}
=\sum_{\mathbf{k}}\frac{M^{22}_{\mathbf{k}}M^{22}_{\mathbf{-k}}J_{zz}^2}{\sqrt{\mid d_A\mid^2+\mid f_A\mid^2}\sqrt{\mid d_B\mid^2+\mid f_B\mid^2}}\left[\left(\sqrt{\mid d_A\mid^2+\mid f_A\mid^2}+d_A\right)\left(\sqrt{\mid d_B\mid^2+\mid f_B\mid^2}+d_B\right)W_{--}\right. \\
&\hspace{8.5cm}\left.+\left(\sqrt{\mid d_A\mid^2+\mid f_A\mid^2}+d_A\right)\left(\sqrt{\mid d_B\mid^2+\mid f_B\mid^2}-d_B\right)W_{-+}\right. \\
&\hspace{8.5cm}\left.+\left(\sqrt{\mid d_A\mid^2+\mid f_A\mid^2}-d_A\right)\left(\sqrt{\mid d_B\mid^2+\mid f_B\mid^2}+d_B\right)W_{+-}\right. \\
&\hspace{8.5cm}\left.+\left(\sqrt{\mid d_A\mid^2+\mid f_A\mid^2}-d_A\right)\left(\sqrt{\mid d_B\mid^2+\mid f_B\mid^2}-d_B\right)W_{++}\right] \\
&\begin{pmatrix}
A2\rightarrow A2\\
B1\rightarrow B2
\end{pmatrix}
+\begin{pmatrix}
A2\rightarrow A2\\
B2\rightarrow B1
\end{pmatrix}
=\sum_{\mathbf{k}}\frac{\mathcal{R}e\left(M^{21}_{\mathbf{k}}M^{22}_\mathbf{-k}f_B(\mathbf{k})\right)J_{zz}^2}{\sqrt{\mid d_A\mid^2+\mid f_A\mid^2}\sqrt{\mid d_B\mid^2+\mid f_B\mid^2}}\left[\left(\sqrt{\mid d_A\mid^2+\mid f_A\mid^2}+d_A\right)\left(W_{-+}-W_{--}\right)\right.\\
&\hspace{11cm}\left.+\left(\sqrt{\mid d_A\mid^2+\mid f_A\mid^2}-d_A\right)\left(W_{++}-W_{+-}\right)\right]\\
&\begin{pmatrix}
A1\rightarrow A2\\
B1\rightarrow B2
\end{pmatrix}+
\begin{pmatrix}
A2\rightarrow A1\\
B2\rightarrow B1
\end{pmatrix}
=\sum_{\mathbf{k}}\frac{J_{zz}^2\left(M^{11}_{\mathbf{k}}M^{22}_{\mathbf{-k}}f_Af_B^*+M^{11}_{\mathbf{-k}}M^{22}_{\mathbf{k}}f_A^*f_B\right)}{\sqrt{\mid d_A\mid^2+\mid f_A\mid^2}\sqrt{\mid d_B\mid^2+\mid f_B\mid^2}}\left[W_{--}-W_{-+}-W_{+-}+W_{++}\right]
\end{align*}

There are five other distinct diagrams which can be obtained by replacing $A\rightarrow B$ in the above diagrams (more specifically, second, third and sixth expression). Adding up all the above contributions, we obtain,
\begin{align}
\Sigma_{\zeta^{(e)}}^\pi({\bf q}=&0,i \Omega)=\frac{J_{sp}^{(e)2}}{N}\sum_{\bf k}\left[\mathcal{P}_{1}(\mathbf{k})W_{++}({\bf k},i\Omega)+\mathcal{P}_{2}(\mathbf{k})W_{--}({\bf k},i\Omega)+\mathcal{P}_{3}(\mathbf{k})W_{+-}({\bf k},i\Omega)+\mathcal{P}_{4}(\mathbf{k})W_{-+}({\bf k},i\Omega)\right]
\label{eq_self energy_pi}
\end{align}
where, $\mathcal{P}_1({\bf k})$, $\mathcal{P}_2({\bf k})$, $\mathcal{P}_3({\bf k})$, $\mathcal{P}_4({\bf k})$ are real functions of momentum and 
\begin{align}
    W_{mn}\left({\bf k},i\Omega\right)=-\frac{1}{\beta}\sum_\omega\frac{1}{\left(\Omega+\omega\right)^2+(\epsilon^\pi_m(\mathbf{k}))^2}~\frac{1}{\omega^2+(\epsilon^\pi_n(\mathbf{k}))^2}
\end{align}
The phonon linewidth is obtained from Eq. \ref{eq_self energy_pi} by calculating its imaginary part. Apart from the momentum dependent form factors, the contribution is mostly dominated by the four $W_{mn}$ terms. Calculating their imaginary parts, we obtain,  
\begin{align}
    &\lim_{\delta\rightarrow0}\mathcal{I}m\left(W_{\pm\pm}({\bf k},\Omega+i\delta)\right)=\frac{\pi(1+2n(\epsilon^\pi_\pm(\mathbf{k})))}{4\epsilon^\pi_\pm(\mathbf{k})^2}\left[\delta\left(\Omega+2\epsilon^\pi_\pm(\mathbf{k})\right)-\delta\left(\Omega-2\epsilon^\pi_\pm(\mathbf{k})\right)\right]\nonumber\\
    &\lim_{\delta\rightarrow0}\mathcal{I}m\left(W_{+-}({\bf k},\Omega+i\delta)\right)=\mathcal{I}m\left(W_{-+}({\bf k},\Omega+i\delta)\right)\nonumber\\
    &\hspace{3.3cm}=\frac{\pi(1+n(\epsilon^\pi_+(\mathbf{k}))+n(\epsilon^\pi_-(\mathbf{k})))}{4\epsilon^\pi_+(\mathbf{k})\epsilon^\pi_-(\mathbf{k})}\left[\delta\left(\Omega+\epsilon^\pi_+(\mathbf{k})+\epsilon^\pi_-(\mathbf{k})\right)-\delta\left(\Omega-\epsilon^\pi_+(\mathbf{k})-\epsilon^\pi_-(\mathbf{k})\right)\right]\nonumber\\
    &\hspace{3.3cm}~~+\frac{\pi(n(\epsilon^\pi_+(\mathbf{k}))-n(\epsilon^\pi_-(\mathbf{k})))}{4\epsilon^\pi_+(\mathbf{k})\epsilon^\pi_-(\mathbf{k})}\left[\delta\left(\Omega+\epsilon^\pi_-(\mathbf{k})-\epsilon^\pi_+(\mathbf{k})\right)-\delta\left(\Omega+\epsilon^\pi_+(\mathbf{k})-\epsilon^\pi_-(\mathbf{k})\right)\right]
\end{align} 
Substituting the above expressions in \ref{eq_self energy_pi}, we obtain the linewidth of the phonon in the $\pi-$flux phase due to the phonon-magnetic monopole coupling which is given in Eq. \ref{eq_lw_pi} of the main text. 
\end{widetext}
%%%%%%%%%%%%%%%%%%%%%%%%

\section{\label{appendix_beyond gmft} Effect of Gauge fluctuations for the magnetic monopoles}

The vertex functions of the photon-magnetic monopole interaction of Eq. \ref{eq_gauge fluctuation hamiltonian} are given by, 
\begin{align}
    &\gamma^{\mu\nu}_B(\mathbf{k,k'})=e^{-i\mathbf{k}\cdot \mathbf{d}_{\boldsymbol{\mu}}}e^{i\mathbf{k'}\cdot \mathbf{d}_{\boldsymbol{\nu}}}-e^{-i\mathbf{k}\cdot \mathbf{d}_{\boldsymbol{\nu}}}e^{i\mathbf{k'}\cdot \mathbf{d}_{\boldsymbol{\mu}}}\nonumber\\
    &\gamma^{\mu\nu}_A(\mathbf{k,k'})=e^{-i\mathbf{k}\cdot(\mathbf{d}_{\boldsymbol{\mu}}-\mathbf{d}_{\boldsymbol{\nu}})}-e^{i\mathbf{k'}\cdot(\mathbf{d}_{\boldsymbol{\mu}}-\mathbf{d}_{\boldsymbol{\nu}})}
\label{eq_vertex_gauge fluctuation }
\end{align} 

Due to the photon-monopole interaction, the interaction vertices of phonon-monopole coupling is modified. The leading order contribution to the vertex correction obtained from the perturbative expansion is given by,
\begin{widetext}
\begin{align}
    \delta\alpha^{(\rho)}(\mathbf{q,p},i\Omega,i\omega_m)=-\frac{J_{\pm}^2}{16N^{\frac{3}{2}}}\sum_{\mathbf{k}}\sum_{\mu,\nu}(\alpha^{(\rho)}_\mathbf{k}+\alpha^{(\rho)}_\mathbf{k+q})\gamma_B^\mu(\mathbf{p,k+q})\gamma_A^\nu(\mathbf{k,-q+p})
    \frac{1}{\beta}\sum_{\omega_n}G_\phi(\mathbf{k+q},B,i\Omega+i\omega_n)\nonumber\\
    D_{\mu\nu}(\mathbf{k+q-p},i\Omega+i\omega_n-i\omega_m)G_\phi(\mathbf{k},A,i\omega_n)
    \label{eq_vertex correction}
\end{align}
\end{widetext}
where, $\gamma^\mu_{A,B}(\mathbf{k,k'})=\sum_{\nu(\neq\mu)}\gamma^{\mu\nu}_{A,B}(\mathbf{k,k'})$. 

To further simplify the above expression, we first perform the frequency summation of the above expression using Matsubara method and then the momentum integrals are computed using the several approximations. The monopole band structure is expanded around the minima at $\mathbf{k}=0$ up to first non-zero term.
\begin{align}
    \epsilon^0_\mathbf{k}\approx\Delta+m_0 k^2
    \label{eq_monopole_quadratic}
\end{align}
where $m_0$ is a constant measuring the curvature of the band at $\mathbf{k}=0$. Further, the vertex functions are also expanded in momentum and approximated to the leading term to obtain from Eq. \ref{eq_vertex functions},
\begin{align}
    \mid\alpha_\mathbf{k}^{(e)}\mid\approx 2~, ~~~
    \mid\alpha_{\mathbf{k}}^{(t_2)}\mid=\sqrt{\frac{\sum_{p=1}^3\mid\alpha^{(t_2)}_{p,\mathbf{k}}\mid^2}{3}}\approx\frac{k}{3}
\end{align}
and from Eq. \ref{eq_vertex_gauge fluctuation },
\begin{align}
    \gamma_A^{\mu\nu}(\mathbf{k,k'})\approx i(\mathbf{k+k'})\cdot(\mathbf{d}_{\boldsymbol{\nu}}-\mathbf{d}_{\boldsymbol{\mu}})\nonumber\\
    \gamma_B^{\mu\nu}(\mathbf{k,k'})\approx i(\mathbf{k+k'})\cdot(\mathbf{d}_{\boldsymbol{\nu}}-\mathbf{d}_{\boldsymbol{\mu}})
\end{align}

We substitute the above expressions in Eq. \ref{eq_vertex correction}. Due to Raman criterion, only ${\bf q}=0$ limit is considered. Further, we set $\Omega=\omega_m=0$ to find the frequency independent correction. We redefine the notations as, 

\begin{align}
    &\delta\alpha^{(e)}( 0,{\bf p},0,0)=\delta\alpha_{\bf p}^{(e)}\nonumber\\
    &\delta\alpha^{(t_2)}( 0,{\bf p},0,0)=\delta\alpha_{\bf p}^{(t_2)}\nonumber
\end{align}

Finally, applying all the approximations described above, the leading corrections to the vertex functions are obtained which is given in Eq. \ref{eq_vertex correction_final}.

%%%%%%%%%%%%%%%%%%%%%%%%%%%%%%%%%%%%%%%%%%%%

%%%%%%%%%%%%%%%%%%%%%%%%%%%%%%%%%%%%%%

\section{\label{appendix_quadratic}spin-phonon interaction: quadratic in spin operators}

The detailed form of the quadratic spin-phonon coupling, described in Eq. \ref{eq_quadratic spin phonon coupling_0}, is given by,

\begin{widetext}
\begin{align}
    &H_1=\sum_{{\bf r},\mu,\nu}\left(\frac{\partial J_{zz}}{\partial R^a_{\mu\nu}}\delta^a_{\mu\nu}({\bf r})s^z_{\bf r,r+e_{\boldsymbol{\mu}}}s^z_{\bf r,r+e_{\boldsymbol{\nu}}}-\frac{\partial J_{\pm}}{\partial R^a_{\mu\nu}}\delta^a_{\mu\nu}({\bf r})\left(s^+_{\mathbf {r},\mathbf{r}+\mathbf{e}_{\boldsymbol{\mu}}}s^-_{\mathbf{r},\mathbf{r}+\mathbf{e}_{\boldsymbol{\nu}}}+h.c.\right)\right)
    \label{eq_quadratic spin phonon coupling_1}\\
    &H_2=\sum_{\bf r,\mu,\nu}\left(\frac{1}{2}\frac{\partial^2J_{zz}}{\partial R^a_{\mu\nu}\partial R^b_{\mu\nu}}\delta^a_{\mu\nu}({\bf r})\delta^b_{\mu\nu}({\bf r})s^z_{\mathbf{r},\mathbf{r}+\mathbf{e}_{\boldsymbol{\mu}}}s^z_{\mathbf{r},\mathbf{r}+\mathbf{e}_{\boldsymbol{\nu}}}-\frac{1}{2}\frac{\partial^2J_{\pm}}{\partial R^a_{\mu\nu}\partial R^b_{\mu\nu}}\delta^a_{\mu\nu}({\bf r})\delta^b_{\mu\nu}({\bf r})\left(s^+_{\mathbf{r},\mathbf{r}+\mathbf{e}_{\boldsymbol{\mu}}}s^-_{\mathbf{r},\mathbf{r}+\mathbf{e}_{\boldsymbol{\nu}}}+h.c.\right)\right)
    \label{eq_quadratic spin phonon coupling_2}
\end{align}

Further, the above interactions can be re-written in terms of the fractionalised degrees of freedom in a QSL phase using the parton decomposition of spins described in Sec. \ref{sec_4}. Within GMFT approximation, it is given by,

\begin{align}
    &H_1=\sum_{{\bf r},\mu,\nu}\left(\frac{\partial J_{zz}}{\partial R^a_{\mu\nu}}\delta^a_{\mu\nu}({\bf r},A)B_{\mathbf{r},{\boldsymbol{\mu}}}B_{\mathbf{r},{\boldsymbol{\nu}}}+\frac{\partial J_{zz}}{\partial R^a_{\mu\nu}}\delta^a_{\mu\nu}({\bf r},B)B_{\mathbf{r}-\mathbf{d}_{\boldsymbol{\mu}},{\boldsymbol{\mu}}}B_{\mathbf{r}-\mathbf{d}_{\boldsymbol{\nu}},{\boldsymbol{\nu}}}\right.\nonumber\\
    &\hspace{3cm}\left.-\frac{\partial J_{\pm}}{\partial R^a_{\mu\nu}}\delta^a_{\mu\nu}({\bf r},A)\left(\phi^\dagger_{{\mathbf{r}+\mathbf{d}_{\boldsymbol{\nu}}},B}\phi_{{\mathbf{r}+\mathbf{d}_{\boldsymbol{\mu}}},B}+h.c.\right)-\frac{\partial J_{\pm}}{\partial R^a_{\mu\nu}}\delta^a_{\mu\nu}({\bf r},B)\left(\phi^\dagger_{{\mathbf{r}-\mathbf{d}_{\boldsymbol{\nu}}},A}\phi_{{\mathbf{r}-\mathbf{d}_{\boldsymbol{\mu}}},A}+h.c.\right)\right)
    \label{eq_quadratic_qsl_1}\\
    &H_2=\sum_{{\bf r},\mu,\nu}\left(\frac{1}{2}\frac{\partial^2J_{zz}}{\partial R^a_{\mu\nu}\partial R^b_{\mu\nu}}\delta^a_{\mu\nu}({\bf r},A)\delta^b_{\mu\nu}({\bf r},A)B_{\mathbf{r},{\boldsymbol{\mu}}}B_{\mathbf{r},{\boldsymbol{\nu}}}+\frac{1}{2}\frac{\partial^2J_{zz}}{\partial R^a_{\mu\nu}\partial R^b_{\mu\nu}}\delta^a_{\mu\nu}({\bf r},B)\delta^b_{\mu\nu}({\bf r},B)B_{\mathbf{r}-\mathbf{d}_{\boldsymbol{\mu}},{\boldsymbol{\mu}}}B_{\mathbf{r}-\mathbf{d}_{\boldsymbol{\nu}},{\boldsymbol{\nu}}}\right.\nonumber\\
    &\hspace{7cm}\left.-\frac{1}{2}\frac{\partial^2J_{\pm}}{\partial R^a_{\mu\nu}\partial R^b_{\mu\nu}}\delta^a_{\mu\nu}({\bf r},A)\delta^b_{\mu\nu}({\bf r},A)\left(\phi^\dagger_{{\mathbf{r}+\mathbf{d}_{\boldsymbol{\nu}}},B}\phi_{{\mathbf{ r}+\mathbf{d}_{\boldsymbol{\mu}}},B}+h.c.\right)\right.\nonumber\\
    &\hspace{7cm}\left.-\frac{1}{2}\frac{\partial^2J_{\pm}}{\partial R^a_{\mu\nu}\partial R^b_{\mu\nu}}\delta^a_{\mu\nu}({\bf r},B)\delta^b_{\mu\nu}({\bf r},B)\left(\phi^\dagger_{{\mathbf{ r}-\mathbf{d}_{\boldsymbol{\nu}}},A}\phi_{{\mathbf{ r}-\mathbf{d}_{\boldsymbol{\mu}}},A}+h.c.\right)\right) 
    \label{eq_quadratic_qsl_2}
\end{align}

\end{widetext}

%%%%%%%%%%%%%%%%%%%%%%%%%%%%%%%%%%%%%%

\section{\label{appendix_paramagnetic}Hamiltonian and form factors in the paramagnetic phase}

For the convenience of calculation, we express the Hamiltonian given in Eq. \ref{eq_Eg coupling} and \ref{eq_T2g coupling} in momentum space representation.
\begin{align}
    &H_{sp}^{(e)}=J_{sp}^{(e)}\sum_{\mathbf{k}}\sum_{\mu=0}^3\left(\zeta_{1,g}^{(e)}(\mathbf{k})s^{x}_\mu(-\mathbf{k})+\zeta_{2,g}^{(e)}(\mathbf{k})s^{y}_\mu(-\mathbf{k})\right)\nonumber\\
    &\hspace{5cm}\times\left(1+e^{i\mathbf{k}\cdot \mathbf{d}_{\boldsymbol{\mu}}}\right)\\
    &H_{sp}^{(t_2)}=J_{sp}^{(t_2)}\sum_{\mathbf{k}}\sum_{p=1}^3\sum_{\alpha=x,y}\sum_{\mu=0}^3L^{(t_2)}_{p,\alpha,\mu}\zeta_{p,g}^{(t_2)}(\mathbf{k})s^\alpha_\mu(-\mathbf{k})\nonumber\\
    &\hspace{5cm}\times\left(1+e^{i\mathbf{k}\cdot \mathbf{d}_{\boldsymbol{\mu}}}\right)
\end{align}
From the above Hamiltonians, we can obtain the self-energy of the phonon using similar kind of perturbation theory as applied to QSL phase. Again, the first non-zero contribution comes in the second order($\mathcal{O}(J^{(\rho)2}_{sp})$) in the perturbative series and it is given in Eq. \ref{eq_self energy in paramagnet} of the main text. The form factors in the Eq. \ref{eq_self energy in paramagnet} are given by,

\begin{align}
    &\eta^{(e)}_{\mu\nu}(\mathbf{q})=1+e^{i\mathbf{q}\cdot\mathbf{d}_{\boldsymbol{\nu}}}+e^{-i\mathbf{q}\cdot\mathbf{d}_{\boldsymbol{\mu}}}+e^{i\mathbf{q}\cdot(\mathbf{d}_{\boldsymbol{\mu}}-\mathbf{d}_{\boldsymbol{\nu}})}\nonumber\\
    &\eta^{(t_2)}_{\mu\nu,\alpha}(\mathbf{q})=(1+e^{i\mathbf{q}\cdot\mathbf{d}_{\boldsymbol{\nu}}}+e^{-i\mathbf{q}\cdot\mathbf{d}_{\boldsymbol{\mu}}}+e^{i\mathbf{q}\cdot(\mathbf{d}_{\boldsymbol{\mu}}-\mathbf{d}_{\boldsymbol{\nu}})})\nonumber\\
    &\hspace{5cm}\times\sum_{p=1}^3L^{(t_2)}_{p,\alpha,\mu}L^{(t_2)}_{p,\alpha,\nu}
\end{align}

%%%%%%%%%%%%%%%%%%%%%%%%%%%%%%%%%%%%%%%%%%

\section{\label{appendix_loudon-fleury}Phonon mediated Loudon-Fleury vertex}

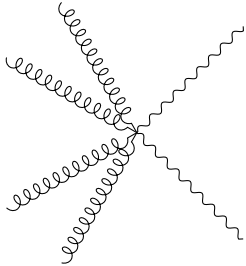
\begin{figure}
    \centering
    \begin{tikzpicture}
    \begin{feynman}
    \vertex (a) at (0, 0);
    \vertex (f1) at (-1, 1.732);
    \vertex (f2) at (-1.732,1);
    \vertex (f3) at (-1.732,-1);
    \vertex (f4) at (-1, -1.732);
    \vertex (i1) at (1.414,1.414);
    \vertex (i2) at (1.414,-1.414);    
    \diagram* {
    (f1) -- [gluon] (a) -- [photon] (i1),
    (f2) -- [gluon] (a) -- [photon] (i2),
    (f3) -- [gluon] (a),
    (f4) -- [gluon] (a),
    };
    \end{feynman}
    \end{tikzpicture}
    \caption{\small { Feynman diagram for {\it phonon mediated Loudon-Fleury} vertex for emergent photons}}
    \label{fig_LF vertex2}
\end{figure}

The phonon mediated Loudon-Fleury vertex between external photons and magnetic monopoles (emergent photons) is obtained by integrating out the phonons from Eq. \ref{eq_eg_phonon monopole coupling}, \ref{eq_t2g_phonon monopole coupling} and \ref{eq_ramver_appen} (Eq. \ref{eq_phonon-photon coupling} and \ref{eq_ramver_appen}). The leading order interaction vertices are then given by,
\begin{align}
    &H^{\phi}_{LF}=\langle H_{Raman}H_{sp}\rangle_{\zeta}\\
    &H^{A}_{LF}=\langle H_{Raman}^2H_{phonon-photon}\rangle_{\zeta}\nonumber\\
    &\hspace{3cm}-\langle H^2_{Raman}\rangle_\zeta\langle H_{phonon-photon}\rangle_{\zeta}
\end{align}
where $\langle \hat{O}\rangle_\zeta=\frac{\int D\zeta~ \hat{O}~ e^{-\beta H_\zeta}}{\int D\zeta~ e^{-\beta H_\zeta}}$ and $H_{sp}=H^{(e)}_{sp}+H^{(t_2)}_{sp}$. Simplifying the above expressions, we get,
\begin{widetext}
\begin{align}
    H^{\phi}_{LF}=\frac{J^{(\rho)}_{sp}}{2\omega_0}\int\prod_{i=1}^4d^3k_i\left[\nabla_{\zeta^{(\rho)}}\Lambda\right]^{ij}_{\zeta^{(\rho)}=0}\omega^{in}_{\bf k_1}\omega^{out}_{\bf k_2}(\alpha_{\bf k_3}+\alpha_{\bf k_4}) \mathcal{A}^{in}_i({\bf k_1})\mathcal{A}^{out}_j({\bf k_2})\phi^\dagger_{{\bf k_3},A}\phi_{{\bf k_4},B}\delta({\bf k_1+k_2+k_3+k_4})+h.c.
    \label{eq_LF1}
\end{align}
\begin{align}
    H^A_{LF}=\frac{1}{2\omega_0^2}\int\prod_{i=1}^8d^3k_i\left[\nabla_{\zeta^{(\rho)}}\Lambda\right]^{ij}_{\zeta^{(\rho)}=0}\left[\nabla_{\zeta^{(\rho)}}\Lambda\right]^{mn}_{\zeta^{(\rho)}=0}\mathcal{G}^{\alpha\beta}({\bf k_1,k_2,k_3,k_4})A^\alpha_{\bf k_3}A^\beta_{\bf k_4}\mathcal{A}^{in}_i({\bf k_5})\mathcal{A}^{out}_j({\bf k_6})\mathcal{A}^{in}_m({\bf k_7})\mathcal{A}^{out}_n({\bf k_8})\nonumber\\
    \times\delta({\bf k_5-k_6+k_1})\delta({\bf k_7-k_8+k_2})
    \label{eq_LF2}
\end{align}
\end{widetext}
where the optical phonon band structure is approximated as $\omega_{\bf q}\approx\omega_0$. Clearly, the above contributions are suppressed by the optical phonon energy scale compared to the usual Loudon-Fleury vertex~\cite{rau2017}. Feynman diagram for Eqs. \ref{eq_LF1} and \ref{eq_LF2} are shown in Figs. \ref{fig_LF vertex} and \ref{fig_LF vertex2}, respectively.

\bibliography{reference}

\providecommand{\noopsort}[1]{}\providecommand{\singleletter}[1]{#1}%
\begin{thebibliography}{118}%
\makeatletter
\providecommand \@ifxundefined [1]{%
 \@ifx{#1\undefined}
}%
\providecommand \@ifnum [1]{%
 \ifnum #1\expandafter \@firstoftwo
 \else \expandafter \@secondoftwo
 \fi
}%
\providecommand \@ifx [1]{%
 \ifx #1\expandafter \@firstoftwo
 \else \expandafter \@secondoftwo
 \fi
}%
\providecommand \natexlab [1]{#1}%
\providecommand \enquote  [1]{``#1''}%
\providecommand \bibnamefont  [1]{#1}%
\providecommand \bibfnamefont [1]{#1}%
\providecommand \citenamefont [1]{#1}%
\providecommand \href@noop [0]{\@secondoftwo}%
\providecommand \href [0]{\begingroup \@sanitize@url \@href}%
\providecommand \@href[1]{\@@startlink{#1}\@@href}%
\providecommand \@@href[1]{\endgroup#1\@@endlink}%
\providecommand \@sanitize@url [0]{\catcode `\\12\catcode `\$12\catcode
  `\&12\catcode `\#12\catcode `\^12\catcode `\_12\catcode `\%12\relax}%
\providecommand \@@startlink[1]{}%
\providecommand \@@endlink[0]{}%
\providecommand \url  [0]{\begingroup\@sanitize@url \@url }%
\providecommand \@url [1]{\endgroup\@href {#1}{\urlprefix }}%
\providecommand \urlprefix  [0]{URL }%
\providecommand \Eprint [0]{\href }%
\providecommand \doibase [0]{https://doi.org/}%
\providecommand \selectlanguage [0]{\@gobble}%
\providecommand \bibinfo  [0]{\@secondoftwo}%
\providecommand \bibfield  [0]{\@secondoftwo}%
\providecommand \translation [1]{[#1]}%
\providecommand \BibitemOpen [0]{}%
\providecommand \bibitemStop [0]{}%
\providecommand \bibitemNoStop [0]{.\EOS\space}%
\providecommand \EOS [0]{\spacefactor3000\relax}%
\providecommand \BibitemShut  [1]{\csname bibitem#1\endcsname}%
\let\auto@bib@innerbib\@empty
%</preamble>
\bibitem [{\citenamefont {Anderson}(1987)}]{anderson1987resonating}%
  \BibitemOpen
  \bibfield  {author} {\bibinfo {author} {\bibfnamefont {P.~W.}\ \bibnamefont
  {Anderson}},\ }\bibfield  {title} {\bibinfo {title} {The resonating valence
  bond state in la2cuo4 and superconductivity},\ }\href
  {https://doi.org/10.1126/science.235.4793.1196} {\bibfield  {journal}
  {\bibinfo  {journal} {science}\ }\textbf {\bibinfo {volume} {235}},\ \bibinfo
  {pages} {1196} (\bibinfo {year} {1987})}\BibitemShut {NoStop}%
\bibitem [{\citenamefont {Anderson}(1973)}]{anderson1973resonating}%
  \BibitemOpen
  \bibfield  {author} {\bibinfo {author} {\bibfnamefont {P.}~\bibnamefont
  {Anderson}},\ }\bibfield  {title} {\bibinfo {title} {Resonating valence
  bonds: A new kind of insulator?},\ }\href
  {https://doi.org/https://doi.org/10.1016/0025-5408(73)90167-0} {\bibfield
  {journal} {\bibinfo  {journal} {Materials Research Bulletin}\ }\textbf
  {\bibinfo {volume} {8}},\ \bibinfo {pages} {153} (\bibinfo {year}
  {1973})}\BibitemShut {NoStop}%
\bibitem [{\citenamefont {Moessner}\ and\ \citenamefont
  {Sondhi}(2001)}]{PhysRevLett.86.1881}%
  \BibitemOpen
  \bibfield  {author} {\bibinfo {author} {\bibfnamefont {R.}~\bibnamefont
  {Moessner}}\ and\ \bibinfo {author} {\bibfnamefont {S.~L.}\ \bibnamefont
  {Sondhi}},\ }\bibfield  {title} {\bibinfo {title} {Resonating valence bond
  phase in the triangular lattice quantum dimer model},\ }\href
  {https://doi.org/10.1103/PhysRevLett.86.1881} {\bibfield  {journal} {\bibinfo
   {journal} {Phys. Rev. Lett.}\ }\textbf {\bibinfo {volume} {86}},\ \bibinfo
  {pages} {1881} (\bibinfo {year} {2001})}\BibitemShut {NoStop}%
\bibitem [{\citenamefont {Wen}(2002)}]{wen2002quantum}%
  \BibitemOpen
  \bibfield  {author} {\bibinfo {author} {\bibfnamefont {X.-G.}\ \bibnamefont
  {Wen}},\ }\bibfield  {title} {\bibinfo {title} {Quantum orders and symmetric
  spin liquids},\ }\href {https://doi.org/10.1103/PhysRevB.65.165113}
  {\bibfield  {journal} {\bibinfo  {journal} {Phys. Rev. B}\ }\textbf {\bibinfo
  {volume} {65}},\ \bibinfo {pages} {165113} (\bibinfo {year}
  {2002})}\BibitemShut {NoStop}%
\bibitem [{\citenamefont {Kitaev}(2006)}]{kitaev2006anyons}%
  \BibitemOpen
  \bibfield  {author} {\bibinfo {author} {\bibfnamefont {A.}~\bibnamefont
  {Kitaev}},\ }\bibfield  {title} {\bibinfo {title} {Anyons in an exactly
  solved model and beyond},\ }\href
  {https://doi.org/https://doi.org/10.1016/j.aop.2005.10.005} {\bibfield
  {journal} {\bibinfo  {journal} {Annals of Physics}\ }\textbf {\bibinfo
  {volume} {321}},\ \bibinfo {pages} {2} (\bibinfo {year} {2006})}\BibitemShut
  {NoStop}%
\bibitem [{\citenamefont {Balents}(2010)}]{balents2010spin}%
  \BibitemOpen
  \bibfield  {author} {\bibinfo {author} {\bibfnamefont {L.}~\bibnamefont
  {Balents}},\ }\bibfield  {title} {\bibinfo {title} {Spin liquids in
  frustrated magnets},\ }\href
  {https://doi.org/https://doi.org/10.1038/nature08917} {\bibfield  {journal}
  {\bibinfo  {journal} {Nature}\ }\textbf {\bibinfo {volume} {464}},\ \bibinfo
  {pages} {199} (\bibinfo {year} {2010})}\BibitemShut {NoStop}%
\bibitem [{\citenamefont {Wen}(2017)}]{wen2017colloquium}%
  \BibitemOpen
  \bibfield  {author} {\bibinfo {author} {\bibfnamefont {X.-G.}\ \bibnamefont
  {Wen}},\ }\bibfield  {title} {\bibinfo {title} {Colloquium: Zoo of
  quantum-topological phases of matter},\ }\href
  {https://doi.org/10.1103/RevModPhys.89.041004} {\bibfield  {journal}
  {\bibinfo  {journal} {Rev. Mod. Phys.}\ }\textbf {\bibinfo {volume} {89}},\
  \bibinfo {pages} {041004} (\bibinfo {year} {2017})}\BibitemShut {NoStop}%
\bibitem [{\citenamefont {Lee}(2008)}]{lee2008end}%
  \BibitemOpen
  \bibfield  {author} {\bibinfo {author} {\bibfnamefont {P.~A.}\ \bibnamefont
  {Lee}},\ }\bibfield  {title} {\bibinfo {title} {An end to the drought of
  quantum spin liquids},\ }\href {https://doi.org/10.1126/science.1163196}
  {\bibfield  {journal} {\bibinfo  {journal} {Science}\ }\textbf {\bibinfo
  {volume} {321}},\ \bibinfo {pages} {1306} (\bibinfo {year}
  {2008})}\BibitemShut {NoStop}%
\bibitem [{\citenamefont {Broholm}\ \emph {et~al.}(2020)\citenamefont
  {Broholm}, \citenamefont {Cava}, \citenamefont {Kivelson}, \citenamefont
  {Nocera}, \citenamefont {Norman},\ and\ \citenamefont
  {Senthil}}]{broholm2020quantum}%
  \BibitemOpen
  \bibfield  {author} {\bibinfo {author} {\bibfnamefont {C.}~\bibnamefont
  {Broholm}}, \bibinfo {author} {\bibfnamefont {R.}~\bibnamefont {Cava}},
  \bibinfo {author} {\bibfnamefont {S.}~\bibnamefont {Kivelson}}, \bibinfo
  {author} {\bibfnamefont {D.}~\bibnamefont {Nocera}}, \bibinfo {author}
  {\bibfnamefont {M.}~\bibnamefont {Norman}},\ and\ \bibinfo {author}
  {\bibfnamefont {T.}~\bibnamefont {Senthil}},\ }\bibfield  {title} {\bibinfo
  {title} {Quantum spin liquids},\ }\bibfield  {journal} {\bibinfo  {journal}
  {Science}\ }\textbf {\bibinfo {volume} {367}},\ \href
  {https://doi.org/https://doi.org/10.1126/science.aay0668}
  {https://doi.org/10.1126/science.aay0668} (\bibinfo {year}
  {2020})\BibitemShut {NoStop}%
\bibitem [{\citenamefont {Knolle}\ and\ \citenamefont
  {Moessner}(2019)}]{Knolle_ARCMP_2019}%
  \BibitemOpen
  \bibfield  {author} {\bibinfo {author} {\bibfnamefont {J.}~\bibnamefont
  {Knolle}}\ and\ \bibinfo {author} {\bibfnamefont {R.}~\bibnamefont
  {Moessner}},\ }\bibfield  {title} {\bibinfo {title} {A field guide to spin
  liquids},\ }\href {https://doi.org/10.1146/annurev-conmatphys-031218-013401}
  {\bibfield  {journal} {\bibinfo  {journal} {Annual Review of Condensed Matter
  Physics}\ }\textbf {\bibinfo {volume} {10}},\ \bibinfo {pages} {451}
  (\bibinfo {year} {2019})},\ \Eprint
  {https://arxiv.org/abs/https://doi.org/10.1146/annurev-conmatphys-031218-013401}
  {https://doi.org/10.1146/annurev-conmatphys-031218-013401} \BibitemShut
  {NoStop}%
\bibitem [{\citenamefont {Takagi}\ \emph {et~al.}(2019)\citenamefont {Takagi},
  \citenamefont {Takayama}, \citenamefont {Jackeli}, \citenamefont
  {Khaliullin},\ and\ \citenamefont {Nagler}}]{takagi2019concept}%
  \BibitemOpen
  \bibfield  {author} {\bibinfo {author} {\bibfnamefont {H.}~\bibnamefont
  {Takagi}}, \bibinfo {author} {\bibfnamefont {T.}~\bibnamefont {Takayama}},
  \bibinfo {author} {\bibfnamefont {G.}~\bibnamefont {Jackeli}}, \bibinfo
  {author} {\bibfnamefont {G.}~\bibnamefont {Khaliullin}},\ and\ \bibinfo
  {author} {\bibfnamefont {S.~E.}\ \bibnamefont {Nagler}},\ }\bibfield  {title}
  {\bibinfo {title} {Concept and realization of kitaev quantum spin liquids},\
  }\href {https://www.nature.com/articles/s42254-019-0038-2} {\bibfield
  {journal} {\bibinfo  {journal} {Nature Reviews Physics}\ }\textbf {\bibinfo
  {volume} {1}},\ \bibinfo {pages} {264} (\bibinfo {year} {2019})}\BibitemShut
  {NoStop}%
\bibitem [{\citenamefont {Zhou}\ and\ \citenamefont {Lee}(2011)}]{lee2011}%
  \BibitemOpen
  \bibfield  {author} {\bibinfo {author} {\bibfnamefont {Y.}~\bibnamefont
  {Zhou}}\ and\ \bibinfo {author} {\bibfnamefont {P.~A.}\ \bibnamefont {Lee}},\
  }\bibfield  {title} {\bibinfo {title} {Spinon phonon interaction and
  ultrasonic attenuation in quantum spin liquids},\ }\href
  {https://doi.org/10.1103/PhysRevLett.106.056402} {\bibfield  {journal}
  {\bibinfo  {journal} {Phys.\ Rev.\ Lett.}\ }\textbf {\bibinfo {volume}
  {106}},\ \bibinfo {pages} {056402} (\bibinfo {year} {2011})}\BibitemShut
  {NoStop}%
\bibitem [{\citenamefont {Shiralieva}\ \emph {et~al.}(2021)\citenamefont
  {Shiralieva}, \citenamefont {Prokoshin},\ and\ \citenamefont
  {Perkins}}]{shiralieva2021magnetoelastic}%
  \BibitemOpen
  \bibfield  {author} {\bibinfo {author} {\bibfnamefont {A.}~\bibnamefont
  {Shiralieva}}, \bibinfo {author} {\bibfnamefont {A.}~\bibnamefont
  {Prokoshin}},\ and\ \bibinfo {author} {\bibfnamefont {N.~B.}\ \bibnamefont
  {Perkins}},\ }\bibfield  {title} {\bibinfo {title} {Magnetoelastic effects in
  the hyperhoneycomb kitaev spin liquid},\ }\href
  {https://aip.scitation.org/doi/10.1063/10.0005801} {\bibfield  {journal}
  {\bibinfo  {journal} {Low Temperature Physics}\ }\textbf {\bibinfo {volume}
  {47}},\ \bibinfo {pages} {784} (\bibinfo {year} {2021})}\BibitemShut
  {NoStop}%
\bibitem [{\citenamefont {Ye}\ \emph {et~al.}(2020)\citenamefont {Ye},
  \citenamefont {Fernandes},\ and\ \citenamefont
  {Perkins}}]{PhysRevResearch.2.033180}%
  \BibitemOpen
  \bibfield  {author} {\bibinfo {author} {\bibfnamefont {M.}~\bibnamefont
  {Ye}}, \bibinfo {author} {\bibfnamefont {R.~M.}\ \bibnamefont {Fernandes}},\
  and\ \bibinfo {author} {\bibfnamefont {N.~B.}\ \bibnamefont {Perkins}},\
  }\bibfield  {title} {\bibinfo {title} {Phonon dynamics in the kitaev spin
  liquid},\ }\href {https://doi.org/10.1103/PhysRevResearch.2.033180}
  {\bibfield  {journal} {\bibinfo  {journal} {Phys. Rev. Research}\ }\textbf
  {\bibinfo {volume} {2}},\ \bibinfo {pages} {033180} (\bibinfo {year}
  {2020})}\BibitemShut {NoStop}%
\bibitem [{\citenamefont {Li}\ \emph {et~al.}(2021)\citenamefont {Li},
  \citenamefont {Zhang}, \citenamefont {Said}, \citenamefont {Fabbris},
  \citenamefont {Mazzone}, \citenamefont {Yan}, \citenamefont {Mandrus},
  \citenamefont {Halász}, \citenamefont {Okamoto}, \citenamefont {Murakami},
  \citenamefont {Dean}, \citenamefont {Lee}, \citenamefont {Miao},\ and\
  \citenamefont {H.}}]{rucl32021}%
  \BibitemOpen
  \bibfield  {author} {\bibinfo {author} {\bibfnamefont {H.}~\bibnamefont
  {Li}}, \bibinfo {author} {\bibfnamefont {T.~T.}\ \bibnamefont {Zhang}},
  \bibinfo {author} {\bibfnamefont {A.}~\bibnamefont {Said}}, \bibinfo {author}
  {\bibfnamefont {G.}~\bibnamefont {Fabbris}}, \bibinfo {author} {\bibfnamefont
  {D.~G.}\ \bibnamefont {Mazzone}}, \bibinfo {author} {\bibfnamefont {J.~Q.}\
  \bibnamefont {Yan}}, \bibinfo {author} {\bibfnamefont {D.}~\bibnamefont
  {Mandrus}}, \bibinfo {author} {\bibfnamefont {G.~B.}\ \bibnamefont
  {Halász}}, \bibinfo {author} {\bibfnamefont {S.}~\bibnamefont {Okamoto}},
  \bibinfo {author} {\bibfnamefont {S.}~\bibnamefont {Murakami}}, \bibinfo
  {author} {\bibfnamefont {M.~P.~M.}\ \bibnamefont {Dean}}, \bibinfo {author}
  {\bibfnamefont {H.~N.}\ \bibnamefont {Lee}}, \bibinfo {author} {\bibnamefont
  {Miao}},\ and\ \bibinfo {author} {\bibnamefont {H.}},\ }\bibfield  {title}
  {\bibinfo {title} {Giant phonon anomalies in the proximate kitaev quantum
  spin liquid ${\mathrm{\alpha}}$-${\mathrm{rucl}}_{3}$},\ }\href
  {https://doi.org/10.1038/s41467-021-23826-1} {\bibfield  {journal} {\bibinfo
  {journal} {Nat. Commun.}\ }\textbf {\bibinfo {volume} {12}} (\bibinfo {year}
  {2021})}\BibitemShut {NoStop}%
\bibitem [{\citenamefont {Mross}\ and\ \citenamefont
  {Senthil}(2011)}]{PhysRevB.84.041102}%
  \BibitemOpen
  \bibfield  {author} {\bibinfo {author} {\bibfnamefont {D.~F.}\ \bibnamefont
  {Mross}}\ and\ \bibinfo {author} {\bibfnamefont {T.}~\bibnamefont
  {Senthil}},\ }\bibfield  {title} {\bibinfo {title} {Charge friedel
  oscillations in a mott insulator},\ }\href
  {https://doi.org/10.1103/PhysRevB.84.041102} {\bibfield  {journal} {\bibinfo
  {journal} {Phys. Rev. B}\ }\textbf {\bibinfo {volume} {84}},\ \bibinfo
  {pages} {041102} (\bibinfo {year} {2011})}\BibitemShut {NoStop}%
\bibitem [{\citenamefont {Grissonnanche}\ \emph {et~al.}(2019)\citenamefont
  {Grissonnanche}, \citenamefont {Legros}, \citenamefont {Badoux},
  \citenamefont {Lefran{\c{c}}ois}, \citenamefont {Zatko}, \citenamefont
  {Lizaire}, \citenamefont {Lalibert{\'e}}, \citenamefont {Gourgout},
  \citenamefont {Zhou}, \citenamefont {Pyon} \emph
  {et~al.}}]{grissonnanche2019giant}%
  \BibitemOpen
  \bibfield  {author} {\bibinfo {author} {\bibfnamefont {G.}~\bibnamefont
  {Grissonnanche}}, \bibinfo {author} {\bibfnamefont {A.}~\bibnamefont
  {Legros}}, \bibinfo {author} {\bibfnamefont {S.}~\bibnamefont {Badoux}},
  \bibinfo {author} {\bibfnamefont {E.}~\bibnamefont {Lefran{\c{c}}ois}},
  \bibinfo {author} {\bibfnamefont {V.}~\bibnamefont {Zatko}}, \bibinfo
  {author} {\bibfnamefont {M.}~\bibnamefont {Lizaire}}, \bibinfo {author}
  {\bibfnamefont {F.}~\bibnamefont {Lalibert{\'e}}}, \bibinfo {author}
  {\bibfnamefont {A.}~\bibnamefont {Gourgout}}, \bibinfo {author}
  {\bibfnamefont {J.-S.}\ \bibnamefont {Zhou}}, \bibinfo {author}
  {\bibfnamefont {S.}~\bibnamefont {Pyon}}, \emph {et~al.},\ }\bibfield
  {title} {\bibinfo {title} {Giant thermal hall conductivity in the pseudogap
  phase of cuprate superconductors},\ }\href
  {https://www.nature.com/articles/s41586-019-1375-0} {\bibfield  {journal}
  {\bibinfo  {journal} {Nature}\ }\textbf {\bibinfo {volume} {571}},\ \bibinfo
  {pages} {376} (\bibinfo {year} {2019})}\BibitemShut {NoStop}%
\bibitem [{\citenamefont {Kasahara}\ \emph
  {et~al.}(2018{\natexlab{a}})\citenamefont {Kasahara}, \citenamefont {Sugii},
  \citenamefont {Ohnishi}, \citenamefont {Shimozawa}, \citenamefont
  {Yamashita}, \citenamefont {Kurita}, \citenamefont {Tanaka}, \citenamefont
  {Nasu}, \citenamefont {Motome}, \citenamefont {Shibauchi},\ and\
  \citenamefont {Matsuda}}]{kasahara2018unusual}%
  \BibitemOpen
  \bibfield  {author} {\bibinfo {author} {\bibfnamefont {Y.}~\bibnamefont
  {Kasahara}}, \bibinfo {author} {\bibfnamefont {K.}~\bibnamefont {Sugii}},
  \bibinfo {author} {\bibfnamefont {T.}~\bibnamefont {Ohnishi}}, \bibinfo
  {author} {\bibfnamefont {M.}~\bibnamefont {Shimozawa}}, \bibinfo {author}
  {\bibfnamefont {M.}~\bibnamefont {Yamashita}}, \bibinfo {author}
  {\bibfnamefont {N.}~\bibnamefont {Kurita}}, \bibinfo {author} {\bibfnamefont
  {H.}~\bibnamefont {Tanaka}}, \bibinfo {author} {\bibfnamefont
  {J.}~\bibnamefont {Nasu}}, \bibinfo {author} {\bibfnamefont {Y.}~\bibnamefont
  {Motome}}, \bibinfo {author} {\bibfnamefont {T.}~\bibnamefont {Shibauchi}},\
  and\ \bibinfo {author} {\bibfnamefont {Y.}~\bibnamefont {Matsuda}},\
  }\bibfield  {title} {\bibinfo {title} {Unusual thermal hall effect in a
  kitaev spin liquid candidate $\alpha$-rucl$_3$},\ }\href
  {https://journals.aps.org/prl/abstract/10.1103/PhysRevLett.120.217205}
  {\bibfield  {journal} {\bibinfo  {journal} {Phys. Rev. Lett.}\ }\textbf
  {\bibinfo {volume} {120}},\ \bibinfo {pages} {217205} (\bibinfo {year}
  {2018}{\natexlab{a}})}\BibitemShut {NoStop}%
\bibitem [{\citenamefont {Nasu}\ \emph {et~al.}(2016)\citenamefont {Nasu},
  \citenamefont {Knolle}, \citenamefont {Kovrizhin}, \citenamefont {Motome},\
  and\ \citenamefont {Moessner}}]{nasu2016fermionic}%
  \BibitemOpen
  \bibfield  {author} {\bibinfo {author} {\bibfnamefont {J.}~\bibnamefont
  {Nasu}}, \bibinfo {author} {\bibfnamefont {J.}~\bibnamefont {Knolle}},
  \bibinfo {author} {\bibfnamefont {D.~L.}\ \bibnamefont {Kovrizhin}}, \bibinfo
  {author} {\bibfnamefont {Y.}~\bibnamefont {Motome}},\ and\ \bibinfo {author}
  {\bibfnamefont {R.}~\bibnamefont {Moessner}},\ }\bibfield  {title} {\bibinfo
  {title} {Fermionic response from fractionalization in an insulating
  two-dimensional magnet},\ }\href {https://www.nature.com/articles/nphys3809}
  {\bibfield  {journal} {\bibinfo  {journal} {Nat. Phys.}\ }\textbf {\bibinfo
  {volume} {12}},\ \bibinfo {pages} {912} (\bibinfo {year} {2016})}\BibitemShut
  {NoStop}%
\bibitem [{\citenamefont {Banerjee}\ \emph {et~al.}(2016)\citenamefont
  {Banerjee}, \citenamefont {Bridges}, \citenamefont {Yan}, \citenamefont
  {Aczel}, \citenamefont {Li}, \citenamefont {Stone}, \citenamefont {Granroth},
  \citenamefont {Lumsden}, \citenamefont {Yiu}, \citenamefont {Knolle} \emph
  {et~al.}}]{banerjee2016proximate}%
  \BibitemOpen
  \bibfield  {author} {\bibinfo {author} {\bibfnamefont {A.}~\bibnamefont
  {Banerjee}}, \bibinfo {author} {\bibfnamefont {C.}~\bibnamefont {Bridges}},
  \bibinfo {author} {\bibfnamefont {J.-Q.}\ \bibnamefont {Yan}}, \bibinfo
  {author} {\bibfnamefont {A.}~\bibnamefont {Aczel}}, \bibinfo {author}
  {\bibfnamefont {L.}~\bibnamefont {Li}}, \bibinfo {author} {\bibfnamefont
  {M.}~\bibnamefont {Stone}}, \bibinfo {author} {\bibfnamefont
  {G.}~\bibnamefont {Granroth}}, \bibinfo {author} {\bibfnamefont
  {M.}~\bibnamefont {Lumsden}}, \bibinfo {author} {\bibfnamefont
  {Y.}~\bibnamefont {Yiu}}, \bibinfo {author} {\bibfnamefont {J.}~\bibnamefont
  {Knolle}}, \emph {et~al.},\ }\bibfield  {title} {\bibinfo {title} {Proximate
  kitaev quantum spin liquid behaviour in a honeycomb magnet},\ }\href
  {https://www.nature.com/articles/nmat4604} {\bibfield  {journal} {\bibinfo
  {journal} {Nat. Mater.}\ }\textbf {\bibinfo {volume} {15}},\ \bibinfo {pages}
  {733} (\bibinfo {year} {2016})}\BibitemShut {NoStop}%
\bibitem [{\citenamefont {Banerjee}\ \emph {et~al.}(2017)\citenamefont
  {Banerjee}, \citenamefont {Yan}, \citenamefont {Knolle}, \citenamefont
  {Bridges}, \citenamefont {Stone}, \citenamefont {Lumsden}, \citenamefont
  {Mandrus}, \citenamefont {Tennant}, \citenamefont {Moessner},\ and\
  \citenamefont {Nagler}}]{banerjee2017neutron}%
  \BibitemOpen
  \bibfield  {author} {\bibinfo {author} {\bibfnamefont {A.}~\bibnamefont
  {Banerjee}}, \bibinfo {author} {\bibfnamefont {J.}~\bibnamefont {Yan}},
  \bibinfo {author} {\bibfnamefont {J.}~\bibnamefont {Knolle}}, \bibinfo
  {author} {\bibfnamefont {C.~A.}\ \bibnamefont {Bridges}}, \bibinfo {author}
  {\bibfnamefont {M.~B.}\ \bibnamefont {Stone}}, \bibinfo {author}
  {\bibfnamefont {M.~D.}\ \bibnamefont {Lumsden}}, \bibinfo {author}
  {\bibfnamefont {D.~G.}\ \bibnamefont {Mandrus}}, \bibinfo {author}
  {\bibfnamefont {D.~A.}\ \bibnamefont {Tennant}}, \bibinfo {author}
  {\bibfnamefont {R.}~\bibnamefont {Moessner}},\ and\ \bibinfo {author}
  {\bibfnamefont {S.~E.}\ \bibnamefont {Nagler}},\ }\bibfield  {title}
  {\bibinfo {title} {Neutron scattering in the proximate quantum spin liquid
  $\alpha$-rucl3},\ }\href
  {https://www.science.org/doi/10.1126/science.aah6015} {\bibfield  {journal}
  {\bibinfo  {journal} {Science}\ }\textbf {\bibinfo {volume} {356}},\ \bibinfo
  {pages} {1055} (\bibinfo {year} {2017})}\BibitemShut {NoStop}%
\bibitem [{\citenamefont {Ye}\ \emph {et~al.}(2018)\citenamefont {Ye},
  \citenamefont {Hal\'asz}, \citenamefont {Savary},\ and\ \citenamefont
  {Balents}}]{PhysRevLett.121.147201}%
  \BibitemOpen
  \bibfield  {author} {\bibinfo {author} {\bibfnamefont {M.}~\bibnamefont
  {Ye}}, \bibinfo {author} {\bibfnamefont {G.~B.}\ \bibnamefont {Hal\'asz}},
  \bibinfo {author} {\bibfnamefont {L.}~\bibnamefont {Savary}},\ and\ \bibinfo
  {author} {\bibfnamefont {L.}~\bibnamefont {Balents}},\ }\bibfield  {title}
  {\bibinfo {title} {Quantization of the thermal hall conductivity at small
  hall angles},\ }\href {https://doi.org/10.1103/PhysRevLett.121.147201}
  {\bibfield  {journal} {\bibinfo  {journal} {Phys. Rev. Lett.}\ }\textbf
  {\bibinfo {volume} {121}},\ \bibinfo {pages} {147201} (\bibinfo {year}
  {2018})}\BibitemShut {NoStop}%
\bibitem [{\citenamefont {Vinkler-Aviv}\ and\ \citenamefont
  {Rosch}(2018)}]{vinkler2018approximately}%
  \BibitemOpen
  \bibfield  {author} {\bibinfo {author} {\bibfnamefont {Y.}~\bibnamefont
  {Vinkler-Aviv}}\ and\ \bibinfo {author} {\bibfnamefont {A.}~\bibnamefont
  {Rosch}},\ }\bibfield  {title} {\bibinfo {title} {Approximately quantized
  thermal hall effect of chiral liquids coupled to phonons},\ }\href
  {https://journals.aps.org/prx/abstract/10.1103/PhysRevX.8.031032} {\bibfield
  {journal} {\bibinfo  {journal} {Phys. Rev. X}\ }\textbf {\bibinfo {volume}
  {8}},\ \bibinfo {pages} {031032} (\bibinfo {year} {2018})}\BibitemShut
  {NoStop}%
\bibitem [{\citenamefont {Kasahara}\ \emph
  {et~al.}(2018{\natexlab{b}})\citenamefont {Kasahara}, \citenamefont
  {Ohnishi}, \citenamefont {Mizukami}, \citenamefont {Tanaka}, \citenamefont
  {Ma}, \citenamefont {Sugii}, \citenamefont {Kurita}, \citenamefont {Tanaka},
  \citenamefont {Nasu}, \citenamefont {Motome} \emph
  {et~al.}}]{kasahara2018majorana}%
  \BibitemOpen
  \bibfield  {author} {\bibinfo {author} {\bibfnamefont {Y.}~\bibnamefont
  {Kasahara}}, \bibinfo {author} {\bibfnamefont {T.}~\bibnamefont {Ohnishi}},
  \bibinfo {author} {\bibfnamefont {Y.}~\bibnamefont {Mizukami}}, \bibinfo
  {author} {\bibfnamefont {O.}~\bibnamefont {Tanaka}}, \bibinfo {author}
  {\bibfnamefont {S.}~\bibnamefont {Ma}}, \bibinfo {author} {\bibfnamefont
  {K.}~\bibnamefont {Sugii}}, \bibinfo {author} {\bibfnamefont
  {N.}~\bibnamefont {Kurita}}, \bibinfo {author} {\bibfnamefont
  {H.}~\bibnamefont {Tanaka}}, \bibinfo {author} {\bibfnamefont
  {J.}~\bibnamefont {Nasu}}, \bibinfo {author} {\bibfnamefont {Y.}~\bibnamefont
  {Motome}}, \emph {et~al.},\ }\bibfield  {title} {\bibinfo {title} {Majorana
  quantization and half-integer thermal quantum hall effect in a kitaev spin
  liquid},\ }\href {https://doi.org/https://doi.org/10.1038/s41586-018-0274-0}
  {\bibfield  {journal} {\bibinfo  {journal} {Nature}\ }\textbf {\bibinfo
  {volume} {559}},\ \bibinfo {pages} {227} (\bibinfo {year}
  {2018}{\natexlab{b}})}\BibitemShut {NoStop}%
\bibitem [{\citenamefont {Yamashita}\ \emph {et~al.}(2020)\citenamefont
  {Yamashita}, \citenamefont {Gouchi}, \citenamefont {Uwatoko}, \citenamefont
  {Kurita},\ and\ \citenamefont {Tanaka}}]{yamashita2020sample}%
  \BibitemOpen
  \bibfield  {author} {\bibinfo {author} {\bibfnamefont {M.}~\bibnamefont
  {Yamashita}}, \bibinfo {author} {\bibfnamefont {J.}~\bibnamefont {Gouchi}},
  \bibinfo {author} {\bibfnamefont {Y.}~\bibnamefont {Uwatoko}}, \bibinfo
  {author} {\bibfnamefont {N.}~\bibnamefont {Kurita}},\ and\ \bibinfo {author}
  {\bibfnamefont {H.}~\bibnamefont {Tanaka}},\ }\bibfield  {title} {\bibinfo
  {title} {Sample dependence of half-integer quantized thermal hall effect in
  the kitaev spin-liquid candidate $\alpha$- rucl 3},\ }\href
  {https://journals.aps.org/prb/abstract/10.1103/PhysRevB.102.220404}
  {\bibfield  {journal} {\bibinfo  {journal} {Phys. Rev. B}\ }\textbf {\bibinfo
  {volume} {102}},\ \bibinfo {pages} {220404} (\bibinfo {year}
  {2020})}\BibitemShut {NoStop}%
\bibitem [{\citenamefont {Yokoi}\ \emph {et~al.}(2021)\citenamefont {Yokoi},
  \citenamefont {Ma}, \citenamefont {Kasahara}, \citenamefont {Kasahara},
  \citenamefont {Shibauchi}, \citenamefont {Kurita}, \citenamefont {Tanaka},
  \citenamefont {Nasu}, \citenamefont {Motome}, \citenamefont {Hickey} \emph
  {et~al.}}]{yokoi2021half}%
  \BibitemOpen
  \bibfield  {author} {\bibinfo {author} {\bibfnamefont {T.}~\bibnamefont
  {Yokoi}}, \bibinfo {author} {\bibfnamefont {S.}~\bibnamefont {Ma}}, \bibinfo
  {author} {\bibfnamefont {Y.}~\bibnamefont {Kasahara}}, \bibinfo {author}
  {\bibfnamefont {S.}~\bibnamefont {Kasahara}}, \bibinfo {author}
  {\bibfnamefont {T.}~\bibnamefont {Shibauchi}}, \bibinfo {author}
  {\bibfnamefont {N.}~\bibnamefont {Kurita}}, \bibinfo {author} {\bibfnamefont
  {H.}~\bibnamefont {Tanaka}}, \bibinfo {author} {\bibfnamefont
  {J.}~\bibnamefont {Nasu}}, \bibinfo {author} {\bibfnamefont {Y.}~\bibnamefont
  {Motome}}, \bibinfo {author} {\bibfnamefont {C.}~\bibnamefont {Hickey}},
  \emph {et~al.},\ }\bibfield  {title} {\bibinfo {title} {Half-integer
  quantized anomalous thermal hall effect in the kitaev material candidate
  $\alpha$-rucl3},\ }\href
  {https://www.science.org/doi/10.1126/science.aay5551} {\bibfield  {journal}
  {\bibinfo  {journal} {Science}\ }\textbf {\bibinfo {volume} {373}},\ \bibinfo
  {pages} {568} (\bibinfo {year} {2021})}\BibitemShut {NoStop}%
\bibitem [{\citenamefont {Czajka}\ \emph {et~al.}(2021)\citenamefont {Czajka},
  \citenamefont {Gao}, \citenamefont {Hirschberger}, \citenamefont
  {Lampen-Kelley}, \citenamefont {Banerjee}, \citenamefont {Yan}, \citenamefont
  {Mandrus}, \citenamefont {Nagler},\ and\ \citenamefont
  {Ong}}]{czajka2021oscillations}%
  \BibitemOpen
  \bibfield  {author} {\bibinfo {author} {\bibfnamefont {P.}~\bibnamefont
  {Czajka}}, \bibinfo {author} {\bibfnamefont {T.}~\bibnamefont {Gao}},
  \bibinfo {author} {\bibfnamefont {M.}~\bibnamefont {Hirschberger}}, \bibinfo
  {author} {\bibfnamefont {P.}~\bibnamefont {Lampen-Kelley}}, \bibinfo {author}
  {\bibfnamefont {A.}~\bibnamefont {Banerjee}}, \bibinfo {author}
  {\bibfnamefont {J.}~\bibnamefont {Yan}}, \bibinfo {author} {\bibfnamefont
  {D.~G.}\ \bibnamefont {Mandrus}}, \bibinfo {author} {\bibfnamefont {S.~E.}\
  \bibnamefont {Nagler}},\ and\ \bibinfo {author} {\bibfnamefont
  {N.}~\bibnamefont {Ong}},\ }\bibfield  {title} {\bibinfo {title}
  {Oscillations of the thermal conductivity in the spin-liquid state of
  $\alpha$-rucl 3},\ }\href
  {https://www.nature.com/articles/s41567-021-01243-x} {\bibfield  {journal}
  {\bibinfo  {journal} {Nat. Phys.}\ ,\ \bibinfo {pages} {915}} (\bibinfo
  {year} {2021})}\BibitemShut {NoStop}%
\bibitem [{\citenamefont {Pal}\ \emph {et~al.}(2021)\citenamefont {Pal},
  \citenamefont {Seth}, \citenamefont {Sakrikar}, \citenamefont {Ali},
  \citenamefont {Bhattacharjee}, \citenamefont {Muthu}, \citenamefont {Singh},\
  and\ \citenamefont {Sood}}]{pal2020probing}%
  \BibitemOpen
  \bibfield  {author} {\bibinfo {author} {\bibfnamefont {S.}~\bibnamefont
  {Pal}}, \bibinfo {author} {\bibfnamefont {A.}~\bibnamefont {Seth}}, \bibinfo
  {author} {\bibfnamefont {P.}~\bibnamefont {Sakrikar}}, \bibinfo {author}
  {\bibfnamefont {A.}~\bibnamefont {Ali}}, \bibinfo {author} {\bibfnamefont
  {S.}~\bibnamefont {Bhattacharjee}}, \bibinfo {author} {\bibfnamefont
  {D.}~\bibnamefont {Muthu}}, \bibinfo {author} {\bibfnamefont
  {Y.}~\bibnamefont {Singh}},\ and\ \bibinfo {author} {\bibfnamefont
  {A.}~\bibnamefont {Sood}},\ }\bibfield  {title} {\bibinfo {title} {Probing
  signatures of fractionalisation in candidate quantum spin liquid
  cu$_2$iro$_3$ via anomalous raman scattering},\ }\href
  {https://doi.org/https://doi.org/10.1103/PhysRevB.104.184420} {\bibfield
  {journal} {\bibinfo  {journal} {Phys.\ Rev. \ B}\ }\textbf {\bibinfo {volume}
  {104}},\ \bibinfo {pages} {184420} (\bibinfo {year} {2021})}\BibitemShut
  {NoStop}%
\bibitem [{\citenamefont {Sandilands}\ \emph {et~al.}(2015)\citenamefont
  {Sandilands}, \citenamefont {Tian}, \citenamefont {Plumb}, \citenamefont
  {Kim},\ and\ \citenamefont {Burch}}]{sandilands}%
  \BibitemOpen
  \bibfield  {author} {\bibinfo {author} {\bibfnamefont {L.~J.}\ \bibnamefont
  {Sandilands}}, \bibinfo {author} {\bibfnamefont {Y.}~\bibnamefont {Tian}},
  \bibinfo {author} {\bibfnamefont {K.~W.}\ \bibnamefont {Plumb}}, \bibinfo
  {author} {\bibfnamefont {Y.-J.}\ \bibnamefont {Kim}},\ and\ \bibinfo {author}
  {\bibfnamefont {K.~S.}\ \bibnamefont {Burch}},\ }\bibfield  {title} {\bibinfo
  {title} {Scattering continuum and possible fractionalized excitations in
  $\alpha-$rucl$_3$},\ }\href {https://doi.org/10.1103/PhysRevLett.114.147201}
  {\bibfield  {journal} {\bibinfo  {journal} {Phys.\ Rev.\ Lett.}\ }\textbf
  {\bibinfo {volume} {114}},\ \bibinfo {pages} {147201} (\bibinfo {year}
  {2015})}\BibitemShut {NoStop}%
\bibitem [{\citenamefont {Glamazda}\ \emph {et~al.}(2016)\citenamefont
  {Glamazda}, \citenamefont {Lemmens}, \citenamefont {Do}, \citenamefont
  {Choi},\ and\ \citenamefont {Choi}}]{LiIrO32016}%
  \BibitemOpen
  \bibfield  {author} {\bibinfo {author} {\bibfnamefont {A.}~\bibnamefont
  {Glamazda}}, \bibinfo {author} {\bibfnamefont {P.}~\bibnamefont {Lemmens}},
  \bibinfo {author} {\bibfnamefont {S.-H.}\ \bibnamefont {Do}}, \bibinfo
  {author} {\bibfnamefont {Y.}~\bibnamefont {Choi}},\ and\ \bibinfo {author}
  {\bibfnamefont {K.-Y.}\ \bibnamefont {Choi}},\ }\bibfield  {title} {\bibinfo
  {title} {Raman spectroscopic signature of fractionalized excitations in the
  harmonic-honeycomb iridates $\beta-$ and $\gamma-$ li$_2$iro$_3$},\ }\href
  {https://doi.org/10.1038/ncomms12286} {\bibfield  {journal} {\bibinfo
  {journal} {Nat. Commun.}\ }\textbf {\bibinfo {volume} {7}} (\bibinfo {year}
  {2016})}\BibitemShut {NoStop}%
\bibitem [{\citenamefont {Xu}\ \emph {et~al.}(2021{\natexlab{a}})\citenamefont
  {Xu}, \citenamefont {Man}, \citenamefont {Tang}, \citenamefont {Baidya},
  \citenamefont {Zhang}, \citenamefont {Nakatsuji}, \citenamefont
  {Vanderbilt},\ and\ \citenamefont {Drichko}}]{satoru}%
  \BibitemOpen
  \bibfield  {author} {\bibinfo {author} {\bibfnamefont {Y.}~\bibnamefont
  {Xu}}, \bibinfo {author} {\bibfnamefont {H.}~\bibnamefont {Man}}, \bibinfo
  {author} {\bibfnamefont {N.}~\bibnamefont {Tang}}, \bibinfo {author}
  {\bibfnamefont {S.}~\bibnamefont {Baidya}}, \bibinfo {author} {\bibfnamefont
  {H.}~\bibnamefont {Zhang}}, \bibinfo {author} {\bibfnamefont
  {S.}~\bibnamefont {Nakatsuji}}, \bibinfo {author} {\bibfnamefont
  {D.}~\bibnamefont {Vanderbilt}},\ and\ \bibinfo {author} {\bibfnamefont
  {N.}~\bibnamefont {Drichko}},\ }\bibfield  {title} {\bibinfo {title}
  {Importance of dynamic lattice effects for crystal field excitations in
  quantum spin ice candidate pr$_2$zr$_2$o$_7$},\ }\href
  {https://doi.org/https://doi.org/10.1103/PhysRevB.104.075125} {\bibfield
  {journal} {\bibinfo  {journal} {Phys.\ Rev.\ B}\ }\textbf {\bibinfo {volume}
  {104}},\ \bibinfo {pages} {075125} (\bibinfo {year}
  {2021}{\natexlab{a}})}\BibitemShut {NoStop}%
\bibitem [{\citenamefont {Xu}\ \emph {et~al.}(2021{\natexlab{b}})\citenamefont
  {Xu}, \citenamefont {Man}, \citenamefont {Ohtsuki}, \citenamefont {Baidya},
  \citenamefont {Zhang}, \citenamefont {Nakatsuji}, \citenamefont
  {Vanderbilt},\ and\ \citenamefont {Drichko}}]{natalia2021}%
  \BibitemOpen
  \bibfield  {author} {\bibinfo {author} {\bibfnamefont {Y.}~\bibnamefont
  {Xu}}, \bibinfo {author} {\bibfnamefont {H.}~\bibnamefont {Man}}, \bibinfo
  {author} {\bibfnamefont {T.}~\bibnamefont {Ohtsuki}}, \bibinfo {author}
  {\bibfnamefont {S.}~\bibnamefont {Baidya}}, \bibinfo {author} {\bibfnamefont
  {H.}~\bibnamefont {Zhang}}, \bibinfo {author} {\bibfnamefont
  {S.}~\bibnamefont {Nakatsuji}}, \bibinfo {author} {\bibfnamefont
  {D.}~\bibnamefont {Vanderbilt}},\ and\ \bibinfo {author} {\bibfnamefont
  {N.}~\bibnamefont {Drichko}},\ }\bibfield  {title} {\bibinfo {title} {Phonon
  spectrum of pr$_2$zr$_2$o$_7$ and pr$_2$ir$_2$o$_7$ as an evidence of
  coupling of the lattice with electronic and magnetic degrees of freedom},\
  }\href {https://arxiv.org/abs/2108.01664} {\bibfield  {journal} {\bibinfo
  {journal} {\textit{arXiv}:2108.01664v1}\ } (\bibinfo {year}
  {2021}{\natexlab{b}})}\BibitemShut {NoStop}%
\bibitem [{\citenamefont {L{\"u}thi}(2007)}]{luthi2007physical}%
  \BibitemOpen
  \bibfield  {author} {\bibinfo {author} {\bibfnamefont {B.}~\bibnamefont
  {L{\"u}thi}},\ }\href@noop {} {\emph {\bibinfo {title} {Physical acoustics in
  the solid state}}},\ Vol.\ \bibinfo {volume} {148}\ (\bibinfo  {publisher}
  {Springer Science \& Business Media},\ \bibinfo {year} {2007})\BibitemShut
  {NoStop}%
\bibitem [{\citenamefont {T{\'o}th}\ \emph {et~al.}(2016)\citenamefont
  {T{\'o}th}, \citenamefont {Wehinger}, \citenamefont {Rolfs}, \citenamefont
  {Birol}, \citenamefont {Stuhr}, \citenamefont {Takatsu}, \citenamefont
  {Kimura}, \citenamefont {Kimura}, \citenamefont {R{\o}nnow},\ and\
  \citenamefont {R{\"u}egg}}]{toth2016electromagnon}%
  \BibitemOpen
  \bibfield  {author} {\bibinfo {author} {\bibfnamefont {S.}~\bibnamefont
  {T{\'o}th}}, \bibinfo {author} {\bibfnamefont {B.}~\bibnamefont {Wehinger}},
  \bibinfo {author} {\bibfnamefont {K.}~\bibnamefont {Rolfs}}, \bibinfo
  {author} {\bibfnamefont {T.}~\bibnamefont {Birol}}, \bibinfo {author}
  {\bibfnamefont {U.}~\bibnamefont {Stuhr}}, \bibinfo {author} {\bibfnamefont
  {H.}~\bibnamefont {Takatsu}}, \bibinfo {author} {\bibfnamefont
  {K.}~\bibnamefont {Kimura}}, \bibinfo {author} {\bibfnamefont
  {T.}~\bibnamefont {Kimura}}, \bibinfo {author} {\bibfnamefont {H.~M.}\
  \bibnamefont {R{\o}nnow}},\ and\ \bibinfo {author} {\bibfnamefont
  {C.}~\bibnamefont {R{\"u}egg}},\ }\bibfield  {title} {\bibinfo {title}
  {Electromagnon dispersion probed by inelastic x-ray scattering in licro 2},\
  }\href {https://www.nature.com/articles/ncomms13547} {\bibfield  {journal}
  {\bibinfo  {journal} {Nat. Commun.}\ }\textbf {\bibinfo {volume} {7}}
  (\bibinfo {year} {2016})}\BibitemShut {NoStop}%
\bibitem [{\citenamefont {Aynajian}\ \emph {et~al.}(2008)\citenamefont
  {Aynajian}, \citenamefont {Keller}, \citenamefont {Boeri}, \citenamefont
  {Shapiro}, \citenamefont {Habicht},\ and\ \citenamefont
  {Keimer}}]{aynajian2008energy}%
  \BibitemOpen
  \bibfield  {author} {\bibinfo {author} {\bibfnamefont {P.}~\bibnamefont
  {Aynajian}}, \bibinfo {author} {\bibfnamefont {T.}~\bibnamefont {Keller}},
  \bibinfo {author} {\bibfnamefont {L.}~\bibnamefont {Boeri}}, \bibinfo
  {author} {\bibfnamefont {S.}~\bibnamefont {Shapiro}}, \bibinfo {author}
  {\bibfnamefont {K.}~\bibnamefont {Habicht}},\ and\ \bibinfo {author}
  {\bibfnamefont {B.}~\bibnamefont {Keimer}},\ }\bibfield  {title} {\bibinfo
  {title} {Energy gaps and kohn anomalies in elemental superconductors},\
  }\href {https://www.science.org/doi/10.1126/science.1154115} {\bibfield
  {journal} {\bibinfo  {journal} {Science}\ }\textbf {\bibinfo {volume}
  {319}},\ \bibinfo {pages} {1509} (\bibinfo {year} {2008})}\BibitemShut
  {NoStop}%
\bibitem [{\citenamefont {Bhattacharjee}\ \emph {et~al.}(2012)\citenamefont
  {Bhattacharjee}, \citenamefont {Lee},\ and\ \citenamefont
  {Kim}}]{bhattacharjee2012spin}%
  \BibitemOpen
  \bibfield  {author} {\bibinfo {author} {\bibfnamefont {S.}~\bibnamefont
  {Bhattacharjee}}, \bibinfo {author} {\bibfnamefont {S.-S.}\ \bibnamefont
  {Lee}},\ and\ \bibinfo {author} {\bibfnamefont {Y.~B.}\ \bibnamefont {Kim}},\
  }\bibfield  {title} {\bibinfo {title} {Spin--orbital locking, emergent
  pseudo-spin and magnetic order in honeycomb lattice iridates},\ }\href
  {https://iopscience.iop.org/article/10.1088/1367-2630/14/7/073015} {\bibfield
   {journal} {\bibinfo  {journal} {New J. Phys.}\ }\textbf {\bibinfo {volume}
  {14}},\ \bibinfo {pages} {073015} (\bibinfo {year} {2012})}\BibitemShut
  {NoStop}%
\bibitem [{\citenamefont {Witczak-Krempa}\ \emph {et~al.}(2014)\citenamefont
  {Witczak-Krempa}, \citenamefont {Chen}, \citenamefont {Kim},\ and\
  \citenamefont {Balents}}]{witczak2014correlated}%
  \BibitemOpen
  \bibfield  {author} {\bibinfo {author} {\bibfnamefont {W.}~\bibnamefont
  {Witczak-Krempa}}, \bibinfo {author} {\bibfnamefont {G.}~\bibnamefont
  {Chen}}, \bibinfo {author} {\bibfnamefont {Y.~B.}\ \bibnamefont {Kim}},\ and\
  \bibinfo {author} {\bibfnamefont {L.}~\bibnamefont {Balents}},\ }\bibfield
  {title} {\bibinfo {title} {Correlated quantum phenomena in the strong
  spin-orbit regime},\ }\href
  {https://doi.org/https://doi.org/10.1146/annurev-conmatphys-020911-125138}
  {\bibfield  {journal} {\bibinfo  {journal} {Annu. Rev. Condens. Matter
  Phys.}\ }\textbf {\bibinfo {volume} {5}},\ \bibinfo {pages} {57} (\bibinfo
  {year} {2014})}\BibitemShut {NoStop}%
\bibitem [{\citenamefont {{Nussinov}}\ and\ \citenamefont {{van den
  Brink}}(2015)}]{NussinovJeroenRMP}%
  \BibitemOpen
  \bibfield  {author} {\bibinfo {author} {\bibfnamefont {Z.}~\bibnamefont
  {{Nussinov}}}\ and\ \bibinfo {author} {\bibfnamefont {J.}~\bibnamefont {{van
  den Brink}}},\ }\bibfield  {title} {\bibinfo {title} {{Compass models: Theory
  and physical motivations}},\ }\href {https://doi.org/10.1103/RevModPhys.87.1}
  {\bibfield  {journal} {\bibinfo  {journal} {Rev. Mod. Phys.}\ }\textbf
  {\bibinfo {volume} {87}},\ \bibinfo {pages} {1} (\bibinfo {year}
  {2015})}\BibitemShut {NoStop}%
\bibitem [{\citenamefont {{Hermanns}}\ \emph {et~al.}(2018)\citenamefont
  {{Hermanns}}, \citenamefont {{Kimchi}},\ and\ \citenamefont
  {{Knolle}}}]{HKK_kitaev}%
  \BibitemOpen
  \bibfield  {author} {\bibinfo {author} {\bibfnamefont {M.}~\bibnamefont
  {{Hermanns}}}, \bibinfo {author} {\bibfnamefont {I.}~\bibnamefont
  {{Kimchi}}},\ and\ \bibinfo {author} {\bibfnamefont {J.}~\bibnamefont
  {{Knolle}}},\ }\bibfield  {title} {\bibinfo {title} {{Physics of the Kitaev
  Model: Fractionalization, Dynamic Correlations, and Material Connections}},\
  }\href {https://doi.org/10.1146/annurev-conmatphys-033117-053934} {\bibfield
  {journal} {\bibinfo  {journal} {Annu. Rev. Condens. Matter Phys.}\ }\textbf
  {\bibinfo {volume} {9}},\ \bibinfo {pages} {17} (\bibinfo {year} {2018})},\
  \Eprint {https://arxiv.org/abs/1705.01740} {arXiv:1705.01740
  [cond-mat.str-el]} \BibitemShut {NoStop}%
\bibitem [{\citenamefont {Metavitsiadis}\ \emph {et~al.}(2021)\citenamefont
  {Metavitsiadis}, \citenamefont {Natori}, \citenamefont {Knolle},\ and\
  \citenamefont {Brenig}}]{metavitsiadis2021optical}%
  \BibitemOpen
  \bibfield  {author} {\bibinfo {author} {\bibfnamefont {A.}~\bibnamefont
  {Metavitsiadis}}, \bibinfo {author} {\bibfnamefont {W.}~\bibnamefont
  {Natori}}, \bibinfo {author} {\bibfnamefont {J.}~\bibnamefont {Knolle}},\
  and\ \bibinfo {author} {\bibfnamefont {W.}~\bibnamefont {Brenig}},\
  }\bibfield  {title} {\bibinfo {title} {Optical phonons coupled to a kitaev
  spin liquid},\ }\href {https://arxiv.org/abs/2103.09828} {\bibfield
  {journal} {\bibinfo  {journal} {arXiv preprint arXiv:2103.09828}\ } (\bibinfo
  {year} {2021})}\BibitemShut {NoStop}%
\bibitem [{\citenamefont {Perreault}\ \emph {et~al.}(2015)\citenamefont
  {Perreault}, \citenamefont {Knolle}, \citenamefont {Perkins},\ and\
  \citenamefont {Burnell}}]{PhysRevB.92.094439}%
  \BibitemOpen
  \bibfield  {author} {\bibinfo {author} {\bibfnamefont {B.}~\bibnamefont
  {Perreault}}, \bibinfo {author} {\bibfnamefont {J.}~\bibnamefont {Knolle}},
  \bibinfo {author} {\bibfnamefont {N.~B.}\ \bibnamefont {Perkins}},\ and\
  \bibinfo {author} {\bibfnamefont {F.~J.}\ \bibnamefont {Burnell}},\
  }\bibfield  {title} {\bibinfo {title} {Theory of raman response in
  three-dimensional kitaev spin liquids: Application to $\ensuremath{\beta}$-
  and $\ensuremath{\gamma}\ensuremath{-}{\mathrm{li}}_{2}{\mathrm{iro}}_{3}$
  compounds},\ }\href {https://doi.org/10.1103/PhysRevB.92.094439} {\bibfield
  {journal} {\bibinfo  {journal} {Phys. Rev. B}\ }\textbf {\bibinfo {volume}
  {92}},\ \bibinfo {pages} {094439} (\bibinfo {year} {2015})}\BibitemShut
  {NoStop}%
\bibitem [{\citenamefont {{Bramwell}}\ and\ \citenamefont
  {{Gingras}}(2001)}]{BramwellGingras2001}%
  \BibitemOpen
  \bibfield  {author} {\bibinfo {author} {\bibfnamefont {S.~T.}\ \bibnamefont
  {{Bramwell}}}\ and\ \bibinfo {author} {\bibfnamefont {M.~J.~P.}\ \bibnamefont
  {{Gingras}}},\ }\bibfield  {title} {\bibinfo {title} {{Spin Ice State in
  Frustrated Magnetic Pyrochlore Materials}},\ }\href
  {https://doi.org/10.1126/science.1064761} {\bibfield  {journal} {\bibinfo
  {journal} {Science}\ }\textbf {\bibinfo {volume} {294}},\ \bibinfo {pages}
  {1495} (\bibinfo {year} {2001})},\ \Eprint
  {https://arxiv.org/abs/cond-mat/0201427} {arXiv:cond-mat/0201427
  [cond-mat.dis-nn]} \BibitemShut {NoStop}%
\bibitem [{\citenamefont {Castelnovo}\ \emph {et~al.}(2012)\citenamefont
  {Castelnovo}, \citenamefont {Moessner},\ and\ \citenamefont
  {Sondhi}}]{castelnovo2012}%
  \BibitemOpen
  \bibfield  {author} {\bibinfo {author} {\bibfnamefont {C.}~\bibnamefont
  {Castelnovo}}, \bibinfo {author} {\bibfnamefont {R.}~\bibnamefont
  {Moessner}},\ and\ \bibinfo {author} {\bibfnamefont {S.}~\bibnamefont
  {Sondhi}},\ }\bibfield  {title} {\bibinfo {title} {Spin ice,
  fractionalization, and topological order},\ }\href
  {https://doi.org/10.1146/annurev-conmatphys-020911-125058} {\bibfield
  {journal} {\bibinfo  {journal} {Annu.\ Rev.\ Condens.\ Matter\ Phys.}\
  }\textbf {\bibinfo {volume} {3}},\ \bibinfo {pages} {35} (\bibinfo {year}
  {2012})}\BibitemShut {NoStop}%
\bibitem [{\citenamefont {Gingras}\ and\ \citenamefont
  {McClarty}(2014)}]{gingrass2014}%
  \BibitemOpen
  \bibfield  {author} {\bibinfo {author} {\bibfnamefont {M.~J.~P.}\
  \bibnamefont {Gingras}}\ and\ \bibinfo {author} {\bibfnamefont {P.~A.}\
  \bibnamefont {McClarty}},\ }\bibfield  {title} {\bibinfo {title} {Quantum
  spin ice: a search for gapless quantum spin liquids in pyrochlore magnets},\
  }\href {https://doi.org/10.1088/0034-4885/77/5/056501} {\bibfield  {journal}
  {\bibinfo  {journal} {Rep.\ Prog.\ Phys.}\ }\textbf {\bibinfo {volume}
  {77}},\ \bibinfo {pages} {056501} (\bibinfo {year} {2014})}\BibitemShut
  {NoStop}%
\bibitem [{\citenamefont {Rau}\ and\ \citenamefont {Gingras}(2019)}]{rau}%
  \BibitemOpen
  \bibfield  {author} {\bibinfo {author} {\bibfnamefont {J.~G.}\ \bibnamefont
  {Rau}}\ and\ \bibinfo {author} {\bibfnamefont {M.~J.~P.}\ \bibnamefont
  {Gingras}},\ }\bibfield  {title} {\bibinfo {title} {Frustrated quantum
  rare-earth pyrochlores},\ }\href
  {https://doi.org/10.1146/annurev-conmatphys-022317-110520} {\bibfield
  {journal} {\bibinfo  {journal} {Annu.\ Rev.\ Condens.\ Matter Phys.}\
  }\textbf {\bibinfo {volume} {10}},\ \bibinfo {pages} {357} (\bibinfo {year}
  {2019})}\BibitemShut {NoStop}%
\bibitem [{\citenamefont {Bramwell}\ and\ \citenamefont
  {Harris}(2020)}]{bramwell2020}%
  \BibitemOpen
  \bibfield  {author} {\bibinfo {author} {\bibfnamefont {S.~T.}\ \bibnamefont
  {Bramwell}}\ and\ \bibinfo {author} {\bibfnamefont {M.~J.}\ \bibnamefont
  {Harris}},\ }\bibfield  {title} {\bibinfo {title} {The history of spin ice},\
  }\href {https://doi.org/10.1088/1361-648X/ab8423} {\bibfield  {journal}
  {\bibinfo  {journal} {J.\ Phys.\ : \ Condens.\ Matter}\ }\textbf {\bibinfo
  {volume} {32}},\ \bibinfo {pages} {374010} (\bibinfo {year}
  {2020})}\BibitemShut {NoStop}%
\bibitem [{\citenamefont {Ramirez}\ \emph {et~al.}(1999)\citenamefont
  {Ramirez}, \citenamefont {Hayashi}, \citenamefont {Cava}, \citenamefont
  {Siddharthan},\ and\ \citenamefont {Shastry}}]{ramirez1999zero}%
  \BibitemOpen
  \bibfield  {author} {\bibinfo {author} {\bibfnamefont {A.~P.}\ \bibnamefont
  {Ramirez}}, \bibinfo {author} {\bibfnamefont {A.}~\bibnamefont {Hayashi}},
  \bibinfo {author} {\bibfnamefont {R.~J.}\ \bibnamefont {Cava}}, \bibinfo
  {author} {\bibfnamefont {R.}~\bibnamefont {Siddharthan}},\ and\ \bibinfo
  {author} {\bibfnamefont {B.}~\bibnamefont {Shastry}},\ }\bibfield  {title}
  {\bibinfo {title} {Zero-point entropy in ‘spin ice’},\ }\href
  {https://www.nature.com/articles/20619} {\bibfield  {journal} {\bibinfo
  {journal} {Nature}\ }\textbf {\bibinfo {volume} {399}},\ \bibinfo {pages}
  {333} (\bibinfo {year} {1999})}\BibitemShut {NoStop}%
\bibitem [{\citenamefont {Harris}\ \emph {et~al.}(1997)\citenamefont {Harris},
  \citenamefont {Bramwell}, \citenamefont {McMorrow}, \citenamefont {Zeiske},\
  and\ \citenamefont {Godfrey}}]{PhysRevLett.79.2554}%
  \BibitemOpen
  \bibfield  {author} {\bibinfo {author} {\bibfnamefont {M.~J.}\ \bibnamefont
  {Harris}}, \bibinfo {author} {\bibfnamefont {S.~T.}\ \bibnamefont
  {Bramwell}}, \bibinfo {author} {\bibfnamefont {D.~F.}\ \bibnamefont
  {McMorrow}}, \bibinfo {author} {\bibfnamefont {T.}~\bibnamefont {Zeiske}},\
  and\ \bibinfo {author} {\bibfnamefont {K.~W.}\ \bibnamefont {Godfrey}},\
  }\bibfield  {title} {\bibinfo {title} {Geometrical frustration in the
  ferromagnetic pyrochlore ${\mathrm{ho}}_{2}{\mathrm{ti}}_{2}{O}_{7}$},\
  }\href {https://doi.org/10.1103/PhysRevLett.79.2554} {\bibfield  {journal}
  {\bibinfo  {journal} {Phys. Rev. Lett.}\ }\textbf {\bibinfo {volume} {79}},\
  \bibinfo {pages} {2554} (\bibinfo {year} {1997})}\BibitemShut {NoStop}%
\bibitem [{\citenamefont {Erfanifam}\ \emph {et~al.}(2014)\citenamefont
  {Erfanifam}, \citenamefont {Zherlitsyn}, \citenamefont {Yasin}, \citenamefont
  {Skourski}, \citenamefont {Wosnitza}, \citenamefont {Zvyagin}, \citenamefont
  {McClarty}, \citenamefont {Moessner}, \citenamefont {Balakrishnan},\ and\
  \citenamefont {Petrenko}}]{erfanifam}%
  \BibitemOpen
  \bibfield  {author} {\bibinfo {author} {\bibfnamefont {S.}~\bibnamefont
  {Erfanifam}}, \bibinfo {author} {\bibfnamefont {S.}~\bibnamefont
  {Zherlitsyn}}, \bibinfo {author} {\bibfnamefont {S.}~\bibnamefont {Yasin}},
  \bibinfo {author} {\bibfnamefont {Y.}~\bibnamefont {Skourski}}, \bibinfo
  {author} {\bibfnamefont {J.}~\bibnamefont {Wosnitza}}, \bibinfo {author}
  {\bibfnamefont {A.~A.}\ \bibnamefont {Zvyagin}}, \bibinfo {author}
  {\bibfnamefont {P.}~\bibnamefont {McClarty}}, \bibinfo {author}
  {\bibfnamefont {R.}~\bibnamefont {Moessner}}, \bibinfo {author}
  {\bibfnamefont {G.}~\bibnamefont {Balakrishnan}},\ and\ \bibinfo {author}
  {\bibfnamefont {O.~A.}\ \bibnamefont {Petrenko}},\ }\bibfield  {title}
  {\bibinfo {title} {Ultrasonic investigations of the spin ices
  dy$_2$ti$_2$o$_7$ and ho$_2$ti$_2$o$_7$ in and out of equilibrium},\ }\href
  {https://doi.org/10.1103/PhysRevB.90.064409} {\bibfield  {journal} {\bibinfo
  {journal} {Phys.\ Rev.\ B}\ }\textbf {\bibinfo {volume} {90}},\ \bibinfo
  {pages} {064409} (\bibinfo {year} {2014})}\BibitemShut {NoStop}%
\bibitem [{\citenamefont {Bhattacharjee}\ \emph {et~al.}(2016)\citenamefont
  {Bhattacharjee}, \citenamefont {Erfanifam}, \citenamefont {Green},
  \citenamefont {Naumann}, \citenamefont {Wang}, \citenamefont {Granovsky},
  \citenamefont {Doerr}, \citenamefont {Wosnitza}, \citenamefont {Zvyagin},
  \citenamefont {Moessner}, \citenamefont {Maljuk}, \citenamefont {Wurmehl},
  \citenamefont {B̈uchner},\ and\ \citenamefont {Zherlitsyn}}]{yb2016}%
  \BibitemOpen
  \bibfield  {author} {\bibinfo {author} {\bibfnamefont {S.}~\bibnamefont
  {Bhattacharjee}}, \bibinfo {author} {\bibfnamefont {S.}~\bibnamefont
  {Erfanifam}}, \bibinfo {author} {\bibfnamefont {E.~L.}\ \bibnamefont
  {Green}}, \bibinfo {author} {\bibfnamefont {M.}~\bibnamefont {Naumann}},
  \bibinfo {author} {\bibfnamefont {Z.}~\bibnamefont {Wang}}, \bibinfo {author}
  {\bibfnamefont {S.}~\bibnamefont {Granovsky}}, \bibinfo {author}
  {\bibfnamefont {M.}~\bibnamefont {Doerr}}, \bibinfo {author} {\bibfnamefont
  {J.}~\bibnamefont {Wosnitza}}, \bibinfo {author} {\bibfnamefont {A.~A.}\
  \bibnamefont {Zvyagin}}, \bibinfo {author} {\bibfnamefont {R.}~\bibnamefont
  {Moessner}}, \bibinfo {author} {\bibfnamefont {A.}~\bibnamefont {Maljuk}},
  \bibinfo {author} {\bibfnamefont {S.}~\bibnamefont {Wurmehl}}, \bibinfo
  {author} {\bibfnamefont {B.}~\bibnamefont {B̈uchner}},\ and\ \bibinfo
  {author} {\bibfnamefont {S.}~\bibnamefont {Zherlitsyn}},\ }\bibfield  {title}
  {\bibinfo {title} {Acoustic signatures of the phases and phase transitions in
  yb$_2$ti$_2$o$_7$},\ }\href {https://doi.org/10.1103/PhysRevB.93.144412}
  {\bibfield  {journal} {\bibinfo  {journal} {Phys.\ Rev.\ B}\ }\textbf
  {\bibinfo {volume} {93}},\ \bibinfo {pages} {144412} (\bibinfo {year}
  {2016})}\BibitemShut {NoStop}%
\bibitem [{\citenamefont {Gardner}\ \emph {et~al.}(2010)\citenamefont
  {Gardner}, \citenamefont {Gingras},\ and\ \citenamefont
  {Greedan}}]{RevModPhys.82.53}%
  \BibitemOpen
  \bibfield  {author} {\bibinfo {author} {\bibfnamefont {J.~S.}\ \bibnamefont
  {Gardner}}, \bibinfo {author} {\bibfnamefont {M.~J.~P.}\ \bibnamefont
  {Gingras}},\ and\ \bibinfo {author} {\bibfnamefont {J.~E.}\ \bibnamefont
  {Greedan}},\ }\bibfield  {title} {\bibinfo {title} {Magnetic pyrochlore
  oxides},\ }\href {https://doi.org/10.1103/RevModPhys.82.53} {\bibfield
  {journal} {\bibinfo  {journal} {Rev. Mod. Phys.}\ }\textbf {\bibinfo {volume}
  {82}},\ \bibinfo {pages} {53} (\bibinfo {year} {2010})}\BibitemShut {NoStop}%
\bibitem [{\citenamefont {Gaudet}\ \emph {et~al.}(2016)\citenamefont {Gaudet},
  \citenamefont {Ross}, \citenamefont {Kermarrec}, \citenamefont {Butch},
  \citenamefont {Ehlers}, \citenamefont {Dabkowska},\ and\ \citenamefont
  {Gaulin}}]{PhysRevB.93.064406}%
  \BibitemOpen
  \bibfield  {author} {\bibinfo {author} {\bibfnamefont {J.}~\bibnamefont
  {Gaudet}}, \bibinfo {author} {\bibfnamefont {K.~A.}\ \bibnamefont {Ross}},
  \bibinfo {author} {\bibfnamefont {E.}~\bibnamefont {Kermarrec}}, \bibinfo
  {author} {\bibfnamefont {N.~P.}\ \bibnamefont {Butch}}, \bibinfo {author}
  {\bibfnamefont {G.}~\bibnamefont {Ehlers}}, \bibinfo {author} {\bibfnamefont
  {H.~A.}\ \bibnamefont {Dabkowska}},\ and\ \bibinfo {author} {\bibfnamefont
  {B.~D.}\ \bibnamefont {Gaulin}},\ }\bibfield  {title} {\bibinfo {title}
  {Gapless quantum excitations from an icelike splayed ferromagnetic ground
  state in stoichiometric
  ${\mathrm{yb}}_{2}{\mathrm{ti}}_{2}{\mathrm{o}}_{7}$},\ }\href
  {https://doi.org/10.1103/PhysRevB.93.064406} {\bibfield  {journal} {\bibinfo
  {journal} {Phys. Rev. B}\ }\textbf {\bibinfo {volume} {93}},\ \bibinfo
  {pages} {064406} (\bibinfo {year} {2016})}\BibitemShut {NoStop}%
\bibitem [{\citenamefont {Thompson}\ \emph {et~al.}(2017)\citenamefont
  {Thompson}, \citenamefont {McClarty}, \citenamefont {Prabhakaran},
  \citenamefont {Cabrera}, \citenamefont {Guidi},\ and\ \citenamefont
  {Coldea}}]{PhysRevLett.119.057203}%
  \BibitemOpen
  \bibfield  {author} {\bibinfo {author} {\bibfnamefont {J.~D.}\ \bibnamefont
  {Thompson}}, \bibinfo {author} {\bibfnamefont {P.~A.}\ \bibnamefont
  {McClarty}}, \bibinfo {author} {\bibfnamefont {D.}~\bibnamefont
  {Prabhakaran}}, \bibinfo {author} {\bibfnamefont {I.}~\bibnamefont
  {Cabrera}}, \bibinfo {author} {\bibfnamefont {T.}~\bibnamefont {Guidi}},\
  and\ \bibinfo {author} {\bibfnamefont {R.}~\bibnamefont {Coldea}},\
  }\bibfield  {title} {\bibinfo {title} {Quasiparticle breakdown and spin
  hamiltonian of the frustrated quantum pyrochlore
  ${\mathrm{yb}}_{2}{\mathrm{ti}}_{2}{\mathrm{o}}_{7}$ in a magnetic field},\
  }\href {https://doi.org/10.1103/PhysRevLett.119.057203} {\bibfield  {journal}
  {\bibinfo  {journal} {Phys. Rev. Lett.}\ }\textbf {\bibinfo {volume} {119}},\
  \bibinfo {pages} {057203} (\bibinfo {year} {2017})}\BibitemShut {NoStop}%
\bibitem [{\citenamefont {Ross}\ \emph {et~al.}(2011)\citenamefont {Ross},
  \citenamefont {Savary}, \citenamefont {Gaulin},\ and\ \citenamefont
  {Balents}}]{PhysRevX.1.021002}%
  \BibitemOpen
  \bibfield  {author} {\bibinfo {author} {\bibfnamefont {K.~A.}\ \bibnamefont
  {Ross}}, \bibinfo {author} {\bibfnamefont {L.}~\bibnamefont {Savary}},
  \bibinfo {author} {\bibfnamefont {B.~D.}\ \bibnamefont {Gaulin}},\ and\
  \bibinfo {author} {\bibfnamefont {L.}~\bibnamefont {Balents}},\ }\bibfield
  {title} {\bibinfo {title} {Quantum excitations in quantum spin ice},\ }\href
  {https://doi.org/10.1103/PhysRevX.1.021002} {\bibfield  {journal} {\bibinfo
  {journal} {Phys. Rev. X}\ }\textbf {\bibinfo {volume} {1}},\ \bibinfo {pages}
  {021002} (\bibinfo {year} {2011})}\BibitemShut {NoStop}%
\bibitem [{\citenamefont {Applegate}\ \emph {et~al.}(2012)\citenamefont
  {Applegate}, \citenamefont {Hayre}, \citenamefont {Singh}, \citenamefont
  {Lin}, \citenamefont {Day},\ and\ \citenamefont
  {Gingras}}]{PhysRevLett.109.097205}%
  \BibitemOpen
  \bibfield  {author} {\bibinfo {author} {\bibfnamefont {R.}~\bibnamefont
  {Applegate}}, \bibinfo {author} {\bibfnamefont {N.~R.}\ \bibnamefont
  {Hayre}}, \bibinfo {author} {\bibfnamefont {R.~R.~P.}\ \bibnamefont {Singh}},
  \bibinfo {author} {\bibfnamefont {T.}~\bibnamefont {Lin}}, \bibinfo {author}
  {\bibfnamefont {A.~G.~R.}\ \bibnamefont {Day}},\ and\ \bibinfo {author}
  {\bibfnamefont {M.~J.~P.}\ \bibnamefont {Gingras}},\ }\bibfield  {title}
  {\bibinfo {title} {Vindication of
  ${\mathrm{yb}}_{2}{\mathrm{ti}}_{2}{\mathrm{o}}_{7}$ as a model exchange
  quantum spin ice},\ }\href {https://doi.org/10.1103/PhysRevLett.109.097205}
  {\bibfield  {journal} {\bibinfo  {journal} {Phys. Rev. Lett.}\ }\textbf
  {\bibinfo {volume} {109}},\ \bibinfo {pages} {097205} (\bibinfo {year}
  {2012})}\BibitemShut {NoStop}%
\bibitem [{\citenamefont {Hayre}\ \emph {et~al.}(2013)\citenamefont {Hayre},
  \citenamefont {Ross}, \citenamefont {Applegate}, \citenamefont {Lin},
  \citenamefont {Singh}, \citenamefont {Gaulin},\ and\ \citenamefont
  {Gingras}}]{PhysRevB.87.184423}%
  \BibitemOpen
  \bibfield  {author} {\bibinfo {author} {\bibfnamefont {N.~R.}\ \bibnamefont
  {Hayre}}, \bibinfo {author} {\bibfnamefont {K.~A.}\ \bibnamefont {Ross}},
  \bibinfo {author} {\bibfnamefont {R.}~\bibnamefont {Applegate}}, \bibinfo
  {author} {\bibfnamefont {T.}~\bibnamefont {Lin}}, \bibinfo {author}
  {\bibfnamefont {R.~R.~P.}\ \bibnamefont {Singh}}, \bibinfo {author}
  {\bibfnamefont {B.~D.}\ \bibnamefont {Gaulin}},\ and\ \bibinfo {author}
  {\bibfnamefont {M.~J.~P.}\ \bibnamefont {Gingras}},\ }\bibfield  {title}
  {\bibinfo {title} {Thermodynamic properties of yb${}_{2}$ti${}_{2}$o${}_{7}$
  pyrochlore as a function of temperature and magnetic field: Validation of a
  quantum spin ice exchange hamiltonian},\ }\href
  {https://doi.org/10.1103/PhysRevB.87.184423} {\bibfield  {journal} {\bibinfo
  {journal} {Phys. Rev. B}\ }\textbf {\bibinfo {volume} {87}},\ \bibinfo
  {pages} {184423} (\bibinfo {year} {2013})}\BibitemShut {NoStop}%
\bibitem [{\citenamefont {Kimura}\ \emph {et~al.}(2013)\citenamefont {Kimura},
  \citenamefont {Nakatsuji}, \citenamefont {Wen}, \citenamefont {Broholm},
  \citenamefont {Stone}, \citenamefont {Nishibori},\ and\ \citenamefont
  {Sawa}}]{kimura}%
  \BibitemOpen
  \bibfield  {author} {\bibinfo {author} {\bibfnamefont {K.}~\bibnamefont
  {Kimura}}, \bibinfo {author} {\bibfnamefont {S.}~\bibnamefont {Nakatsuji}},
  \bibinfo {author} {\bibfnamefont {J.}~\bibnamefont {Wen}}, \bibinfo {author}
  {\bibfnamefont {C.}~\bibnamefont {Broholm}}, \bibinfo {author} {\bibfnamefont
  {M.}~\bibnamefont {Stone}}, \bibinfo {author} {\bibfnamefont
  {E.}~\bibnamefont {Nishibori}},\ and\ \bibinfo {author} {\bibfnamefont
  {H.}~\bibnamefont {Sawa}},\ }\bibfield  {title} {\bibinfo {title} {Quantum
  fluctuations in spin-ice-like pr$_2$zr$_2$o$_7$},\ }\href
  {https://www.nature.com/articles/ncomms2914} {\bibfield  {journal} {\bibinfo
  {journal} {Nat. Commun.}\ }\textbf {\bibinfo {volume} {4}} (\bibinfo {year}
  {2013})}\BibitemShut {NoStop}%
\bibitem [{\citenamefont {Sibille}\ \emph {et~al.}(2016)\citenamefont
  {Sibille}, \citenamefont {Lhotel}, \citenamefont {Hatnean}, \citenamefont
  {Balakrishnan}, \citenamefont {Fak}, \citenamefont {Gauthier}, \citenamefont
  {Fennell},\ and\ \citenamefont {Kenzelmann}}]{prhfo}%
  \BibitemOpen
  \bibfield  {author} {\bibinfo {author} {\bibfnamefont {R.}~\bibnamefont
  {Sibille}}, \bibinfo {author} {\bibfnamefont {E.}~\bibnamefont {Lhotel}},
  \bibinfo {author} {\bibfnamefont {M.~C.}\ \bibnamefont {Hatnean}}, \bibinfo
  {author} {\bibfnamefont {G.}~\bibnamefont {Balakrishnan}}, \bibinfo {author}
  {\bibfnamefont {B.}~\bibnamefont {Fak}}, \bibinfo {author} {\bibfnamefont
  {N.}~\bibnamefont {Gauthier}}, \bibinfo {author} {\bibfnamefont
  {T.}~\bibnamefont {Fennell}},\ and\ \bibinfo {author} {\bibfnamefont
  {M.}~\bibnamefont {Kenzelmann}},\ }\bibfield  {title} {\bibinfo {title}
  {Candidate quantum spin ice in the pyrochlore
  ${\mathrm{pr}}_{2}{\mathrm{hf}}_{2}{\mathrm{o}}_{7}$},\ }\href
  {https://doi.org/10.1103/PhysRevB.94.024436} {\bibfield  {journal} {\bibinfo
  {journal} {Phys. Rev. B}\ }\textbf {\bibinfo {volume} {94}},\ \bibinfo
  {pages} {024436} (\bibinfo {year} {2016})}\BibitemShut {NoStop}%
\bibitem [{\citenamefont {Princep}\ \emph {et~al.}(2015)\citenamefont
  {Princep}, \citenamefont {Walker}, \citenamefont {Adroja}, \citenamefont
  {Prabhakaran},\ and\ \citenamefont {T.}}]{tbtioprincep}%
  \BibitemOpen
  \bibfield  {author} {\bibinfo {author} {\bibfnamefont {A.~J.}\ \bibnamefont
  {Princep}}, \bibinfo {author} {\bibfnamefont {H.~C.}\ \bibnamefont {Walker}},
  \bibinfo {author} {\bibfnamefont {D.~T.}\ \bibnamefont {Adroja}}, \bibinfo
  {author} {\bibfnamefont {D.}~\bibnamefont {Prabhakaran}},\ and\ \bibinfo
  {author} {\bibfnamefont {B.~A.}\ \bibnamefont {T.}},\ }\bibfield  {title}
  {\bibinfo {title} {Crystal field states of ${\mathrm{tb}}^{3+}$ in the
  pyrochlore spin liquid ${\mathrm{tb}}_{2}{\mathrm{ti}}_{2}{\mathrm{o}}_{7}$
  from neutron spectroscopy},\ }\href
  {https://doi.org/10.1103/PhysRevB.91.224430} {\bibfield  {journal} {\bibinfo
  {journal} {Phys. Rev. B}\ }\textbf {\bibinfo {volume} {91}},\ \bibinfo
  {pages} {224430} (\bibinfo {year} {2015})}\BibitemShut {NoStop}%
\bibitem [{\citenamefont {Ruminy}\ \emph
  {et~al.}(2016{\natexlab{a}})\citenamefont {Ruminy}, \citenamefont
  {Pomjakushina}, \citenamefont {Iida}, \citenamefont {Kamazawa}, \citenamefont
  {Adroja}, \citenamefont {Stuhr},\ and\ \citenamefont {T.}}]{HoTbDy}%
  \BibitemOpen
  \bibfield  {author} {\bibinfo {author} {\bibfnamefont {M.}~\bibnamefont
  {Ruminy}}, \bibinfo {author} {\bibfnamefont {E.}~\bibnamefont
  {Pomjakushina}}, \bibinfo {author} {\bibfnamefont {K.}~\bibnamefont {Iida}},
  \bibinfo {author} {\bibfnamefont {K.}~\bibnamefont {Kamazawa}}, \bibinfo
  {author} {\bibfnamefont {D.~T.}\ \bibnamefont {Adroja}}, \bibinfo {author}
  {\bibfnamefont {U.}~\bibnamefont {Stuhr}},\ and\ \bibinfo {author}
  {\bibfnamefont {F.}~\bibnamefont {T.}},\ }\bibfield  {title} {\bibinfo
  {title} {Crystal-field parameters of the rare-earth pyrochlores
  ${\mathrm{r}}_{2}{\mathrm{ti}}_{2}{\mathrm{o}}_{7}$ (${\mathrm{r=tb}}$,
  ${\mathrm{dy}}$, and ${\mathrm{ho}}$)},\ }\href
  {https://doi.org/10.1103/PhysRevB.94.024430} {\bibfield  {journal} {\bibinfo
  {journal} {Phys. Rev. B}\ }\textbf {\bibinfo {volume} {94}},\ \bibinfo
  {pages} {024430} (\bibinfo {year} {2016}{\natexlab{a}})}\BibitemShut
  {NoStop}%
\bibitem [{\citenamefont {Fennell}\ \emph
  {et~al.}(2014{\natexlab{a}})\citenamefont {Fennell}, \citenamefont
  {Kenzelmann}, \citenamefont {Roessli}, \citenamefont {Mutka}, \citenamefont
  {Ollivier}, \citenamefont {Ruminy}, \citenamefont {Stuhr}, \citenamefont
  {Zaharko}, \citenamefont {Bovo}, \citenamefont {Cervellino}, \citenamefont
  {Haas},\ and\ \citenamefont {Cava}}]{ruminy2014}%
  \BibitemOpen
  \bibfield  {author} {\bibinfo {author} {\bibfnamefont {T.}~\bibnamefont
  {Fennell}}, \bibinfo {author} {\bibfnamefont {M.}~\bibnamefont {Kenzelmann}},
  \bibinfo {author} {\bibfnamefont {B.}~\bibnamefont {Roessli}}, \bibinfo
  {author} {\bibfnamefont {H.}~\bibnamefont {Mutka}}, \bibinfo {author}
  {\bibfnamefont {J.}~\bibnamefont {Ollivier}}, \bibinfo {author}
  {\bibfnamefont {M.}~\bibnamefont {Ruminy}}, \bibinfo {author} {\bibfnamefont
  {U.}~\bibnamefont {Stuhr}}, \bibinfo {author} {\bibfnamefont
  {O.}~\bibnamefont {Zaharko}}, \bibinfo {author} {\bibfnamefont
  {L.}~\bibnamefont {Bovo}}, \bibinfo {author} {\bibfnamefont {A.}~\bibnamefont
  {Cervellino}}, \bibinfo {author} {\bibfnamefont {M.~K.}\ \bibnamefont
  {Haas}},\ and\ \bibinfo {author} {\bibfnamefont {R.~J.}\ \bibnamefont
  {Cava}},\ }\bibfield  {title} {\bibinfo {title} {Magnetoelastic excitations
  in the pyrochlore spin liquid tb$_2$ti$_2$o$_7$},\ }\href
  {https://doi.org/10.1103/PhysRevLett.112.017203} {\bibfield  {journal}
  {\bibinfo  {journal} {Phys.\ Rev.\ Lett.}\ }\textbf {\bibinfo {volume}
  {112}},\ \bibinfo {pages} {017203} (\bibinfo {year}
  {2014}{\natexlab{a}})}\BibitemShut {NoStop}%
\bibitem [{\citenamefont {{Moessner}}\ and\ \citenamefont
  {{Chalker}}(1998)}]{MoessnerChalker98}%
  \BibitemOpen
  \bibfield  {author} {\bibinfo {author} {\bibfnamefont {R.}~\bibnamefont
  {{Moessner}}}\ and\ \bibinfo {author} {\bibfnamefont {J.~T.}\ \bibnamefont
  {{Chalker}}},\ }\bibfield  {title} {\bibinfo {title} {{Properties of a
  Classical Spin Liquid: The Heisenberg Pyrochlore Antiferromagnet}},\ }\href
  {https://doi.org/10.1103/PhysRevLett.80.2929} {\bibfield  {journal} {\bibinfo
   {journal} {\prl}\ }\textbf {\bibinfo {volume} {80}},\ \bibinfo {pages}
  {2929} (\bibinfo {year} {1998})},\ \Eprint
  {https://arxiv.org/abs/cond-mat/9712063} {arXiv:cond-mat/9712063
  [cond-mat.stat-mech]} \BibitemShut {NoStop}%
\bibitem [{\citenamefont {Chalker}(2015)}]{chalker}%
  \BibitemOpen
  \bibfield  {author} {\bibinfo {author} {\bibfnamefont {J.}~\bibnamefont
  {Chalker}},\ }\bibfield  {title} {\bibinfo {title} {Spin liquids and
  frustrated magnetism},\ }\href
  {http://topo-houches.pks.mpg.de/wp-content/uploads/2015/07/ChalkerFinal.pdf}
  {\bibfield  {journal} {\bibinfo  {journal} {Oxford University Press}\ }
  (\bibinfo {year} {2015})}\BibitemShut {NoStop}%
\bibitem [{\citenamefont {Henley}(2005)}]{PhysRevB.71.014424}%
  \BibitemOpen
  \bibfield  {author} {\bibinfo {author} {\bibfnamefont {C.~L.}\ \bibnamefont
  {Henley}},\ }\bibfield  {title} {\bibinfo {title} {Power-law spin
  correlations in pyrochlore antiferromagnets},\ }\href
  {https://doi.org/10.1103/PhysRevB.71.014424} {\bibfield  {journal} {\bibinfo
  {journal} {Phys. Rev. B}\ }\textbf {\bibinfo {volume} {71}},\ \bibinfo
  {pages} {014424} (\bibinfo {year} {2005})}\BibitemShut {NoStop}%
\bibitem [{\citenamefont {Henley}(2010)}]{henley2010coulomb}%
  \BibitemOpen
  \bibfield  {author} {\bibinfo {author} {\bibfnamefont {C.~L.}\ \bibnamefont
  {Henley}},\ }\bibfield  {title} {\bibinfo {title} {The “coulomb phase” in
  frustrated systems},\ }\href
  {https://www.annualreviews.org/doi/10.1146/annurev-conmatphys-070909-104138}
  {\bibfield  {journal} {\bibinfo  {journal} {Annu. Rev. Condens. Matter
  Phys.}\ }\textbf {\bibinfo {volume} {1}},\ \bibinfo {pages} {179} (\bibinfo
  {year} {2010})}\BibitemShut {NoStop}%
\bibitem [{\citenamefont {Hermele}\ \emph {et~al.}(2004)\citenamefont
  {Hermele}, \citenamefont {Fisher},\ and\ \citenamefont {Balents}}]{hermele}%
  \BibitemOpen
  \bibfield  {author} {\bibinfo {author} {\bibfnamefont {M.}~\bibnamefont
  {Hermele}}, \bibinfo {author} {\bibfnamefont {M.~P.~A.}\ \bibnamefont
  {Fisher}},\ and\ \bibinfo {author} {\bibfnamefont {L.}~\bibnamefont
  {Balents}},\ }\bibfield  {title} {\bibinfo {title} {Pyrochlore photons: The
  u(1) spin liquid in a s=1/2 three-dimensional frustrated magnet},\ }\href
  {https://doi.org/10.1103/PhysRevB.69.064404} {\bibfield  {journal} {\bibinfo
  {journal} {Phys.\ Rev.\ B}\ }\textbf {\bibinfo {volume} {69}},\ \bibinfo
  {pages} {064404} (\bibinfo {year} {2004})}\BibitemShut {NoStop}%
\bibitem [{\citenamefont {{Raman}}\ \emph {et~al.}(2005)\citenamefont
  {{Raman}}, \citenamefont {{Moessner}},\ and\ \citenamefont
  {{Sondhi}}}]{ramandimer2005}%
  \BibitemOpen
  \bibfield  {author} {\bibinfo {author} {\bibfnamefont {K.~S.}\ \bibnamefont
  {{Raman}}}, \bibinfo {author} {\bibfnamefont {R.}~\bibnamefont
  {{Moessner}}},\ and\ \bibinfo {author} {\bibfnamefont {S.~L.}\ \bibnamefont
  {{Sondhi}}},\ }\bibfield  {title} {\bibinfo {title} {{SU(2)-invariant spin-
  (1)/(2) Hamiltonians with resonating and other valence bond phases}},\ }\href
  {https://doi.org/10.1103/PhysRevB.72.064413} {\bibfield  {journal} {\bibinfo
  {journal} {\prb}\ }\textbf {\bibinfo {volume} {72}},\ \bibinfo {eid} {064413}
  (\bibinfo {year} {2005})},\ \Eprint {https://arxiv.org/abs/cond-mat/0502146}
  {arXiv:cond-mat/0502146 [cond-mat.str-el]} \BibitemShut {NoStop}%
\bibitem [{\citenamefont {Huse}\ \emph {et~al.}(2003)\citenamefont {Huse},
  \citenamefont {Krauth}, \citenamefont {Moessner},\ and\ \citenamefont
  {Sondhi}}]{PhysRevLett.91.167004}%
  \BibitemOpen
  \bibfield  {author} {\bibinfo {author} {\bibfnamefont {D.~A.}\ \bibnamefont
  {Huse}}, \bibinfo {author} {\bibfnamefont {W.}~\bibnamefont {Krauth}},
  \bibinfo {author} {\bibfnamefont {R.}~\bibnamefont {Moessner}},\ and\
  \bibinfo {author} {\bibfnamefont {S.~L.}\ \bibnamefont {Sondhi}},\ }\bibfield
   {title} {\bibinfo {title} {Coulomb and liquid dimer models in three
  dimensions},\ }\href {https://doi.org/10.1103/PhysRevLett.91.167004}
  {\bibfield  {journal} {\bibinfo  {journal} {Phys. Rev. Lett.}\ }\textbf
  {\bibinfo {volume} {91}},\ \bibinfo {pages} {167004} (\bibinfo {year}
  {2003})}\BibitemShut {NoStop}%
\bibitem [{\citenamefont {Savary}\ and\ \citenamefont
  {Balents}(2012)}]{savary}%
  \BibitemOpen
  \bibfield  {author} {\bibinfo {author} {\bibfnamefont {L.}~\bibnamefont
  {Savary}}\ and\ \bibinfo {author} {\bibfnamefont {L.}~\bibnamefont
  {Balents}},\ }\bibfield  {title} {\bibinfo {title} {Coulombic quantum liquids
  in spin-1/2 pyrochlores},\ }\href
  {https://doi.org/10.1103/PhysRevLett.108.037202} {\bibfield  {journal}
  {\bibinfo  {journal} {Phys.\ Rev.\ Lett.}\ }\textbf {\bibinfo {volume}
  {108}},\ \bibinfo {pages} {037202} (\bibinfo {year} {2012})}\BibitemShut
  {NoStop}%
\bibitem [{\citenamefont {Lee}\ \emph {et~al.}(2012)\citenamefont {Lee},
  \citenamefont {Onoda},\ and\ \citenamefont {Balents}}]{sungbin2012}%
  \BibitemOpen
  \bibfield  {author} {\bibinfo {author} {\bibfnamefont {S.}~\bibnamefont
  {Lee}}, \bibinfo {author} {\bibfnamefont {S.}~\bibnamefont {Onoda}},\ and\
  \bibinfo {author} {\bibfnamefont {L.}~\bibnamefont {Balents}},\ }\bibfield
  {title} {\bibinfo {title} {Generic quantum spin ice},\ }\href
  {https://doi.org/10.1103/PhysRevB.86.104412} {\bibfield  {journal} {\bibinfo
  {journal} {Phys.\ Rev.\ B}\ }\textbf {\bibinfo {volume} {86}},\ \bibinfo
  {pages} {104412} (\bibinfo {year} {2012})}\BibitemShut {NoStop}%
\bibitem [{\citenamefont {Shannon}\ \emph {et~al.}(2012)\citenamefont
  {Shannon}, \citenamefont {Sikora}, \citenamefont {Pollmann}, \citenamefont
  {Penc},\ and\ \citenamefont {Fulde}}]{PhysRevLett.108.067204}%
  \BibitemOpen
  \bibfield  {author} {\bibinfo {author} {\bibfnamefont {N.}~\bibnamefont
  {Shannon}}, \bibinfo {author} {\bibfnamefont {O.}~\bibnamefont {Sikora}},
  \bibinfo {author} {\bibfnamefont {F.}~\bibnamefont {Pollmann}}, \bibinfo
  {author} {\bibfnamefont {K.}~\bibnamefont {Penc}},\ and\ \bibinfo {author}
  {\bibfnamefont {P.}~\bibnamefont {Fulde}},\ }\bibfield  {title} {\bibinfo
  {title} {Quantum ice: A quantum monte carlo study},\ }\href
  {https://doi.org/10.1103/PhysRevLett.108.067204} {\bibfield  {journal}
  {\bibinfo  {journal} {Phys. Rev. Lett.}\ }\textbf {\bibinfo {volume} {108}},\
  \bibinfo {pages} {067204} (\bibinfo {year} {2012})}\BibitemShut {NoStop}%
\bibitem [{\citenamefont {Benton}\ \emph {et~al.}(2012)\citenamefont {Benton},
  \citenamefont {Sikora},\ and\ \citenamefont {Shannon}}]{benton}%
  \BibitemOpen
  \bibfield  {author} {\bibinfo {author} {\bibfnamefont {O.}~\bibnamefont
  {Benton}}, \bibinfo {author} {\bibfnamefont {O.}~\bibnamefont {Sikora}},\
  and\ \bibinfo {author} {\bibfnamefont {N.}~\bibnamefont {Shannon}},\
  }\bibfield  {title} {\bibinfo {title} {Seeing the light: Experimental
  signatures of emergent electromagnetism in a quantum spin ice},\ }\href
  {https://doi.org/10.1103/PhysRevB.86.075154} {\bibfield  {journal} {\bibinfo
  {journal} {Phys. Rev. B}\ }\textbf {\bibinfo {volume} {86}},\ \bibinfo
  {pages} {075154} (\bibinfo {year} {2012})}\BibitemShut {NoStop}%
\bibitem [{\citenamefont {Banerjee}\ \emph {et~al.}(2008)\citenamefont
  {Banerjee}, \citenamefont {Isakov}, \citenamefont {Damle},\ and\
  \citenamefont {Kim}}]{PhysRevLett.100.047208}%
  \BibitemOpen
  \bibfield  {author} {\bibinfo {author} {\bibfnamefont {A.}~\bibnamefont
  {Banerjee}}, \bibinfo {author} {\bibfnamefont {S.~V.}\ \bibnamefont
  {Isakov}}, \bibinfo {author} {\bibfnamefont {K.}~\bibnamefont {Damle}},\ and\
  \bibinfo {author} {\bibfnamefont {Y.~B.}\ \bibnamefont {Kim}},\ }\bibfield
  {title} {\bibinfo {title} {Unusual liquid state of hard-core bosons on the
  pyrochlore lattice},\ }\href {https://doi.org/10.1103/PhysRevLett.100.047208}
  {\bibfield  {journal} {\bibinfo  {journal} {Phys. Rev. Lett.}\ }\textbf
  {\bibinfo {volume} {100}},\ \bibinfo {pages} {047208} (\bibinfo {year}
  {2008})}\BibitemShut {NoStop}%
\bibitem [{\citenamefont {Huang}\ \emph {et~al.}(2014)\citenamefont {Huang},
  \citenamefont {Chen},\ and\ \citenamefont
  {Hermele}}]{PhysRevLett.112.167203}%
  \BibitemOpen
  \bibfield  {author} {\bibinfo {author} {\bibfnamefont {Y.-P.}\ \bibnamefont
  {Huang}}, \bibinfo {author} {\bibfnamefont {G.}~\bibnamefont {Chen}},\ and\
  \bibinfo {author} {\bibfnamefont {M.}~\bibnamefont {Hermele}},\ }\bibfield
  {title} {\bibinfo {title} {Quantum spin ices and topological phases from
  dipolar-octupolar doublets on the pyrochlore lattice},\ }\href
  {https://doi.org/10.1103/PhysRevLett.112.167203} {\bibfield  {journal}
  {\bibinfo  {journal} {Phys. Rev. Lett.}\ }\textbf {\bibinfo {volume} {112}},\
  \bibinfo {pages} {167203} (\bibinfo {year} {2014})}\BibitemShut {NoStop}%
\bibitem [{\citenamefont {Rau}\ and\ \citenamefont
  {Gingras}(2015)}]{PhysRevB.92.144417}%
  \BibitemOpen
  \bibfield  {author} {\bibinfo {author} {\bibfnamefont {J.~G.}\ \bibnamefont
  {Rau}}\ and\ \bibinfo {author} {\bibfnamefont {M.~J.~P.}\ \bibnamefont
  {Gingras}},\ }\bibfield  {title} {\bibinfo {title} {Magnitude of quantum
  effects in classical spin ices},\ }\href
  {https://doi.org/10.1103/PhysRevB.92.144417} {\bibfield  {journal} {\bibinfo
  {journal} {Phys. Rev. B}\ }\textbf {\bibinfo {volume} {92}},\ \bibinfo
  {pages} {144417} (\bibinfo {year} {2015})}\BibitemShut {NoStop}%
\bibitem [{\citenamefont {Chang}\ \emph {et~al.}(2012)\citenamefont {Chang},
  \citenamefont {Onoda}, \citenamefont {Su}, \citenamefont {Kao}, \citenamefont
  {Tsuei}, \citenamefont {Yasui}, \citenamefont {Kakurai},\ and\ \citenamefont
  {Lees}}]{chang2012higgs}%
  \BibitemOpen
  \bibfield  {author} {\bibinfo {author} {\bibfnamefont {L.-J.}\ \bibnamefont
  {Chang}}, \bibinfo {author} {\bibfnamefont {S.}~\bibnamefont {Onoda}},
  \bibinfo {author} {\bibfnamefont {Y.}~\bibnamefont {Su}}, \bibinfo {author}
  {\bibfnamefont {Y.-J.}\ \bibnamefont {Kao}}, \bibinfo {author} {\bibfnamefont
  {K.-D.}\ \bibnamefont {Tsuei}}, \bibinfo {author} {\bibfnamefont
  {Y.}~\bibnamefont {Yasui}}, \bibinfo {author} {\bibfnamefont
  {K.}~\bibnamefont {Kakurai}},\ and\ \bibinfo {author} {\bibfnamefont {M.~R.}\
  \bibnamefont {Lees}},\ }\bibfield  {title} {\bibinfo {title} {Higgs
  transition from a magnetic coulomb liquid to a ferromagnet in yb 2 ti 2 o
  7},\ }\href {https://www.nature.com/articles/ncomms1989} {\bibfield
  {journal} {\bibinfo  {journal} {Nat. Commun.}\ }\textbf {\bibinfo {volume}
  {3}} (\bibinfo {year} {2012})}\BibitemShut {NoStop}%
\bibitem [{\citenamefont {Kato}\ and\ \citenamefont
  {Onoda}(2015)}]{PhysRevLett.115.077202}%
  \BibitemOpen
  \bibfield  {author} {\bibinfo {author} {\bibfnamefont {Y.}~\bibnamefont
  {Kato}}\ and\ \bibinfo {author} {\bibfnamefont {S.}~\bibnamefont {Onoda}},\
  }\bibfield  {title} {\bibinfo {title} {Numerical evidence of quantum melting
  of spin ice: Quantum-to-classical crossover},\ }\href
  {https://doi.org/10.1103/PhysRevLett.115.077202} {\bibfield  {journal}
  {\bibinfo  {journal} {Phys. Rev. Lett.}\ }\textbf {\bibinfo {volume} {115}},\
  \bibinfo {pages} {077202} (\bibinfo {year} {2015})}\BibitemShut {NoStop}%
\bibitem [{\citenamefont {Petit}\ \emph
  {et~al.}(2016{\natexlab{a}})\citenamefont {Petit}, \citenamefont {Lhotel},
  \citenamefont {Guitteny}, \citenamefont {Florea}, \citenamefont {Robert},
  \citenamefont {Bonville}, \citenamefont {Mirebeau}, \citenamefont {Ollivier},
  \citenamefont {Mutka}, \citenamefont {Ressouche}, \citenamefont {Decorse},
  \citenamefont {Ciomaga~Hatnean},\ and\ \citenamefont
  {Balakrishnan}}]{PhysRevB.94.165153}%
  \BibitemOpen
  \bibfield  {author} {\bibinfo {author} {\bibfnamefont {S.}~\bibnamefont
  {Petit}}, \bibinfo {author} {\bibfnamefont {E.}~\bibnamefont {Lhotel}},
  \bibinfo {author} {\bibfnamefont {S.}~\bibnamefont {Guitteny}}, \bibinfo
  {author} {\bibfnamefont {O.}~\bibnamefont {Florea}}, \bibinfo {author}
  {\bibfnamefont {J.}~\bibnamefont {Robert}}, \bibinfo {author} {\bibfnamefont
  {P.}~\bibnamefont {Bonville}}, \bibinfo {author} {\bibfnamefont
  {I.}~\bibnamefont {Mirebeau}}, \bibinfo {author} {\bibfnamefont
  {J.}~\bibnamefont {Ollivier}}, \bibinfo {author} {\bibfnamefont
  {H.}~\bibnamefont {Mutka}}, \bibinfo {author} {\bibfnamefont
  {E.}~\bibnamefont {Ressouche}}, \bibinfo {author} {\bibfnamefont
  {C.}~\bibnamefont {Decorse}}, \bibinfo {author} {\bibfnamefont
  {M.}~\bibnamefont {Ciomaga~Hatnean}},\ and\ \bibinfo {author} {\bibfnamefont
  {G.}~\bibnamefont {Balakrishnan}},\ }\bibfield  {title} {\bibinfo {title}
  {Antiferroquadrupolar correlations in the quantum spin ice candidate
  ${\mathrm{pr}}_{2}{\mathrm{zr}}_{2}{\mathrm{o}}_{7}$},\ }\href
  {https://doi.org/10.1103/PhysRevB.94.165153} {\bibfield  {journal} {\bibinfo
  {journal} {Phys. Rev. B}\ }\textbf {\bibinfo {volume} {94}},\ \bibinfo
  {pages} {165153} (\bibinfo {year} {2016}{\natexlab{a}})}\BibitemShut
  {NoStop}%
\bibitem [{\citenamefont {Petit}\ \emph
  {et~al.}(2016{\natexlab{b}})\citenamefont {Petit}, \citenamefont {Lhotel},
  \citenamefont {Guitteny}, \citenamefont {Florea}, \citenamefont {Robert},
  \citenamefont {Bonville}, \citenamefont {Mirebeau}, \citenamefont {Ollivier},
  \citenamefont {Mutka}, \citenamefont {Ressouche}, \citenamefont {Decorse},
  \citenamefont {Hatnean},\ and\ \citenamefont {Balakrishnan}}]{PZO_doublet}%
  \BibitemOpen
  \bibfield  {author} {\bibinfo {author} {\bibfnamefont {S.}~\bibnamefont
  {Petit}}, \bibinfo {author} {\bibfnamefont {E.}~\bibnamefont {Lhotel}},
  \bibinfo {author} {\bibfnamefont {S.}~\bibnamefont {Guitteny}}, \bibinfo
  {author} {\bibfnamefont {O.}~\bibnamefont {Florea}}, \bibinfo {author}
  {\bibfnamefont {J.}~\bibnamefont {Robert}}, \bibinfo {author} {\bibfnamefont
  {P.}~\bibnamefont {Bonville}}, \bibinfo {author} {\bibfnamefont
  {I.}~\bibnamefont {Mirebeau}}, \bibinfo {author} {\bibfnamefont
  {J.}~\bibnamefont {Ollivier}}, \bibinfo {author} {\bibfnamefont
  {H.}~\bibnamefont {Mutka}}, \bibinfo {author} {\bibfnamefont
  {E.}~\bibnamefont {Ressouche}}, \bibinfo {author} {\bibfnamefont
  {C.}~\bibnamefont {Decorse}}, \bibinfo {author} {\bibfnamefont {M.~C.}\
  \bibnamefont {Hatnean}},\ and\ \bibinfo {author} {\bibfnamefont
  {G.}~\bibnamefont {Balakrishnan}},\ }\bibfield  {title} {\bibinfo {title}
  {Antiferroquadrupolar correlations in the quantum spin ice candidate
  pr$_2$zr$_2$o$_7$},\ }\href {https://doi.org/10.1103/PhysRevB.94.165153}
  {\bibfield  {journal} {\bibinfo  {journal} {Phys. Rev. B}\ }\textbf {\bibinfo
  {volume} {94}},\ \bibinfo {pages} {165153} (\bibinfo {year}
  {2016}{\natexlab{b}})}\BibitemShut {NoStop}%
\bibitem [{\citenamefont {Nakatsuji}()}]{Comsatoru}%
  \BibitemOpen
  \bibfield  {author} {\bibinfo {author} {\bibfnamefont {S.}~\bibnamefont
  {Nakatsuji}},\ }\href@noop {} {\bibinfo {title} {Private
  communications}}\BibitemShut {NoStop}%
\bibitem [{\citenamefont {Castelnovo}\ \emph {et~al.}(2008)\citenamefont
  {Castelnovo}, \citenamefont {Moessner},\ and\ \citenamefont
  {Sondhi}}]{sondhi}%
  \BibitemOpen
  \bibfield  {author} {\bibinfo {author} {\bibfnamefont {C.}~\bibnamefont
  {Castelnovo}}, \bibinfo {author} {\bibfnamefont {R.}~\bibnamefont
  {Moessner}},\ and\ \bibinfo {author} {\bibfnamefont {.~S.}\ \bibnamefont
  {Sondhi}},\ }\bibfield  {title} {\bibinfo {title} {Magnetic monopoles in spin
  ice},\ }\href {https://www.nature.com/articles/nature06433?proof=t}
  {\bibfield  {journal} {\bibinfo  {journal} {Nat. Commun.}\ }\textbf {\bibinfo
  {volume} {451}} (\bibinfo {year} {2008})}\BibitemShut {NoStop}%
\bibitem [{\citenamefont {Feynman}(1988)}]{feynman1988behavior}%
  \BibitemOpen
  \bibfield  {author} {\bibinfo {author} {\bibfnamefont {R.~P.}\ \bibnamefont
  {Feynman}},\ }\bibfield  {title} {\bibinfo {title} {The behavior of hadron
  collisions at extreme energies},\ }in\ \href@noop {} {\emph {\bibinfo
  {booktitle} {Special Relativity and Quantum Theory}}}\ (\bibinfo  {publisher}
  {Springer},\ \bibinfo {year} {1988})\ pp.\ \bibinfo {pages}
  {289--304}\BibitemShut {NoStop}%
\bibitem [{\citenamefont {Bjorken}\ and\ \citenamefont
  {Paschos}(1969)}]{PhysRev.185.1975}%
  \BibitemOpen
  \bibfield  {author} {\bibinfo {author} {\bibfnamefont {J.~D.}\ \bibnamefont
  {Bjorken}}\ and\ \bibinfo {author} {\bibfnamefont {E.~A.}\ \bibnamefont
  {Paschos}},\ }\bibfield  {title} {\bibinfo {title} {Inelastic electron-proton
  and $\ensuremath{\gamma}$-proton scattering and the structure of the
  nucleon},\ }\href {https://doi.org/10.1103/PhysRev.185.1975} {\bibfield
  {journal} {\bibinfo  {journal} {Phys. Rev.}\ }\textbf {\bibinfo {volume}
  {185}},\ \bibinfo {pages} {1975} (\bibinfo {year} {1969})}\BibitemShut
  {NoStop}%
\bibitem [{\citenamefont {Bloom}\ \emph {et~al.}(1969)\citenamefont {Bloom},
  \citenamefont {Coward}, \citenamefont {DeStaebler}, \citenamefont {Drees},
  \citenamefont {Miller}, \citenamefont {Mo}, \citenamefont {Taylor},
  \citenamefont {Breidenbach}, \citenamefont {Friedman}, \citenamefont
  {Hartmann},\ and\ \citenamefont {Kendall}}]{PhysRevLett.23.930}%
  \BibitemOpen
  \bibfield  {author} {\bibinfo {author} {\bibfnamefont {E.~D.}\ \bibnamefont
  {Bloom}}, \bibinfo {author} {\bibfnamefont {D.~H.}\ \bibnamefont {Coward}},
  \bibinfo {author} {\bibfnamefont {H.}~\bibnamefont {DeStaebler}}, \bibinfo
  {author} {\bibfnamefont {J.}~\bibnamefont {Drees}}, \bibinfo {author}
  {\bibfnamefont {G.}~\bibnamefont {Miller}}, \bibinfo {author} {\bibfnamefont
  {L.~W.}\ \bibnamefont {Mo}}, \bibinfo {author} {\bibfnamefont {R.~E.}\
  \bibnamefont {Taylor}}, \bibinfo {author} {\bibfnamefont {M.}~\bibnamefont
  {Breidenbach}}, \bibinfo {author} {\bibfnamefont {J.~I.}\ \bibnamefont
  {Friedman}}, \bibinfo {author} {\bibfnamefont {G.~C.}\ \bibnamefont
  {Hartmann}},\ and\ \bibinfo {author} {\bibfnamefont {H.~W.}\ \bibnamefont
  {Kendall}},\ }\bibfield  {title} {\bibinfo {title} {High-energy inelastic
  $e\ensuremath{-}p$ scattering at 6\ifmmode^\circ\else\textdegree\fi{} and
  10\ifmmode^\circ\else\textdegree\fi{}},\ }\href
  {https://doi.org/10.1103/PhysRevLett.23.930} {\bibfield  {journal} {\bibinfo
  {journal} {Phys. Rev. Lett.}\ }\textbf {\bibinfo {volume} {23}},\ \bibinfo
  {pages} {930} (\bibinfo {year} {1969})}\BibitemShut {NoStop}%
\bibitem [{\citenamefont {Breidenbach}\ \emph {et~al.}(1969)\citenamefont
  {Breidenbach}, \citenamefont {Friedman}, \citenamefont {Kendall},
  \citenamefont {Bloom}, \citenamefont {Coward}, \citenamefont {DeStaebler},
  \citenamefont {Drees}, \citenamefont {Mo},\ and\ \citenamefont
  {Taylor}}]{PhysRevLett.23.935}%
  \BibitemOpen
  \bibfield  {author} {\bibinfo {author} {\bibfnamefont {M.}~\bibnamefont
  {Breidenbach}}, \bibinfo {author} {\bibfnamefont {J.~I.}\ \bibnamefont
  {Friedman}}, \bibinfo {author} {\bibfnamefont {H.~W.}\ \bibnamefont
  {Kendall}}, \bibinfo {author} {\bibfnamefont {E.~D.}\ \bibnamefont {Bloom}},
  \bibinfo {author} {\bibfnamefont {D.~H.}\ \bibnamefont {Coward}}, \bibinfo
  {author} {\bibfnamefont {H.}~\bibnamefont {DeStaebler}}, \bibinfo {author}
  {\bibfnamefont {J.}~\bibnamefont {Drees}}, \bibinfo {author} {\bibfnamefont
  {L.~W.}\ \bibnamefont {Mo}},\ and\ \bibinfo {author} {\bibfnamefont {R.~E.}\
  \bibnamefont {Taylor}},\ }\bibfield  {title} {\bibinfo {title} {Observed
  behavior of highly inelastic electron-proton scattering},\ }\href
  {https://doi.org/10.1103/PhysRevLett.23.935} {\bibfield  {journal} {\bibinfo
  {journal} {Phys. Rev. Lett.}\ }\textbf {\bibinfo {volume} {23}},\ \bibinfo
  {pages} {935} (\bibinfo {year} {1969})}\BibitemShut {NoStop}%
\bibitem [{\citenamefont {Shastry}\ and\ \citenamefont
  {Shraiman}(1990)}]{shastry}%
  \BibitemOpen
  \bibfield  {author} {\bibinfo {author} {\bibfnamefont {B.~S.}\ \bibnamefont
  {Shastry}}\ and\ \bibinfo {author} {\bibfnamefont {B.~I.}\ \bibnamefont
  {Shraiman}},\ }\bibfield  {title} {\bibinfo {title} {Theory of raman
  scattering in mott-hubbard systems},\ }\href
  {https://doi.org/10.1103/PhysRevLett.65.1068} {\bibfield  {journal} {\bibinfo
   {journal} {Phys.\ Rev.\ Lett.}\ }\textbf {\bibinfo {volume} {65}},\ \bibinfo
  {pages} {1068} (\bibinfo {year} {1990})}\BibitemShut {NoStop}%
\bibitem [{\citenamefont {Knolle}\ \emph {et~al.}(2014)\citenamefont {Knolle},
  \citenamefont {Chern}, \citenamefont {Kovrizhin}, \citenamefont {Moessner},\
  and\ \citenamefont {Perkins}}]{PhysRevLett.113.187201}%
  \BibitemOpen
  \bibfield  {author} {\bibinfo {author} {\bibfnamefont {J.}~\bibnamefont
  {Knolle}}, \bibinfo {author} {\bibfnamefont {G.-W.}\ \bibnamefont {Chern}},
  \bibinfo {author} {\bibfnamefont {D.~L.}\ \bibnamefont {Kovrizhin}}, \bibinfo
  {author} {\bibfnamefont {R.}~\bibnamefont {Moessner}},\ and\ \bibinfo
  {author} {\bibfnamefont {N.~B.}\ \bibnamefont {Perkins}},\ }\bibfield
  {title} {\bibinfo {title} {Raman scattering signatures of kitaev spin liquids
  in ${A}_{2}{\mathrm{iro}}_{3}$ iridates with $a=\mathrm{Na}$ or li},\ }\href
  {https://doi.org/10.1103/PhysRevLett.113.187201} {\bibfield  {journal}
  {\bibinfo  {journal} {Phys. Rev. Lett.}\ }\textbf {\bibinfo {volume} {113}},\
  \bibinfo {pages} {187201} (\bibinfo {year} {2014})}\BibitemShut {NoStop}%
\bibitem [{\citenamefont {Fu}\ \emph {et~al.}(2017)\citenamefont {Fu},
  \citenamefont {Rau}, \citenamefont {Gingras},\ and\ \citenamefont
  {Perkins}}]{rau2017}%
  \BibitemOpen
  \bibfield  {author} {\bibinfo {author} {\bibfnamefont {J.}~\bibnamefont
  {Fu}}, \bibinfo {author} {\bibfnamefont {J.~G.}\ \bibnamefont {Rau}},
  \bibinfo {author} {\bibfnamefont {M.~J.~P.}\ \bibnamefont {Gingras}},\ and\
  \bibinfo {author} {\bibfnamefont {N.~B.}\ \bibnamefont {Perkins}},\
  }\bibfield  {title} {\bibinfo {title} {Fingerprints of quantum spin ice in
  raman scattering},\ }\href {https://doi.org/10.1103/PhysRevB.96.035136}
  {\bibfield  {journal} {\bibinfo  {journal} {Phys.\ Rev.\ B}\ }\textbf
  {\bibinfo {volume} {96}},\ \bibinfo {pages} {035136} (\bibinfo {year}
  {2017})}\BibitemShut {NoStop}%
\bibitem [{\citenamefont {Ko}\ \emph {et~al.}(2010)\citenamefont {Ko},
  \citenamefont {Liu}, \citenamefont {Ng},\ and\ \citenamefont
  {Lee}}]{PhysRevB.81.024414}%
  \BibitemOpen
  \bibfield  {author} {\bibinfo {author} {\bibfnamefont {W.-H.}\ \bibnamefont
  {Ko}}, \bibinfo {author} {\bibfnamefont {Z.-X.}\ \bibnamefont {Liu}},
  \bibinfo {author} {\bibfnamefont {T.-K.}\ \bibnamefont {Ng}},\ and\ \bibinfo
  {author} {\bibfnamefont {P.~A.}\ \bibnamefont {Lee}},\ }\bibfield  {title}
  {\bibinfo {title} {Raman signature of the u(1) dirac spin-liquid state in the
  spin-$\frac{1}{2}$ kagome system},\ }\href
  {https://doi.org/10.1103/PhysRevB.81.024414} {\bibfield  {journal} {\bibinfo
  {journal} {Phys. Rev. B}\ }\textbf {\bibinfo {volume} {81}},\ \bibinfo
  {pages} {024414} (\bibinfo {year} {2010})}\BibitemShut {NoStop}%
\bibitem [{\citenamefont {Fleury}\ and\ \citenamefont
  {Loudon}(1968)}]{PhysRev.166.514}%
  \BibitemOpen
  \bibfield  {author} {\bibinfo {author} {\bibfnamefont {P.~A.}\ \bibnamefont
  {Fleury}}\ and\ \bibinfo {author} {\bibfnamefont {R.}~\bibnamefont
  {Loudon}},\ }\bibfield  {title} {\bibinfo {title} {Scattering of light by
  one- and two-magnon excitations},\ }\href
  {https://doi.org/10.1103/PhysRev.166.514} {\bibfield  {journal} {\bibinfo
  {journal} {Phys. Rev.}\ }\textbf {\bibinfo {volume} {166}},\ \bibinfo {pages}
  {514} (\bibinfo {year} {1968})}\BibitemShut {NoStop}%
\bibitem [{\citenamefont {C\'epas}\ \emph {et~al.}(2008)\citenamefont
  {C\'epas}, \citenamefont {Haerter},\ and\ \citenamefont
  {Lhuillier}}]{PhysRevB.77.172406}%
  \BibitemOpen
  \bibfield  {author} {\bibinfo {author} {\bibfnamefont {O.}~\bibnamefont
  {C\'epas}}, \bibinfo {author} {\bibfnamefont {J.~O.}\ \bibnamefont
  {Haerter}},\ and\ \bibinfo {author} {\bibfnamefont {C.}~\bibnamefont
  {Lhuillier}},\ }\bibfield  {title} {\bibinfo {title} {Detection of weak
  emergent broken-symmetries of the kagome antiferromagnet by raman
  spectroscopy},\ }\href {https://doi.org/10.1103/PhysRevB.77.172406}
  {\bibfield  {journal} {\bibinfo  {journal} {Phys. Rev. B}\ }\textbf {\bibinfo
  {volume} {77}},\ \bibinfo {pages} {172406} (\bibinfo {year}
  {2008})}\BibitemShut {NoStop}%
\bibitem [{\citenamefont {Pace}\ \emph {et~al.}(2021)\citenamefont {Pace},
  \citenamefont {Morampudi}, \citenamefont {Moessner},\ and\ \citenamefont
  {Laumann}}]{PhysRevLett.127.117205}%
  \BibitemOpen
  \bibfield  {author} {\bibinfo {author} {\bibfnamefont {S.~D.}\ \bibnamefont
  {Pace}}, \bibinfo {author} {\bibfnamefont {S.~C.}\ \bibnamefont {Morampudi}},
  \bibinfo {author} {\bibfnamefont {R.}~\bibnamefont {Moessner}},\ and\
  \bibinfo {author} {\bibfnamefont {C.~R.}\ \bibnamefont {Laumann}},\
  }\bibfield  {title} {\bibinfo {title} {Emergent fine structure constant of
  quantum spin ice is large},\ }\href
  {https://doi.org/10.1103/PhysRevLett.127.117205} {\bibfield  {journal}
  {\bibinfo  {journal} {Phys. Rev. Lett.}\ }\textbf {\bibinfo {volume} {127}},\
  \bibinfo {pages} {117205} (\bibinfo {year} {2021})}\BibitemShut {NoStop}%
\bibitem [{\citenamefont {Ruminy}\ \emph
  {et~al.}(2016{\natexlab{b}})\citenamefont {Ruminy}, \citenamefont {Valdez},
  \citenamefont {Wehinger}, \citenamefont {Bosak}, \citenamefont {Adroja},
  \citenamefont {Stuhr}, \citenamefont {Iida}, \citenamefont {Kamazawa},
  \citenamefont {Pomjakushina}, \citenamefont {Prabakharan}, \citenamefont
  {Haas}, \citenamefont {Bovo}, \citenamefont {Sheptyakov}, \citenamefont
  {Cervellino}, \citenamefont {Cava}, \citenamefont {Kenzelmann}, \citenamefont
  {Spaldin}, ,\ and\ \citenamefont {Fennell}}]{ruminy2016}%
  \BibitemOpen
  \bibfield  {author} {\bibinfo {author} {\bibfnamefont {M.}~\bibnamefont
  {Ruminy}}, \bibinfo {author} {\bibfnamefont {M.~N.}\ \bibnamefont {Valdez}},
  \bibinfo {author} {\bibfnamefont {B.}~\bibnamefont {Wehinger}}, \bibinfo
  {author} {\bibfnamefont {A.}~\bibnamefont {Bosak}}, \bibinfo {author}
  {\bibfnamefont {D.~T.}\ \bibnamefont {Adroja}}, \bibinfo {author}
  {\bibfnamefont {U.}~\bibnamefont {Stuhr}}, \bibinfo {author} {\bibfnamefont
  {K.}~\bibnamefont {Iida}}, \bibinfo {author} {\bibfnamefont {K.}~\bibnamefont
  {Kamazawa}}, \bibinfo {author} {\bibfnamefont {E.}~\bibnamefont
  {Pomjakushina}}, \bibinfo {author} {\bibfnamefont {D.}~\bibnamefont
  {Prabakharan}}, \bibinfo {author} {\bibfnamefont {M.~K.}\ \bibnamefont
  {Haas}}, \bibinfo {author} {\bibfnamefont {L.}~\bibnamefont {Bovo}}, \bibinfo
  {author} {\bibfnamefont {D.}~\bibnamefont {Sheptyakov}}, \bibinfo {author}
  {\bibfnamefont {A.}~\bibnamefont {Cervellino}}, \bibinfo {author}
  {\bibfnamefont {R.~J.}\ \bibnamefont {Cava}}, \bibinfo {author}
  {\bibfnamefont {M.}~\bibnamefont {Kenzelmann}}, \bibinfo {author}
  {\bibfnamefont {N.~A.}\ \bibnamefont {Spaldin}}, ,\ and\ \bibinfo {author}
  {\bibfnamefont {T.}~\bibnamefont {Fennell}},\ }\bibfield  {title} {\bibinfo
  {title} {First-principles calculation and experimental investigation of
  lattice dynamics in the rare-earth pyrochlores r$2$ti$2$o$7$(r=tb,dy,ho)},\
  }\href {https://doi.org/10.1103/PhysRevB.93.214308} {\bibfield  {journal}
  {\bibinfo  {journal} {Phys. Rev. B}\ }\textbf {\bibinfo {volume} {93}},\
  \bibinfo {pages} {214308} (\bibinfo {year} {2016}{\natexlab{b}})}\BibitemShut
  {NoStop}%
\bibitem [{\citenamefont {Curnoe}(2008)}]{PhysRevB.78.094418}%
  \BibitemOpen
  \bibfield  {author} {\bibinfo {author} {\bibfnamefont {S.~H.}\ \bibnamefont
  {Curnoe}},\ }\bibfield  {title} {\bibinfo {title} {Structural distortion and
  the spin liquid state in ${\text{tb}}_{2}{\text{ti}}_{2}{\text{o}}_{7}$},\
  }\href {https://doi.org/10.1103/PhysRevB.78.094418} {\bibfield  {journal}
  {\bibinfo  {journal} {Phys. Rev. B}\ }\textbf {\bibinfo {volume} {78}},\
  \bibinfo {pages} {094418} (\bibinfo {year} {2008})}\BibitemShut {NoStop}%
\bibitem [{\citenamefont {Gardner}\ \emph {et~al.}(2001)\citenamefont
  {Gardner}, \citenamefont {Gaulin}, \citenamefont {Berlinsky}, \citenamefont
  {Waldron}, \citenamefont {Dunsiger}, \citenamefont {Raju},\ and\
  \citenamefont {Greedan}}]{PhysRevB.64.224416}%
  \BibitemOpen
  \bibfield  {author} {\bibinfo {author} {\bibfnamefont {J.~S.}\ \bibnamefont
  {Gardner}}, \bibinfo {author} {\bibfnamefont {B.~D.}\ \bibnamefont {Gaulin}},
  \bibinfo {author} {\bibfnamefont {A.~J.}\ \bibnamefont {Berlinsky}}, \bibinfo
  {author} {\bibfnamefont {P.}~\bibnamefont {Waldron}}, \bibinfo {author}
  {\bibfnamefont {S.~R.}\ \bibnamefont {Dunsiger}}, \bibinfo {author}
  {\bibfnamefont {N.~P.}\ \bibnamefont {Raju}},\ and\ \bibinfo {author}
  {\bibfnamefont {J.~E.}\ \bibnamefont {Greedan}},\ }\bibfield  {title}
  {\bibinfo {title} {Neutron scattering studies of the cooperative paramagnet
  pyrochlore ${\mathrm{tb}}_{2}{\mathrm{ti}}_{2}{\mathrm{o}}_{7}$},\ }\href
  {https://doi.org/10.1103/PhysRevB.64.224416} {\bibfield  {journal} {\bibinfo
  {journal} {Phys. Rev. B}\ }\textbf {\bibinfo {volume} {64}},\ \bibinfo
  {pages} {224416} (\bibinfo {year} {2001})}\BibitemShut {NoStop}%
\bibitem [{\citenamefont {Fennell}\ \emph
  {et~al.}(2014{\natexlab{b}})\citenamefont {Fennell}, \citenamefont
  {Kenzelmann}, \citenamefont {Roessli}, \citenamefont {Mutka}, \citenamefont
  {Ollivier}, \citenamefont {Ruminy}, \citenamefont {Stuhr}, \citenamefont
  {Zaharko}, \citenamefont {Bovo}, \citenamefont {Cervellino}, \citenamefont
  {Haas},\ and\ \citenamefont {Cava}}]{PhysRevLett.112.017203}%
  \BibitemOpen
  \bibfield  {author} {\bibinfo {author} {\bibfnamefont {T.}~\bibnamefont
  {Fennell}}, \bibinfo {author} {\bibfnamefont {M.}~\bibnamefont {Kenzelmann}},
  \bibinfo {author} {\bibfnamefont {B.}~\bibnamefont {Roessli}}, \bibinfo
  {author} {\bibfnamefont {H.}~\bibnamefont {Mutka}}, \bibinfo {author}
  {\bibfnamefont {J.}~\bibnamefont {Ollivier}}, \bibinfo {author}
  {\bibfnamefont {M.}~\bibnamefont {Ruminy}}, \bibinfo {author} {\bibfnamefont
  {U.}~\bibnamefont {Stuhr}}, \bibinfo {author} {\bibfnamefont
  {O.}~\bibnamefont {Zaharko}}, \bibinfo {author} {\bibfnamefont
  {L.}~\bibnamefont {Bovo}}, \bibinfo {author} {\bibfnamefont {A.}~\bibnamefont
  {Cervellino}}, \bibinfo {author} {\bibfnamefont {M.~K.}\ \bibnamefont
  {Haas}},\ and\ \bibinfo {author} {\bibfnamefont {R.~J.}\ \bibnamefont
  {Cava}},\ }\bibfield  {title} {\bibinfo {title} {Magnetoelastic excitations
  in the pyrochlore spin liquid
  ${\mathrm{tb}}_{2}{\mathrm{ti}}_{2}{\mathbf{o}}_{7}$},\ }\href
  {https://doi.org/10.1103/PhysRevLett.112.017203} {\bibfield  {journal}
  {\bibinfo  {journal} {Phys. Rev. Lett.}\ }\textbf {\bibinfo {volume} {112}},\
  \bibinfo {pages} {017203} (\bibinfo {year} {2014}{\natexlab{b}})}\BibitemShut
  {NoStop}%
\bibitem [{\citenamefont {Ruff}\ \emph {et~al.}(2007)\citenamefont {Ruff},
  \citenamefont {Gaulin}, \citenamefont {Castellan}, \citenamefont {Rule},
  \citenamefont {Clancy}, \citenamefont {Rodriguez},\ and\ \citenamefont
  {Dabkowska}}]{PhysRevLett.99.237202}%
  \BibitemOpen
  \bibfield  {author} {\bibinfo {author} {\bibfnamefont {J.~P.~C.}\
  \bibnamefont {Ruff}}, \bibinfo {author} {\bibfnamefont {B.~D.}\ \bibnamefont
  {Gaulin}}, \bibinfo {author} {\bibfnamefont {J.~P.}\ \bibnamefont
  {Castellan}}, \bibinfo {author} {\bibfnamefont {K.~C.}\ \bibnamefont {Rule}},
  \bibinfo {author} {\bibfnamefont {J.~P.}\ \bibnamefont {Clancy}}, \bibinfo
  {author} {\bibfnamefont {J.}~\bibnamefont {Rodriguez}},\ and\ \bibinfo
  {author} {\bibfnamefont {H.~A.}\ \bibnamefont {Dabkowska}},\ }\bibfield
  {title} {\bibinfo {title} {Structural fluctuations in the spin-liquid state
  of ${\mathrm{tb}}_{2}{\mathrm{ti}}_{2}{\mathrm{o}}_{7}$},\ }\href
  {https://doi.org/10.1103/PhysRevLett.99.237202} {\bibfield  {journal}
  {\bibinfo  {journal} {Phys. Rev. Lett.}\ }\textbf {\bibinfo {volume} {99}},\
  \bibinfo {pages} {237202} (\bibinfo {year} {2007})}\BibitemShut {NoStop}%
\bibitem [{\citenamefont {Gingras}\ \emph {et~al.}(2000)\citenamefont
  {Gingras}, \citenamefont {den Hertog}, \citenamefont {Faucher}, \citenamefont
  {Gardner}, \citenamefont {Dunsiger}, \citenamefont {Chang}, \citenamefont
  {Gaulin}, \citenamefont {Raju},\ and\ \citenamefont
  {Greedan}}]{PhysRevB.62.6496}%
  \BibitemOpen
  \bibfield  {author} {\bibinfo {author} {\bibfnamefont {M.~J.~P.}\
  \bibnamefont {Gingras}}, \bibinfo {author} {\bibfnamefont {B.~C.}\
  \bibnamefont {den Hertog}}, \bibinfo {author} {\bibfnamefont
  {M.}~\bibnamefont {Faucher}}, \bibinfo {author} {\bibfnamefont {J.~S.}\
  \bibnamefont {Gardner}}, \bibinfo {author} {\bibfnamefont {S.~R.}\
  \bibnamefont {Dunsiger}}, \bibinfo {author} {\bibfnamefont {L.~J.}\
  \bibnamefont {Chang}}, \bibinfo {author} {\bibfnamefont {B.~D.}\ \bibnamefont
  {Gaulin}}, \bibinfo {author} {\bibfnamefont {N.~P.}\ \bibnamefont {Raju}},\
  and\ \bibinfo {author} {\bibfnamefont {J.~E.}\ \bibnamefont {Greedan}},\
  }\bibfield  {title} {\bibinfo {title} {Thermodynamic and single-ion
  properties of ${\mathrm{tb}}^{3+}$ within the collective paramagnetic-spin
  liquid state of the frustrated pyrochlore antiferromagnet
  ${\mathrm{tb}}_{2}{\mathrm{ti}}_{2}{\mathrm{o}}_{7}$},\ }\href
  {https://doi.org/10.1103/PhysRevB.62.6496} {\bibfield  {journal} {\bibinfo
  {journal} {Phys. Rev. B}\ }\textbf {\bibinfo {volume} {62}},\ \bibinfo
  {pages} {6496} (\bibinfo {year} {2000})}\BibitemShut {NoStop}%
\bibitem [{\citenamefont {Gardner}\ \emph {et~al.}(1999)\citenamefont
  {Gardner}, \citenamefont {Dunsiger}, \citenamefont {Gaulin}, \citenamefont
  {Gingras}, \citenamefont {Greedan}, \citenamefont {Kiefl}, \citenamefont
  {Lumsden}, \citenamefont {MacFarlane}, \citenamefont {Raju}, \citenamefont
  {Sonier}, \citenamefont {Swainson},\ and\ \citenamefont
  {Tun}}]{PhysRevLett.82.1012}%
  \BibitemOpen
  \bibfield  {author} {\bibinfo {author} {\bibfnamefont {J.~S.}\ \bibnamefont
  {Gardner}}, \bibinfo {author} {\bibfnamefont {S.~R.}\ \bibnamefont
  {Dunsiger}}, \bibinfo {author} {\bibfnamefont {B.~D.}\ \bibnamefont
  {Gaulin}}, \bibinfo {author} {\bibfnamefont {M.~J.~P.}\ \bibnamefont
  {Gingras}}, \bibinfo {author} {\bibfnamefont {J.~E.}\ \bibnamefont
  {Greedan}}, \bibinfo {author} {\bibfnamefont {R.~F.}\ \bibnamefont {Kiefl}},
  \bibinfo {author} {\bibfnamefont {M.~D.}\ \bibnamefont {Lumsden}}, \bibinfo
  {author} {\bibfnamefont {W.~A.}\ \bibnamefont {MacFarlane}}, \bibinfo
  {author} {\bibfnamefont {N.~P.}\ \bibnamefont {Raju}}, \bibinfo {author}
  {\bibfnamefont {J.~E.}\ \bibnamefont {Sonier}}, \bibinfo {author}
  {\bibfnamefont {I.}~\bibnamefont {Swainson}},\ and\ \bibinfo {author}
  {\bibfnamefont {Z.}~\bibnamefont {Tun}},\ }\bibfield  {title} {\bibinfo
  {title} {Cooperative paramagnetism in the geometrically frustrated pyrochlore
  antiferromagnet ${\mathrm{tb}}_{2}{\mathrm{ti}}_{2}{\mathrm{o}}_{7}$},\
  }\href {https://doi.org/10.1103/PhysRevLett.82.1012} {\bibfield  {journal}
  {\bibinfo  {journal} {Phys. Rev. Lett.}\ }\textbf {\bibinfo {volume} {82}},\
  \bibinfo {pages} {1012} (\bibinfo {year} {1999})}\BibitemShut {NoStop}%
\bibitem [{\citenamefont {Wen}\ \emph {et~al.}(2017)\citenamefont {Wen},
  \citenamefont {Koohpayeh}, \citenamefont {Ross}, \citenamefont {Trump},
  \citenamefont {McQueen}, \citenamefont {Kimura}, \citenamefont {Nakatsuji},
  \citenamefont {Qiu}, \citenamefont {Pajerowski}, \citenamefont {Copley},\
  and\ \citenamefont {Broholm}}]{wen-disorder}%
  \BibitemOpen
  \bibfield  {author} {\bibinfo {author} {\bibfnamefont {J.-J.}\ \bibnamefont
  {Wen}}, \bibinfo {author} {\bibfnamefont {S.~M.}\ \bibnamefont {Koohpayeh}},
  \bibinfo {author} {\bibfnamefont {K.~A.}\ \bibnamefont {Ross}}, \bibinfo
  {author} {\bibfnamefont {B.~A.}\ \bibnamefont {Trump}}, \bibinfo {author}
  {\bibfnamefont {T.~M.}\ \bibnamefont {McQueen}}, \bibinfo {author}
  {\bibfnamefont {K.}~\bibnamefont {Kimura}}, \bibinfo {author} {\bibfnamefont
  {S.}~\bibnamefont {Nakatsuji}}, \bibinfo {author} {\bibfnamefont
  {Y.}~\bibnamefont {Qiu}}, \bibinfo {author} {\bibfnamefont {D.~M.}\
  \bibnamefont {Pajerowski}}, \bibinfo {author} {\bibfnamefont {J.~R.~D.}\
  \bibnamefont {Copley}},\ and\ \bibinfo {author} {\bibfnamefont {C.~L.}\
  \bibnamefont {Broholm}},\ }\bibfield  {title} {\bibinfo {title} {Disordered
  route to the coulomb quantum spin liquid: Random transverse fields on spin
  ice in pr$_2$zr$_2$o$_7$},\ }\href
  {https://doi.org/10.1103/PhysRevLett.118.107206} {\bibfield  {journal}
  {\bibinfo  {journal} {Phys.\ Rev.\ Lett}\ }\textbf {\bibinfo {volume}
  {118}},\ \bibinfo {pages} {107206} (\bibinfo {year} {2017})}\BibitemShut
  {NoStop}%
\bibitem [{\citenamefont {Princep}\ \emph {et~al.}(2013)\citenamefont
  {Princep}, \citenamefont {Prabhakaran}, \citenamefont {Boothroyd},\ and\
  \citenamefont {Adroja}}]{PhysRevB.88.104421}%
  \BibitemOpen
  \bibfield  {author} {\bibinfo {author} {\bibfnamefont {A.~J.}\ \bibnamefont
  {Princep}}, \bibinfo {author} {\bibfnamefont {D.}~\bibnamefont
  {Prabhakaran}}, \bibinfo {author} {\bibfnamefont {A.~T.}\ \bibnamefont
  {Boothroyd}},\ and\ \bibinfo {author} {\bibfnamefont {D.~T.}\ \bibnamefont
  {Adroja}},\ }\bibfield  {title} {\bibinfo {title} {Crystal-field states of
  pr${}^{3+}$ in the candidate quantum spin ice
  pr${}_{2}$sn${}_{2}$o${}_{7}$},\ }\href
  {https://doi.org/10.1103/PhysRevB.88.104421} {\bibfield  {journal} {\bibinfo
  {journal} {Phys. Rev. B}\ }\textbf {\bibinfo {volume} {88}},\ \bibinfo
  {pages} {104421} (\bibinfo {year} {2013})}\BibitemShut {NoStop}%
\bibitem [{\citenamefont {Mirebeau}\ \emph {et~al.}(2005)\citenamefont
  {Mirebeau}, \citenamefont {Apetrei}, \citenamefont {Rodr\'{\i}guez-Carvajal},
  \citenamefont {Bonville}, \citenamefont {Forget}, \citenamefont {Colson},
  \citenamefont {Glazkov}, \citenamefont {Sanchez}, \citenamefont {Isnard},\
  and\ \citenamefont {Suard}}]{PhysRevLett.94.246402}%
  \BibitemOpen
  \bibfield  {author} {\bibinfo {author} {\bibfnamefont {I.}~\bibnamefont
  {Mirebeau}}, \bibinfo {author} {\bibfnamefont {A.}~\bibnamefont {Apetrei}},
  \bibinfo {author} {\bibfnamefont {J.}~\bibnamefont
  {Rodr\'{\i}guez-Carvajal}}, \bibinfo {author} {\bibfnamefont
  {P.}~\bibnamefont {Bonville}}, \bibinfo {author} {\bibfnamefont
  {A.}~\bibnamefont {Forget}}, \bibinfo {author} {\bibfnamefont
  {D.}~\bibnamefont {Colson}}, \bibinfo {author} {\bibfnamefont
  {V.}~\bibnamefont {Glazkov}}, \bibinfo {author} {\bibfnamefont {J.~P.}\
  \bibnamefont {Sanchez}}, \bibinfo {author} {\bibfnamefont {O.}~\bibnamefont
  {Isnard}},\ and\ \bibinfo {author} {\bibfnamefont {E.}~\bibnamefont
  {Suard}},\ }\bibfield  {title} {\bibinfo {title} {Ordered spin ice state and
  magnetic fluctuations in
  ${\mathrm{tb}}_{2}{\mathrm{sn}}_{2}{\mathrm{o}}_{7}$},\ }\href
  {https://doi.org/10.1103/PhysRevLett.94.246402} {\bibfield  {journal}
  {\bibinfo  {journal} {Phys. Rev. Lett.}\ }\textbf {\bibinfo {volume} {94}},\
  \bibinfo {pages} {246402} (\bibinfo {year} {2005})}\BibitemShut {NoStop}%
\bibitem [{\citenamefont {Onoda}\ and\ \citenamefont
  {Tanaka}(2011)}]{PhysRevB.83.094411}%
  \BibitemOpen
  \bibfield  {author} {\bibinfo {author} {\bibfnamefont {S.}~\bibnamefont
  {Onoda}}\ and\ \bibinfo {author} {\bibfnamefont {Y.}~\bibnamefont {Tanaka}},\
  }\bibfield  {title} {\bibinfo {title} {Quantum fluctuations in the effective
  pseudospin-$\frac{1}{2}$ model for magnetic pyrochlore oxides},\ }\href
  {https://doi.org/10.1103/PhysRevB.83.094411} {\bibfield  {journal} {\bibinfo
  {journal} {Phys. Rev. B}\ }\textbf {\bibinfo {volume} {83}},\ \bibinfo
  {pages} {094411} (\bibinfo {year} {2011})}\BibitemShut {NoStop}%
\bibitem [{\citenamefont {Onoda}\ and\ \citenamefont
  {Tanaka}(2010)}]{PhysRevLett.105.047201}%
  \BibitemOpen
  \bibfield  {author} {\bibinfo {author} {\bibfnamefont {S.}~\bibnamefont
  {Onoda}}\ and\ \bibinfo {author} {\bibfnamefont {Y.}~\bibnamefont {Tanaka}},\
  }\bibfield  {title} {\bibinfo {title} {Quantum melting of spin ice: Emergent
  cooperative quadrupole and chirality},\ }\href
  {https://doi.org/10.1103/PhysRevLett.105.047201} {\bibfield  {journal}
  {\bibinfo  {journal} {Phys. Rev. Lett.}\ }\textbf {\bibinfo {volume} {105}},\
  \bibinfo {pages} {047201} (\bibinfo {year} {2010})}\BibitemShut {NoStop}%
\bibitem [{\citenamefont {Hao}\ \emph {et~al.}(2014)\citenamefont {Hao},
  \citenamefont {Day},\ and\ \citenamefont {Gingras}}]{PhysRevB.90.214430}%
  \BibitemOpen
  \bibfield  {author} {\bibinfo {author} {\bibfnamefont {Z.}~\bibnamefont
  {Hao}}, \bibinfo {author} {\bibfnamefont {A.~G.~R.}\ \bibnamefont {Day}},\
  and\ \bibinfo {author} {\bibfnamefont {M.~J.~P.}\ \bibnamefont {Gingras}},\
  }\bibfield  {title} {\bibinfo {title} {Bosonic many-body theory of quantum
  spin ice},\ }\href {https://doi.org/10.1103/PhysRevB.90.214430} {\bibfield
  {journal} {\bibinfo  {journal} {Phys. Rev. B}\ }\textbf {\bibinfo {volume}
  {90}},\ \bibinfo {pages} {214430} (\bibinfo {year} {2014})}\BibitemShut
  {NoStop}%
\bibitem [{\citenamefont {Patri}\ \emph {et~al.}(2020)\citenamefont {Patri},
  \citenamefont {Hosoi}, \citenamefont {Lee},\ and\ \citenamefont
  {Kim}}]{adarsh}%
  \BibitemOpen
  \bibfield  {author} {\bibinfo {author} {\bibfnamefont {A.~S.}\ \bibnamefont
  {Patri}}, \bibinfo {author} {\bibfnamefont {M.}~\bibnamefont {Hosoi}},
  \bibinfo {author} {\bibfnamefont {S.}~\bibnamefont {Lee}},\ and\ \bibinfo
  {author} {\bibfnamefont {Y.~B.}\ \bibnamefont {Kim}},\ }\href
  {https://doi.org/10.1103/PhysRevResearch.2.033015} {\bibfield  {journal}
  {\bibinfo  {journal} {Phys.\ Rev.\ Research}\ }\textbf {\bibinfo {volume}
  {3}},\ \bibinfo {pages} {033015} (\bibinfo {year} {2020})}\BibitemShut
  {NoStop}%
\bibitem [{\citenamefont {Seth}\ \emph {et~al.}()\citenamefont {Seth},
  \citenamefont {Bhattacharjee},\ and\ \citenamefont {Moessner}}]{j32}%
  \BibitemOpen
  \bibfield  {author} {\bibinfo {author} {\bibfnamefont {A.}~\bibnamefont
  {Seth}}, \bibinfo {author} {\bibfnamefont {S.}~\bibnamefont
  {Bhattacharjee}},\ and\ \bibinfo {author} {\bibfnamefont {R.}~\bibnamefont
  {Moessner}},\ }\bibinfo {title} {(unpublished)}\BibitemShut {NoStop}%
\bibitem [{\citenamefont {Matsuhira}\ \emph {et~al.}(2009)\citenamefont
  {Matsuhira}, \citenamefont {Sekine}, \citenamefont {Paulsen}, \citenamefont
  {Wakeshima}, \citenamefont {Hinatsu}, \citenamefont {Kitazawa}, \citenamefont
  {Kiuchi}, \citenamefont {Hiroi},\ and\ \citenamefont
  {Takagi}}]{matsuhira2009spin}%
  \BibitemOpen
\bibfield  {title} {  }\bibfield  {author} {\bibinfo {author} {\bibfnamefont
  {K.}~\bibnamefont {Matsuhira}}, \bibinfo {author} {\bibfnamefont
  {C.}~\bibnamefont {Sekine}}, \bibinfo {author} {\bibfnamefont
  {C.}~\bibnamefont {Paulsen}}, \bibinfo {author} {\bibfnamefont
  {M.}~\bibnamefont {Wakeshima}}, \bibinfo {author} {\bibfnamefont
  {Y.}~\bibnamefont {Hinatsu}}, \bibinfo {author} {\bibfnamefont
  {T.}~\bibnamefont {Kitazawa}}, \bibinfo {author} {\bibfnamefont
  {Y.}~\bibnamefont {Kiuchi}}, \bibinfo {author} {\bibfnamefont
  {Z.}~\bibnamefont {Hiroi}},\ and\ \bibinfo {author} {\bibfnamefont
  {S.}~\bibnamefont {Takagi}},\ }\bibfield  {title} {\bibinfo {title} {Spin
  freezing in the pyrochlore antiferromagnet pr2zr2o7},\ }in\ \href
  {https://iopscience.iop.org/article/10.1088/1742-6596/145/1/012031} {\emph
  {\bibinfo {booktitle} {J. Phys.: Conf. Ser.}}},\ Vol.\ \bibinfo {volume}
  {145}\ (\bibinfo {organization} {IOP Publishing},\ \bibinfo {year} {2009})\
  p.\ \bibinfo {pages} {012031}\BibitemShut {NoStop}%
\bibitem [{\citenamefont {Doron L.~Bergman}\ and\ \citenamefont
  {Balents}(2006)}]{doron}%
  \BibitemOpen
  \bibfield  {author} {\bibinfo {author} {\bibfnamefont {G.~A.~F.}\
  \bibnamefont {Doron L.~Bergman}}\ and\ \bibinfo {author} {\bibfnamefont
  {L.}~\bibnamefont {Balents}},\ }\bibfield  {title} {\bibinfo {title}
  {Ordering in a frustrated pyrochlore antiferromagnet proximate to a spin
  liquid},\ }\href {https://doi.org/10.1103/PhysRevB.73.134402} {\bibfield
  {journal} {\bibinfo  {journal} {Phys.\ Rev.\ B}\ }\textbf {\bibinfo {volume}
  {73}},\ \bibinfo {pages} {134402} (\bibinfo {year} {2006})}\BibitemShut
  {NoStop}%
\bibitem [{\citenamefont {Isakov}\ \emph {et~al.}(2005)\citenamefont {Isakov},
  \citenamefont {Moessner},\ and\ \citenamefont
  {Sondhi}}]{PhysRevLett.95.217201}%
  \BibitemOpen
  \bibfield  {author} {\bibinfo {author} {\bibfnamefont {S.~V.}\ \bibnamefont
  {Isakov}}, \bibinfo {author} {\bibfnamefont {R.}~\bibnamefont {Moessner}},\
  and\ \bibinfo {author} {\bibfnamefont {S.~L.}\ \bibnamefont {Sondhi}},\
  }\bibfield  {title} {\bibinfo {title} {Why spin ice obeys the ice rules},\
  }\href {https://doi.org/10.1103/PhysRevLett.95.217201} {\bibfield  {journal}
  {\bibinfo  {journal} {Phys. Rev. Lett.}\ }\textbf {\bibinfo {volume} {95}},\
  \bibinfo {pages} {217201} (\bibinfo {year} {2005})}\BibitemShut {NoStop}%
\bibitem [{\citenamefont {Motrunich}\ and\ \citenamefont
  {Senthil}(2005)}]{motrunich}%
  \BibitemOpen
  \bibfield  {author} {\bibinfo {author} {\bibfnamefont {O.~I.}\ \bibnamefont
  {Motrunich}}\ and\ \bibinfo {author} {\bibfnamefont {T.}~\bibnamefont
  {Senthil}},\ }\bibfield  {title} {\bibinfo {title} {Origin of artificial
  electrodynamics in three-dimensional bosonic models},\ }\href
  {https://doi.org/10.1103/PhysRevB.71.125102} {\bibfield  {journal} {\bibinfo
  {journal} {Phys.\ Rev.\ B}\ }\textbf {\bibinfo {volume} {71}},\ \bibinfo
  {pages} {125102} (\bibinfo {year} {2005})}\BibitemShut {NoStop}%
\bibitem [{\citenamefont {Chen}(2016)}]{gang2016}%
  \BibitemOpen
  \bibfield  {author} {\bibinfo {author} {\bibfnamefont {G.}~\bibnamefont
  {Chen}},\ }\bibfield  {title} {\bibinfo {title} {“magnetic monopole”
  condensation of the pyrochlore ice u(1) quantum spin liquid: Application to
  pr$_2$ir$_2$o$_7$ and yb$_2$ti$_2$o$_7$},\ }\href
  {https://doi.org/10.1103/PhysRevB.94.205107} {\bibfield  {journal} {\bibinfo
  {journal} {Phys.\ Rev.\ B}\ }\textbf {\bibinfo {volume} {94}},\ \bibinfo
  {pages} {205107} (\bibinfo {year} {2016})}\BibitemShut {NoStop}%
\bibitem [{\citenamefont {Étienne Lantagne-Hurtubise}\ \emph
  {et~al.}(2017)\citenamefont {Étienne Lantagne-Hurtubise}, \citenamefont
  {Bhattacharjee},\ and\ \citenamefont {Moessner}}]{etienne}%
  \BibitemOpen
  \bibfield  {author} {\bibinfo {author} {\bibnamefont {Étienne
  Lantagne-Hurtubise}}, \bibinfo {author} {\bibfnamefont {S.}~\bibnamefont
  {Bhattacharjee}},\ and\ \bibinfo {author} {\bibfnamefont {R.}~\bibnamefont
  {Moessner}},\ }\bibfield  {title} {\bibinfo {title} {Electric field control
  of emergent electrodynamics in quantum spin ice},\ }\href
  {https://doi.org/10.1103/PhysRevB.96.125145} {\bibfield  {journal} {\bibinfo
  {journal} {Phys. Rev. B}\ }\textbf {\bibinfo {volume} {96}},\ \bibinfo
  {pages} {125145} (\bibinfo {year} {2017})}\BibitemShut {NoStop}%
\bibitem [{\citenamefont {Porto}\ and\ \citenamefont {Wright}(nger)}]{porto}%
  \BibitemOpen
  \bibfield  {author} {\bibinfo {author} {\bibfnamefont {S.~P.~S.}\
  \bibnamefont {Porto}}\ and\ \bibinfo {author} {\bibfnamefont {G.~B.}\
  \bibnamefont {Wright}},\ }\href@noop {} {\emph {\bibinfo {title} {Light
  Scattering Spectra of Solids}}}\ (\bibinfo  {publisher} {Springer},\ \bibinfo
  {year} {1969, Springer})\BibitemShut {NoStop}%
\bibitem [{\citenamefont {Devereaux}\ and\ \citenamefont
  {Hackl}(2007)}]{devereaux}%
  \BibitemOpen
  \bibfield  {author} {\bibinfo {author} {\bibfnamefont {T.~P.}\ \bibnamefont
  {Devereaux}}\ and\ \bibinfo {author} {\bibfnamefont {R.}~\bibnamefont
  {Hackl}},\ }\bibfield  {title} {\bibinfo {title} {Inelastic light scattering
  from correlated electrons},\ }\href
  {https://doi.org/10.1103/RevModPhys.79.175} {\bibfield  {journal} {\bibinfo
  {journal} {Rev. Mod. Phys.}\ }\textbf {\bibinfo {volume} {79}},\ \bibinfo
  {pages} {175} (\bibinfo {year} {2007})}\BibitemShut {NoStop}%
\bibitem [{\citenamefont {Mahan}(2000)}]{mahan}%
  \BibitemOpen
  \bibfield  {author} {\bibinfo {author} {\bibfnamefont {G.~D.}\ \bibnamefont
  {Mahan}},\ }\href@noop {} {\emph {\bibinfo {title} {Many-Particle Physics}}}\
  (\bibinfo  {publisher} {Springer},\ \bibinfo {year} {2000})\BibitemShut
  {NoStop}%
\bibitem [{\citenamefont {Lee}\ and\ \citenamefont {Moon}(2019)}]{moon}%
  \BibitemOpen
  \bibfield  {author} {\bibinfo {author} {\bibfnamefont {S.}~\bibnamefont
  {Lee}}\ and\ \bibinfo {author} {\bibfnamefont {E.-G.}\ \bibnamefont {Moon}},\
  }\bibfield  {title} {\bibinfo {title} {Spin-lattice coupling in u(1) quantum
  spin liquids},\ }\href {https://doi.org/10.1103/PhysRevB.99.014412}
  {\bibfield  {journal} {\bibinfo  {journal} {Phys.\ Rev.\ B}\ }\textbf
  {\bibinfo {volume} {99}},\ \bibinfo {pages} {014412} (\bibinfo {year}
  {2019})}\BibitemShut {NoStop}%
\bibitem [{\citenamefont {Bhattacharjee}\ \emph {et~al.}(2011)\citenamefont
  {Bhattacharjee}, \citenamefont {Zherlitsyn}, \citenamefont {Chiatti},
  \citenamefont {Sytcheva}, \citenamefont {Wosnitza}, \citenamefont {Moessner},
  \citenamefont {Zhitomirsky}, \citenamefont {Lemmens}, \citenamefont
  {Tsurkan},\ and\ \citenamefont {Loidl}}]{cdcro}%
  \BibitemOpen
  \bibfield  {author} {\bibinfo {author} {\bibfnamefont {S.}~\bibnamefont
  {Bhattacharjee}}, \bibinfo {author} {\bibfnamefont {S.}~\bibnamefont
  {Zherlitsyn}}, \bibinfo {author} {\bibfnamefont {O.}~\bibnamefont {Chiatti}},
  \bibinfo {author} {\bibfnamefont {A.}~\bibnamefont {Sytcheva}}, \bibinfo
  {author} {\bibfnamefont {J.}~\bibnamefont {Wosnitza}}, \bibinfo {author}
  {\bibfnamefont {R.}~\bibnamefont {Moessner}}, \bibinfo {author}
  {\bibfnamefont {M.~E.}\ \bibnamefont {Zhitomirsky}}, \bibinfo {author}
  {\bibfnamefont {P.}~\bibnamefont {Lemmens}}, \bibinfo {author} {\bibfnamefont
  {V.}~\bibnamefont {Tsurkan}},\ and\ \bibinfo {author} {\bibfnamefont
  {A.}~\bibnamefont {Loidl}},\ }\bibfield  {title} {\bibinfo {title} {Interplay
  of spin and lattice degrees of freedom in the frustrated antiferromagnet
  cdcr$_2$o$_4$: High-field and temperature-induced anomalies of the elastic
  constants},\ }\href {https://doi.org/10.1103/PhysRevB.83.184421} {\bibfield
  {journal} {\bibinfo  {journal} {Phys. Rev. B}\ }\textbf {\bibinfo {volume}
  {83}},\ \bibinfo {pages} {184421} (\bibinfo {year} {2011})}\BibitemShut
  {NoStop}%
\end{thebibliography}%
%%%%%%%%%%%%%%%%%%%%%%%%%%%%
\end{document}